Florian Hebenstreit

# Schwinger effect in inhomogeneous electric fields

Dissertation

zur Erlangung des Doktorgrades der Naturwissenschaften

verfasst am Institut für Physik, Fachbereich Theoretische Physik
an der Karl-Franzens-Universität Graz

Betreuer: Univ.-Prof. Dr. rer. nat. Reinhard Alkofer

Graz, 2011

**Abstract**

The vacuum of quantum electrodynamics is unstable against the formation of many-body states in the presence of an external electric field, manifesting itself as the creation of electron-positron pairs (Schwinger effect). This effect has been a long-standing but still unobserved prediction as the generation of the required field strengths has not been feasible so far. However, due to the advent of a new generation of high-intensity laser systems such as the European XFEL or the Extreme Light Infrastructure (ELI), this effect might eventually become observable within the next decades.

Previous investigations of the Schwinger effect led to a good understanding of the general mechanisms behind the pair creation process, however, realistic electric fields as they might be present in upcoming high-intensity laser experiments have not been fully considered yet. Actually, it was only recently that it became possible to study the Schwinger effect in realistic electric fields showing both temporal and spatial variations owing to the theoretical progress as well as the rapid development of computer technology.

Based on the equal-time Wigner formalism, various aspects of the Schwinger effect in such inhomogeneous electric fields are investigated in this thesis. Regarding the Schwinger effect in time-dependent electric fields, analytic expressions for the equal-time Wigner function in the presence of a static as well as a pulsed electric field are derived. Moreover, the pair creation process in the presence of a pulsed electric field with sub-cycle structure, which acts as a model for a realistic laser pulse, is examined. Finally, an ab initio simulation of the Schwinger effect in a simple space- and time-dependent electric field is performed for the first time, allowing for the calculation of the time evolution of various observables like the charge density, the particle number density or the number of created particles.

# Contents

Chapter 1

# Introduction

Ever since the publication of 'Philosophiae Naturalis Principia Mathematica' by Sir Isaac Newton in 1687, people have started to believe that the world around us is governed by the action of forces. Nowadays, more than 300 years later, we are rather convinced that there are four fundamental forces in nature: The *electromagnetic* force, the *weak* force and the *strong* force are properly formulated in terms of quantum field theories whereas the *gravitational* force is still best described by the classical theory of general relativity.

The force of interest in this thesis is the electromagnetic force, which is well known from everyday life: Phenomena ranging from the radiation of light over all home and personal electronic equipment to high-tech devices in industry and research are based upon electromagnetic radiation and electric currents, most notably composed of electrons. It is sometimes appropriate to describe these phenomena by the classical theory of electromagnetism [1], however, a quantum description becomes essential as soon as phenomena such as laser emission are considered.

The quantum theory of electromagnetism is called *quantum electrodynamics* (QED) and has been developed during the first half of the 20th century [2, 3, 4], which finally culminated in the Physics Nobel Price[1] awarded to J. Schwinger, R. Feynman and S. Tomonaga in 1965. Since then, this theory has been tested experimentally to a very high precision in a regime which is characterized by high energies and low intensities. In this high-energy regime, which is realized in particle accelerator experiments, perturbation theory is a well-suited tool in order to calculate observables such as the anomalous magnetic moment of the electron [5, 6], which is one of the most precisely known physical constants nowadays.

There is yet another regime which is characterized by low energies and high intensities and, therefore, experimentally not accessible in accelerator experiments. This strong-field regime, however, might become accessible in the near future by the advent of a new generation of high-intensity laser systems. Physicists pin their hopes on two different technologies: X-ray free electron laser (XFEL) systems such

[1] cf. `http://nobelprize.org/nobel_prizes/physics/laureates/1965/`

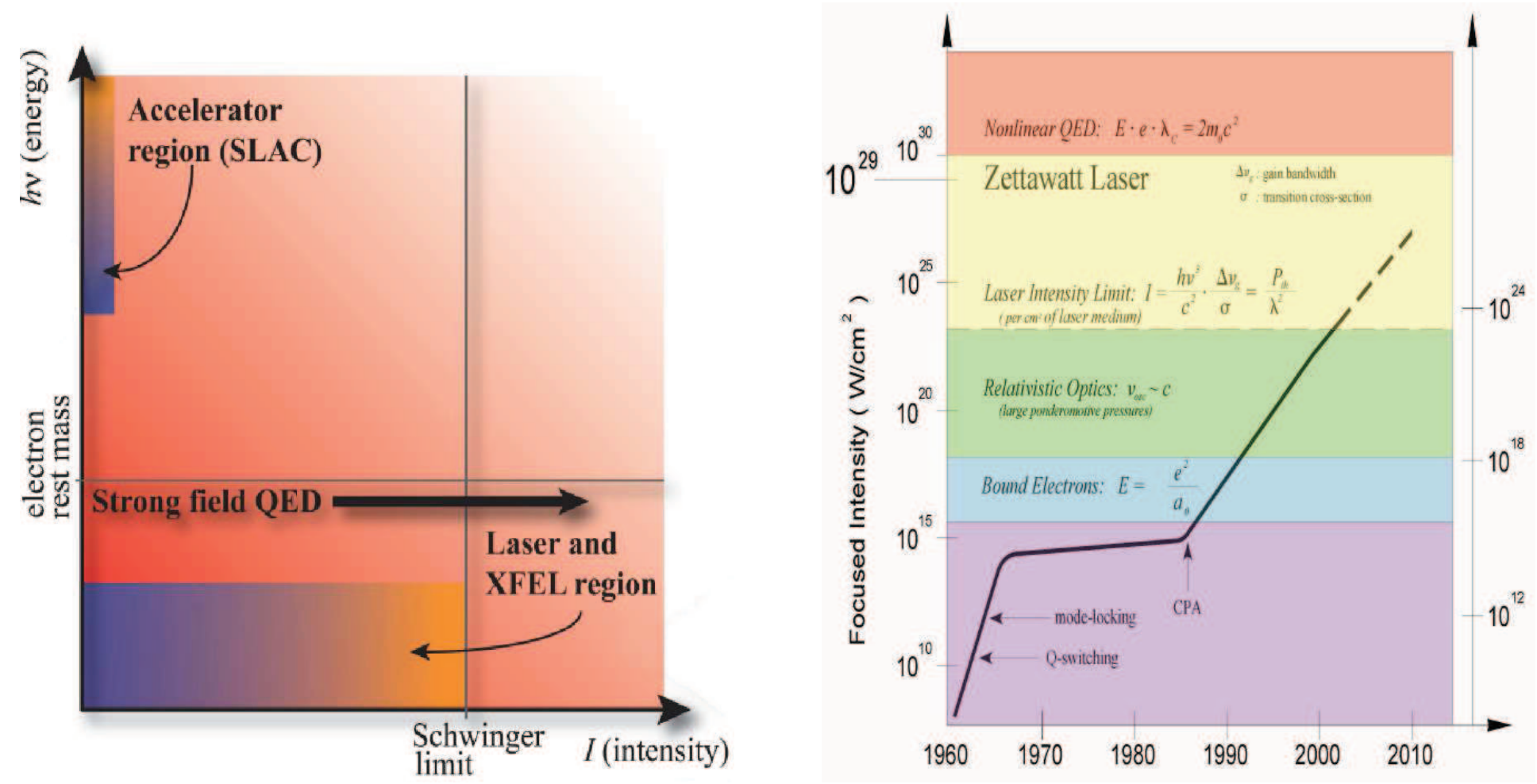

Figure 1.1: *Left:* Accelerator experiments are characterized by high energies but low intensities whereas laser-based experiments operate at low energies but high intensities. *Right:* The attainable intensity in optical high-intensity laser systems has continuously increased since the advent of the chirped pulse amplification (CPA) technology in 1985 [7]. The figures are taken from [8, 9].

as the European XFEL[2] or the Linac Coherent Light Source LCLS[3] on the one hand, and optical high-intensity laser systems such as the Extreme Light Infrastructure ELI[4] or the High Power laser Energy Research facility HiPER[5], which are based on chirped pulse amplification (CPA) technology [7], on the other hand.

The non-trivial and non-perturbative structure of the QED vacuum plays a crucial role in the strong-field regime: Due to the fact that quantum field theories allow for virtual particles being produced according to the energy-time uncertainty principle, the vacuum cannot be regarded as totally empty. In fact, the vacuum has to be considered as a polarizable medium with refractive and absorptive index. Consequences are rather minor in the high-energy regime: The vacuum is considered as the ground state of the system and observables are measured with reference to this state. In the strong-field regime, however, the non-trivial vacuum structure has a significant effect: Whilst the direct interaction between photons is forbidden by $U(1)$ gauge symmetry, photons might interact indirectly via vacuum fluctuations. Consequently, depending on the laser's frequency and intensity, various effects ranging from photon-photon scattering over vacuum birefringence to vacuum pair creation might arise. For recent reviews on various quantum vacuum experiments see [9, 10].

[2]cf. `http://www.xfel.eu/`
[3]cf. `http://lcls.slac.stanford.edu/`
[4]cf. `http://www.extreme-light-infrastructure.eu/`
[5]cf. `http://www.hiper-laser.org/`

In this thesis I focus on vacuum pair creation (Schwinger effect), which has been a long-standing but still unobserved prediction of QED [11, 12, 13]: The QED vacuum becomes unstable against the formation of many-body states in the presence of an external electric field, manifesting itself as the creation of electron-positron pairs. Subsequently, it turned out that the Schwinger effect shows two characteristic features which make its theoretical treatment both challenging and valuable in view of current problems in modern physics: On the one hand, the pair creation process is *non-perturbative* in the electromagnetic coupling so that QED perturbation theory fails. On the other hand, the system's time evolution is a *non-equilibrium* process and belongs as such to the lest-well understood branches of quantum field theory. Accordingly, a careful investigation of the Schwinger effect might entail important lessons for related phenomena such as the formation and thermalization of the so-called quark gluon plasma or facilitate the search for hidden sector particles [14, 15, 16].

In order to cope with this challenging non-equilibrium and non-perturbative problem, various methods have been invented, which might be roughly divided into two groups: On the one hand, there are various *semi-classical methods* such as WKB approximation [17, 18, 19, 20] or instanton techniques [21, 22, 23, 24]. These methods relate the imaginary part of the effective action for a given background field to the vacuum persistence amplitude, which in turn can be related to the pair creation rate [25, 26]. On the other hand, there are various *quantum kinetic methods* such as the quantum Vlasov equation [27, 28, 29, 30] or the Wigner function formalism [31, 32, 33], which account for both the pair creation and the subsequent transport process. In addition to the pair creation rate, quantum kinetic methods yield valuable phase-space information such as the momentum distribution.

Initially, the Schwinger effect has been considered for static electric fields [13] as well as simple time-dependent electric fields [17, 18]. Since then, various investigations have been conducted for more complicated time-dependent electric fields [34, 35, 36, 37, 38, 39, 40], space-dependent electric fields [22, 24, 25, 41, 42] as well as collinear electric and magnetic fields [43, 44, 45], resulting in a good understanding of the general mechanisms behind the pair creation process by now. However, realistic field configurations as they might be present in upcoming high-intensity laser experiments have not been fully considered yet. Most notably, there are two aspects which have not been properly taken into account so far: First, a realistic electric field shows both *spatial and temporal* dependence. Accordingly, a subtle interplay between the different scales might occur, resulting in a modification of both the pair creation and/or the transport behavior. Secondly, the *magnetic field* should be taken into consideration in a realistic manner as well.

This thesis is devoted to take a further step towards an accurate description of the Schwinger effect in upcoming high-intensity laser experiments. To this end, I focus on several aspects of the Schwinger effect in space- and time-dependent electric fields, most notably the interplay between spatial and temporal inhomogeneities. The Wigner function formalism and the closely related quantum Vlasov equation are adopted throughout as they provide adequate methods to study the influence of space- and time-dependent electric fields on the Schwinger effect.

The thesis is organized in the following way: In Chapter 2, I give a short overview of the Schwinger effect including the current state of research and the objectives for this thesis. The theoretical basis for the subsequent investigations is provided in Chapter 3 by reviewing both the covariant and the equal-time Wigner formalism. In Chapter 4, the Schwinger effect in time-dependent electric fields is discussed in some detail. First, I give a short review on the derivation of the quantum Vlasov equation and prove its equivalence to the equal-time Wigner formalism in the limit of a spatially homogeneous, time-dependent electric field. Afterwards, the analytic solution of the quantum Vlasov equation for the constant electric field as well as for the pulsed electric field is given. Moreover, the Schwinger effect in a short pulse with sub-cycle structure is studied. In Chapter 5, the Schwinger effect in space- and time-dependent electric fields is investigated. After introducing a $1+1$ dimensional model of QED, I present some solution strategies for the corresponding equation of motion. The remainder of this chapter is then devoted to the discussion of the results of an ab initio simulation of the Schwinger effect in a simple space- and time-dependent electric field. In the final Chapter 6, some concluding remarks as well as an outlook on potential future investigations are given.

The thesis concludes with two appendices: In Appendix A, the details of diverse derivations and calculations are given which have been deferred from the main text in order to improve readability. In Appendix B, the finite difference scheme for solving hyperbolic PDE systems is presented upon which the numerical simulations are based.

Chapter 2

# Overview and objectives

This chapter is dedicated to giving an overview of the major developments in the field of strong-field QED with a special focus on the Schwinger effect. In Section 2.1, I concentrate on the early milestones which have been the basis for many ensuing investigations. In the subsequent Section 2.2, I focus on quantum kinetic methods and briefly review some main results. In Section 2.3, I draw my attention to some current research topics in the field of vacuum pair creation. Finally, I formulate the objectives of this thesis in Section 2.4.

## 2.1 Schwinger effect

The history of the Schwinger effect started shortly after the invention of quantum mechanics in the mid 1920s. It was P. Dirac who first set up the relativistic wave equation for the electron in 1928 [46]. In the first instance, the negative energy solutions of the Dirac equation exhibited an interpretational difficulty: Why does a positive energy solution not decay into a negative energy solution by continuous emission of photons? This puzzle was first solved by the Dirac sea picture,[1] according to which the vacuum of the theory consists of a completely filled, negative energy band as well as an empty, positive energy band, which are separated by an energy gap of $2m$.[2] In this picture, a hole in the negative energy band is interpreted as the electron's antiparticle, the positron.[3]

Subsequently, the Dirac equation was solved in the presence of a static electric field $\mathbf{E}_0$ by F. Sauter in 1931 [11]. He found that the positive and negative energy bands are bend by the electric field so that a level crossing occurs. In the Dirac sea picture, this means that a negative energy electron is allowed to tunnel through the energy gap to the positive energy band, leaving a hole behind. In modern language, this result states that there is a non-vanishing probability for electron-positron pairs to be created in the presence of an electric field. As the pair creation process might be considered as tunneling process, it is exponentially suppressed though:

$$\mathcal{P}[e^+e^-] \sim \exp\left(-\frac{\pi m^2}{|e\mathbf{E}_0|}\right) , \tag{2.1}$$

with the scale set by the electron mass. Accordingly, electric fields of the order of the *critical field strength* $E_{\rm cr}$ are needed to overcome the exponential supression:

$$E_{\rm cr} = \frac{m^2}{e} \sim 10^{18}\, V/m \; . \tag{2.2}$$

[1] Note, that the Dirac sea picture is not applicable to relativistic theories of scalar particles, even though the negative energy problem exists there as well. Consequently, the notion of the Dirac sea should be dropped in favor of a quantum field theoretical treatment of the negative energy solutions.

[2] Natural units $\hbar = c = 1$ are used throughout the thesis.

[3] Since the positron has not been observed until 1932, it was the proton which was first interpreted as the electron's antiparticle.

These early investigations indicated that the Dirac equation is by no means a single-particle equation but should rather be treated in a quantum field theoretical framework in order to account for the possibility of particle creation. Accordingly, the vacuum should be considered as polarizable medium due to permanent vacuum fluctuations. As a consequence, the linear Maxwell's equations have to be replaced by more complicated *non-linear* equations.

This peculiarity has first been pointed out by W. Heisenberg and H. Euler in 1935 [12]. They derived an integral representation of the one-loop effective Lagrangian $\mathcal{L}^{(1)}_{\rm EH}$ which accounts for the coupling of a static electromagnetic field $F^{\mu\nu}$ to the electron vacuum loop. The leading order correction to Maxwell's Lagrangian is then given by quartic terms in the field strength tensor, resulting in non-linear corrections to Maxwell's equations:

$$\mathcal{L}^{(1)}_{EH} = -\mathcal{F} + \frac{e^4}{360\pi^2 m^4}[4\mathcal{F}^2 + 7\mathcal{G}^2] + \mathcal{O}(F^6) \; . \tag{2.3}$$

Here, $\mathcal{F} = \frac{1}{4}F^{\mu\nu}F_{\mu\nu} = \frac{1}{2}(\mathbf{B_0}^2 - \mathbf{E}_0^2)$ and $\mathcal{G} = \frac{1}{4}F^{\mu\nu}\tilde{F}_{\mu\nu} = -\mathbf{E}_0 \cdot \mathbf{B}_0$ denote the invariants of the electromagnetic field. Since then, the effective Lagrangian became a powerful tool in describing non-linear effects such as photon-photon scattering, vacuum birefringence or vacuum pair creation. Note for completeness that an integral representation of the two-loop effective Lagrangian $\mathcal{L}^{(2)}_{\rm EH}$, accounting for a photon correction to the electron vacuum loop, has been derived as well [47]. For a recent review on the topic of Euler-Heisenberg effective Lagrangians see [48].

In his seminal work, J. Schwinger was the first[4] to calculate the imaginary part of the one-loop effective Lagrangian in the presence of a static electric field $\mathbf{E}_0$ exactly [13]. Since then, vacuum pair creation by electric fields has been widely referred to as the *Schwinger effect*:

$$2\,\mathrm{Im}\,\mathcal{L}^{(1)}_{\mathrm{EH}} = \frac{|e\mathbf{E}_0|^2}{4\pi^3}\sum_{n=1}^{\infty}\frac{1}{n^2}\exp\left(-\frac{n\pi m^2}{|e\mathbf{E}_0|}\right)\,. \tag{2.4}$$

[4] Actually, it had been pointed out by W. Heisenberg and H. Euler that the imaginary part of the effective action could be calculated analytically by taking the pole sum in their expression, however, they did not give the final result [12].

This expression has often been identified with $\dot{\mathcal{N}}[e^+e^-]$, that is the rate at which electron-positron pairs are created in the presence of a static electric field $\mathbf{E}_0$.[5] This statement, however, is not totally correct as this expression should rather be identified with $\mathcal{P}$[vac.], that is the rate at which the vacuum decays. In fact, the vacuum decay rate is always larger than the pair creation rate, which is given by the first term in this series only [25, 26]:

$$\dot{\mathcal{N}}[e^+e^-] = \frac{|e\mathbf{E}_0|^2}{4\pi^3}\exp\left(-\frac{\pi m^2}{|e\mathbf{E}_0|}\right)\,. \tag{2.5}$$

As a matter of fact, the difference between Eq. (2.4) and Eq. (2.5) is rather minor in the sub-critical field strength regime whereas it becomes important for super-critical fields. Note for completeness that Eq. (2.5) can be generalized for arbitrary spins and space dimensions [49].

Subsequently, there has been further research on the Schwinger effect in time-dependent electric fields $\mathbf{E}(t)$ as well. These investigations have been largely motivated by studies on atomic ionization in either static [50] or sinusoidal [51] electric fields. There is indeed a close analogy in spirit between vacuum pair creation and atomic ionization: Most notably, both of them can be described as tunneling process which is facilitated by an applied electric field.[6] Accordingly, methods and notions which were used in the theoretical treatment of atomic ionization have also been applied in the investigation of the Schwinger effect.

Atomic ionization in a sinusoidal electric field $\mathbf{E}(t) = \mathbf{E}_0\sin(\omega t)$ is one prominent example: This is, in fact, a multiple time scale problem, with the relevant scales given by the frequency $\omega$ as well as the tunneling frequency $\omega_T$, which is related to the characteristic time of a tunneling event. The *Keldysh adiabaticity parameter* $\gamma$

[5] Even J. Schwinger referred to this expression as the pair creation rate [13].

[6] Amongst others, this can be seen from the expressions for the pair creation and ionization rate, respectively. The main difference is the scale, which is orders of magnitudes smaller for atomic ionization, whereas the general structure is the same ($\mathcal{E}_b$ denotes the binding energy):

$$\dot{\mathcal{N}}[e^+e^-] \sim \exp\left(-\frac{\pi m^2}{|e\mathbf{E}_0|}\right) \qquad \text{and} \qquad \dot{\mathcal{N}}[\text{ion.}] \sim \exp\left(-\frac{4}{3}\frac{\sqrt{2m\mathcal{E}_b^3}}{|e\mathbf{E}_0|}\right)$$

then determines the dominant scale of the problem:

$$\gamma = \frac{\omega}{\omega_T} = \frac{\omega\sqrt{2m\mathcal{E}_b}}{|e\mathbf{E}_0|} \,, \tag{2.6}$$

with $\mathcal{E}_b$ denoting the binding energy. For $\gamma \ll 1$, the problem is dominated by $\omega_T$ so that the electric field might be well considered in an instantaneous approximation. As the ionization is due to a tunneling event, this type has been termed instantaneous or non-perturbative ionization. For $\gamma \gg 1$, however, the temporal variation of the electric field is too quick so that no tunneling event is possible. In this case, the ionization is due to the absorption of a certain number of photons so that this effect has been named multi-photon ionization.

The notion of the Keldysh parameter has been adopted in strong-field QED as well. Again, $\gamma$ measures the ratio between the characteristic frequency of the electric field $\omega$ and the characteristic tunneling frequency $\omega_T$:

$$\gamma = \frac{\omega}{\omega_T} = \frac{m\omega}{|e\mathbf{E}_0|} \,. \tag{2.7}$$

As for atomic ionization, the non-perturbative regime is characterised by $\gamma \ll 1$ whereas $\gamma \gg 1$ denotes the multi-photon regime.[7] As a matter of fact, an analytic expression for the vacuum decay rate has been derived which interpolates between those two regimes [17, 19]. Considering the pair creation rate which is in the subcritical field strength regime again very close to the vacuum decay rate, one obtains:

$$\dot{\mathcal{N}}[e^+e^-] \simeq \begin{cases} \exp\left(-\frac{\pi m^2}{|e\mathbf{E}_0|}\right) & \gamma \ll 1 \quad \text{(instantaneous)} \,, \\ \left(\frac{|e\mathbf{E}_0|}{2m\omega}\right)^{4m/\omega} & \gamma \gg 1 \quad \text{(dynamical)} \,. \end{cases} \tag{2.8}$$

This result nicely illustrates the meaning of the Keldysh parameter: On the one hand, the characteristic exponential supression of a tunneling event is found for $\gamma \ll 1$. On the other hand, the result for $\gamma \gg 1$ corresponds to n-th order perturbation theory, with $n$ being the minmum number of photons to be absorbed in order to overcome the threshold energy for pair creation $n\omega \geq 2m$.

Moreover, the pulsed electric field $\mathbf{E}(t) = \mathbf{E}_0 \operatorname{sech}^2(\frac{t}{\tau})$, with $\tau$ being a measure for the pulse length, has been of special interest as it describes an electric field which is only switched on for a finite time period [18]. Again, an analytic expression for the vacuum decay rate in terms of an integral representation has been derived for this type of electric field. I will return to this issue in Section 4.3.

[7] In the context of pair creation, this effect is also termed dynamical pair creation.

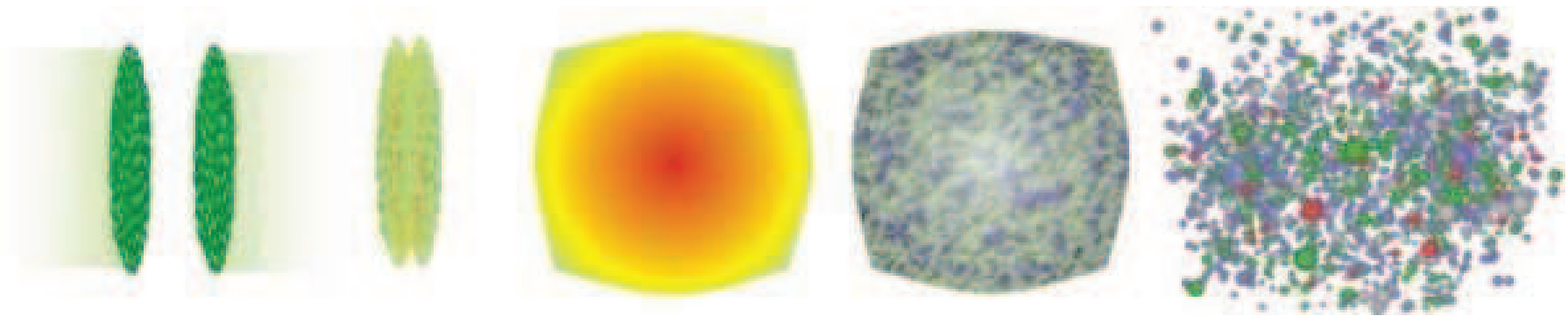

Figure 2.1: Schematic representation of a heavy ion collision. From left to right: Approach of two nuclei; heavy ion collision; formation of the QGP; expansion of the QGP and hadronization; hadronic freeze out. The figure is taken from `http://www.phy.duke.edu/research/NPTheory/QGP/transport/`

## 2.2 Quantum kinetic methods

The Schwinger effect attracted much attention not only in the field of strong-field QED but also in other research areas. One important example is the theoretical description of the pre-equilibrium evolution of the so-called quark gluon plasma (QGP) in ultrarelativistic heavy ion collision at the Relativistic Heavy Ion Collider RHIC[8] or at the Large Hadron Collider LHC[9]. In order to describe the formation of the QGP, one prominent model considers the colliding nuclei as passing through each other, creating a chromoelectric flux tube which breaks via the production of quark-antiquark pairs [52]. Even though this phenomenologically motivated description bears a number of shortcomings, most notably it is by no means clear that a non-Abelian chromoelectric field can be treated on the same ground as an Abelian electric field [53], much effort has been made to find a quantum kinetic description of non-equilibrium systems showing the phenomenon of particle creation.

### 2.2.1 Quantum Vlasov equation

Kinetic theory has been very successful in describing transport phenomena in non-relativistic theories such as fluid dynamics as well as in relativistic applications such as plasma physics. Even the generalization to quantum systems can be done consistently. For a review on kinetic theory see [54].

The kinetic description of non-equilibrium systems is based on the *Boltzmann equation*, which describes the time evolution of the one-particle distribution function $\mathcal{F}(\mathbf{x}, \mathbf{p}, t)$. In general, this equation reads:

$$\dot{\mathcal{F}}(\mathbf{x}, \mathbf{p}, t) = \left[\tfrac{\partial}{\partial t} + \dot{\mathbf{x}} \cdot \nabla_{\mathbf{x}} + \dot{\mathbf{p}} \cdot \nabla_{\mathbf{p}}\right] \mathcal{F}(\mathbf{x}, \mathbf{p}, t) = \mathcal{C}[\mathcal{F}] \,. \tag{2.9}$$

[8]cf. `http://www.bnl.gov/rhic/`

[9]cf. `http://public.web.cern.ch/public/`

Here, $\mathbf{x}$ denotes the spatial variable and $\mathbf{p}$ is the momentum variable. Moreover, the dot represents a total time derivative and $\mathcal{C}[\mathcal{F}]$ denotes the collision term, which is unspecified so far. This equation states that the temporal change of the one-particle distribution function has three reasons: The particle's motion, the action of forces as well as collisions between particles. In order to include pair creation into the kinetic description, one has to account for this by an additional source term $\mathcal{S}[\mathcal{F}]$.

The importance of collisions strongly depends on the problem: On the one hand, the density of particles is expected to be quite high in heavy ion collision so that collisions surely have to be taken into account. There have been various investigations of this effect by means of relaxation time approximations [55, 56, 57] as well as more sophisticated models [58, 59, 60]. On the other hand, the density of particles is expected to be rather low when considering the Schwinger effect in the sub-critical field strength regime so that collisions might be neglected to a good approximation [61]. Taking this into account, the collisionless approximation of the Boltzmann equation (*quantum Vlasov equation*) for the case of a spatially homogeneous, time-dependent electric field $\mathbf{E}(t)$ reads:

$$\dot{\mathcal{F}}(\mathbf{p},t) = \left[\tfrac{\partial}{\partial t} + e\mathbf{E}(t)\cdot\nabla_{\mathbf{p}}\right]\mathcal{F}(\mathbf{p},t) = \mathcal{S}[\mathcal{F}]\ . \tag{2.10}$$

A first ansatz for $\mathcal{S}[\mathcal{F}]$ was based on the Schwinger formula Eq. (2.5) and assumed the electron-positron pairs to be created at rest [28, 52, 62, 63]:

$$\mathcal{S}[\mathcal{F}] = -2|e\mathbf{E}(t)|\log\left[1-\exp\left(-\pi\frac{m^2+\mathbf{p}_\perp^2}{|e\mathbf{E}(t)|}\right)\right]\delta(\,p_\parallel)\ . \tag{2.11}$$

Here, $p_\parallel$ denotes the momentum in the field direction whereas $\mathbf{p}_\perp$ is the momentum in the orthogonal direction. Note that $\mathcal{S}[\mathcal{F}]$ does not depend on $\mathcal{F}(\mathbf{p},t)$ in this ansatz. It is clear that the validity of this approximation is restricted: The Schwinger term had been derived for a static electric field $\mathbf{E}_0$ whereas this ansatz assumes a time-dependent electric field $\mathbf{E}(t)$. Consequently, the applicability in the case of a rapidly varying electric field is not well justified.

It was only in the mid 1990s that the nowadays widely used source term for the Schwinger effect in time-dependent electric fields $\mathbf{E}(t)$ was derived. It was first shown for scalar QED (sQED) [29] and later also for QED [30] that $\mathcal{S}[\mathcal{F}]$ can be found from first principles in *mean field approximation.* In QED, this source term reads:

$$\mathcal{S}[\mathcal{F}] = Q(\mathbf{p},t)\int_{-\infty}^{t} Q(\mathbf{p},t')[1-\mathcal{F}(\mathbf{p},t')]\cos[2\Theta(\mathbf{p},t',t)]\ . \tag{2.12}$$

The detailed derivation of this source term with all quantities properly defined is given in Section 4.1. Note that this source term indeed depends on $\mathcal{F}(\mathbf{p}, t)$ as distinguished from Eq. (2.11). Accordingly, it was pointed out that the Schwinger effect possesses various characteristic features: Most notably, the pair creation process shows non-Markovian behavior because of memory effects and non-locality in time [61, 64]. Additionally, it was shown that the quantum Vlasov equation exhibits the features of quantum statistics, i. e. Bose enhancement for scalar particles as well as Pauli-blocking for spinor particles [30]. For computational reasons, it turned out to be of advantage to rewrite this integro-differential equation as a first order, ordinary differential equation (ODE) system [65]:

$$\dot{\mathcal{F}}(\mathbf{p}, t) = Q(\mathbf{p}, t)\mathcal{G}(\mathbf{p}, t) \ , \tag{2.13}$$

$$\dot{\mathcal{G}}(\mathbf{p}, t) = Q(\mathbf{p}, t)[1 - \mathcal{F}(\mathbf{p}, t)] - 2\omega(\mathbf{p}, t)\mathcal{H}(\mathbf{p}, t) \ , \tag{2.14}$$

$$\dot{\mathcal{H}}(\mathbf{p}, t) = 2\omega(\mathbf{p}, t)\mathcal{G}(\mathbf{p}, t) \ . \tag{2.15}$$

Again, the detailed derivation of this ODE system is given in Section 4.1.

The rather clear and simple structure of the ODE system made detailed analysis of the Schwinger effect in any kind of time-dependent electric field $\mathbf{E}(t)$ feasible. Initially, these investigations dealt with rather simple field configurations such as static [64, 65], sinusoidal [66, 67, 68] or pulsed [57, 69, 70] electric fields. Nonetheless, these studies revealed remarkable features such as a non-trivial momentum dependence of the pair creation process, the phenomenon of pair creation and annihilation in tune with the frequency of the sinusoidal field or the strong enhancement of the Schwinger effect in electric fields with a temporal variation of the order of the Compton time $\tau_C$. Additionally, it was shown that the density of created particles is rather low in the sub-critical field strength regime so that the omission of the collision term $\mathcal{C}[\mathcal{F}]$ seems well justified. Very recently, further investigations of more realistic field configurations such as pulsed electric fields with sub-cycle structure have been carried out [36, 39]. It was shown that the momentum distribution of created particles strongly depends on the detailed form of the electric field.

A further problem which has been studied is the influence of the self-induced electric field due to the pair creation process: The applied electric field $\mathbf{E}_{\text{ext}}(t)$ creates charged particles which in turn produce an opposing internal electric field $\mathbf{E}_{\text{int}}(t)$. Accordingly, one has to consider the Schwinger effect in the resulting electric field $\mathbf{E}(t) = \mathbf{E}_{\text{ext}}(t) + \mathbf{E}_{\text{int}}(t)$ in a self-consistent way. Again, this phenomenon has been investigated in mean field approximation, where an analytic expression for the internal electric field can be derived [27, 65]. Loosely speaking, it consists of three contributions: One contribution because of the motion of existing particles (conduction current), another one due to the continued pair creation (polarization current) and a third one because of the necessity of charge renormalization. As for the influence of collisions, it has been shown that this backreaction mechanism can be well ignored in the sub-critical field strength regime [65, 71].

### 2.2.2 Wigner function formalism

The phase space formulation of quantum mechanics has first been introduced in order to account for quantum corrections to the Boltzmann equation. It turns out that this approach is by no means unique but allows for a large number of different realizations in terms of, most prominently, the Wigner function [72], the Husimi function [73] or the Glauber-Sudarshan function [74, 75]. For a thorough introduction see [76].

I will focus on the *Wigner function* $\mathcal{W}(\mathbf{x}, \mathbf{p}, t)$ here, which has been widely adopted in the treatment of non-relativistic quantum physics. The phase space formalism has the advantage of offering an approach which uses classical language in the description of quantum systems as well as allowing direct access to real-time dynamics. Even though the Wigner function cannot be considered as a classical distribution function, it is in fact an observable which has been measured in various optical and atomic setups [77, 78, 79].

In a non-relativistic context, the Wigner function is defined as the Fourier transform of the density matrix.[10] Accordingly, the relative coordinate $\mathbf{y}$ is traded for the momentum variable $\mathbf{p}$:

$$\mathcal{W}(\mathbf{x}, \mathbf{p}, t) \equiv \int \mathrm{d}^3 y \, e^{-i\mathbf{p}\cdot\mathbf{y}} \langle \mathbf{x} + \tfrac{\mathbf{y}}{2} | \hat{\rho} | \mathbf{x} - \tfrac{\mathbf{y}}{2} \rangle \ . \tag{2.16}$$

The expectation value of any quantum mechanical observable $\hat{\mathcal{O}}$ is then calculated according to:

$$\langle \hat{\mathcal{O}} \rangle = \int \mathrm{d}\Gamma \, \mathcal{O}(\mathbf{x}, \mathbf{p}, t) \mathcal{W}(\mathbf{x}, \mathbf{p}, t) \ , \tag{2.17}$$

with $\mathrm{d}\Gamma = \mathrm{d}^3x \, \mathrm{d}^3p/(2\pi)^3$ being the phase space volume element. As already mentioned, the Wigner function cannot be considered as a distribution function but should rather be termed a *quasi-distribution function* since it can take negative values $\mathcal{W}(\mathbf{x}, \mathbf{p}, t) < 0$. It is especially this peculiarity which makes its interpretation quite challenging. Nevertheless, it still possesses various desired properties, most

[10] For a pure state $|\Psi\rangle$, the density matrix reduces to $\hat{\rho} = |\Psi\rangle\langle\Psi|$ and the Wigner transform reads

$$\mathcal{W}(\mathbf{x}, \mathbf{p}, t) = \int \mathrm{d}^3 y \, e^{-i\mathbf{p}\cdot\mathbf{y}} \Psi^*(\mathbf{x} + \tfrac{\mathbf{y}}{2}) \Psi(\mathbf{x} - \tfrac{\mathbf{y}}{2})$$

notably, it yields the correct quantum-mechanical marginal distributions:

$$\int \frac{\mathrm{d}^3p}{(2\pi)^3}\,\mathcal{W}(\mathbf{x},\mathbf{p},t) = \langle \mathbf{x}|\hat{\rho}|\mathbf{x}\rangle \quad \text{and} \quad \int \frac{\mathrm{d}^3x}{(2\pi)^3}\,\mathcal{W}(\mathbf{x},\mathbf{p},t) = \langle \mathbf{p}|\hat{\rho}|\mathbf{p}\rangle \ . \tag{2.18}$$

The need for an appropriate description of transport processes in astrophysics as well as plasma physics applications led to the generalization of the non-relativistic Wigner function to a relativistic context [80]. The description of both Abelian gauge theories like QED and non-Abelian gauge theories like quantum chromodynamics (QCD), however, entails still another complication: Most quantities such as correlation functions depend on the chosen gauge whereas all observable quantities must not. Due to the fact that the Wigner function should at least in the classical limit coincide with the corresponding classical distribution function, which in fact is an observable quantity, it is necessary to define the covariant Wigner function in a gauge invariant way [81, 82, 83].

As my focus lies on the Schwinger effect, I restrict myself to QED in the following. The *covariant Wigner operator* $\hat{\mathcal{W}}^{(4)}(\mathsf{x},\mathsf{p})$ is then defined in a similar manner as it's non-relativistic counterpart[11]:

$$\hat{\mathcal{W}}^{(4)}_{\alpha\beta}(\mathsf{x},\mathsf{p}) \equiv \int \mathrm{d}^4y\, e^{i\mathsf{p}\cdot\mathsf{y}}\, \bar{\Psi}_\beta(\mathsf{x}-\tfrac{\mathsf{y}}{2})\mathcal{U}(\mathsf{A};\mathsf{x},\mathsf{y})\Psi_\alpha(\mathsf{x}+\tfrac{\mathsf{y}}{2}) \ , \tag{2.19}$$

with $\Psi(\mathsf{x})$, $\bar{\Psi}(\mathsf{x})$ and $A^\mu(\mathsf{x})$ being considered as Heisenberg operators. It has to be emphasized that the path-ordered Wilson line integral $\mathcal{U}(\mathsf{A};\mathsf{x},\mathsf{y})$, which ensures gauge invariance, is by no means unique. A physical sensible interpretation of $\mathbf{p}$ as kinetic momentum, however, forces the integration path to be chosen along the straight line [84]:

$$\mathcal{U}(\mathsf{A};\mathsf{x},\mathsf{y}) = \mathcal{P}\exp\left( ie \int\limits_{\mathsf{x}-\mathsf{y}/2}^{\mathsf{x}+\mathsf{y}/2} \mathrm{d}\mathsf{z}\cdot\mathsf{A}(\mathsf{z}) \right) \ . \tag{2.20}$$

The *covariant Wigner function* $\mathcal{W}^{(4)}(\mathsf{x},\mathsf{p})$ is defined as the ensemble average of the corresponding Wigner operator. Due to the fact that I focus on vacuum pair creation, the ensemble average can be replaced by the vacuum expectation value:

$$\mathcal{W}^{(4)}(\mathsf{x},\mathsf{p}) \equiv \langle\Omega|\hat{\mathcal{W}}^{(4)}(\mathsf{x},\mathsf{p})|\Omega\rangle \ , \tag{2.21}$$

where $|\Omega\rangle$ denotes the Heisenberg vacuum. Note that the resulting equation of motion for $\mathcal{W}^{(4)}(\mathsf{x},\mathsf{p})$ forms a tower of coupled equations due to the operator nature

[11] Note that depending on the situation four-vectors are denoted either by sans serif variables or by explicitly writing the Lorentz indices, for example $\mathsf{p} \equiv p^\mu = (p^0, \mathbf{p})$

of $A^\mu(\mathsf{x})$. This infinite BBGKY hierarchy, named after N. Bogoliubov, M. Born, H. Green, G. Kirkwood and J. Yvon, has to be truncated for practical reasons. Treating the vector potential as classical background, this truncation of the BBGKY hierarchy happens automatically at the one-body level. Note that this formalism is put on the same footing as the quantum Vlasov equation once all photon contributions beyond mean field approximation are neglected. Further details on the covariant Wigner formalism can be found in Section 3.1.

The covariant Wigner formalism provides a clear theoretical description, however, it is rather inconvenient for various reasons. Consider the following situation in an upcoming high-intensity laser experiment:[12] Electron-positron pairs are created in the focus of colliding laser pulses after the system had been prepared in the vacuum state $|\Omega\rangle$. Subsequently, the created particles move apart from each other with the positrons being detected by surrounding detectors.

This idealized experimental situation has implications on the theoretical description: First, in order to trace the system's time evolution, the problem should be posed as an initial value problem. Consequently, it is necessary to choose a distinguished time $t = x^0$ and give up explicit Lorentz covariance. Secondly, it is not required to stick to a gauge invariant approach once the laser pulse is described as a classical electromagnetic field. Accordingly, it is convenient to choose a gauge fixed expression for the vector potential $A^\mu(\mathsf{x})$.

There have been two approaches to derive such an equal-time Wigner formalism: On the one hand, there have been attempts to define the *equal-time Wigner function* $\mathcal{W}^{(3)}(\mathbf{x}, \mathbf{p}, t)$ as an energy average of the covariant Wigner function [85, 86, 87]:

$$\mathcal{W}^{(3)}(\mathbf{x}, \mathbf{p}, t) = \int \frac{\mathrm{d}p_0}{(2\pi)} \mathcal{W}^{(4)}(\mathsf{x}, \mathsf{p}) \ . \tag{2.22}$$

On the other hand, the equation of motion for the equal-time Wigner function has also been derived in a non-covariant and gauge-fixed manner from the very beginning [31]. In fact, both approaches yield the same equation of motion:

$$D_t \mathcal{W}^{(3)} = -\tfrac{1}{2}\mathbf{D}^i \left[\gamma^0 \gamma^i, \mathcal{W}^{(3)}\right] - im \left[\gamma^0, \mathcal{W}^{(3)}\right] - i\mathbf{\Pi}^i \left\{\gamma^0 \gamma^i, \mathcal{W}^{(3)}\right\} \ , \tag{2.23}$$

where $D_t$, $\mathbf{D}$ and $\mathbf{\Pi}$ denote pseudo-differential operators to be defined later. Further details on the equal-time Wigner formalism can be found in Section 3.2. To be complete note that this formalism has been applied to sQED as well [88, 89].

[12] Note that there are even other proposals to detect the Schwinger effect, for instance in photon-nucleus collisions.

There have been various investigations on the equal-time Wigner formalism since then: First, it has been shown that this formalism gives a transport equation of the form of Eq. (2.9) in its classical limit [90, 91, 92, 93, 94]. Further investigations included loop corrections to the classical limit and showed the necessity of renormalization at higher orders [31, 95]. Additionally, there have been first attempts to tackle the problem of pair creation within this formalism [31, 61, 89, 96]. These investigations were closely related to studies of the quantum Vlasov equation with a source term Eq. (2.11). The close connection of the equal-time Wigner formalism to the non-Markovian source term Eq. (2.12) has been pointed out only recently [33].

In addition, there have been more general investigations of strong-field effects by means of the Wigner formalism in the light-front approach lately. These studies showed that the Wigner function incorporates all characteristic features of strong-field QED, most notably the mass shift of a particle in a plane wave background as well as effects due to multi-photon absorption and emission [97].

## 2.3 Current research topics

After giving this brief overview of the theoretical description of the Schwinger effect by means of quantum kinetic methods, I now turn to some topics which are of special interest in the current research.

### 2.3.1 Momentum distribution in realistic laser pulses

The upcoming high-intensity laser facilities will be capable of doing fundamental physics in the strong-field regime [9, 10, 14, 16]. Accordingly, it is necessary to improve both experimental techniques and theoretical predictions.

Theoretical studies focused on the general mechanisms behind vacuum pair creation in the past. As those mechanisms are quite well understood by now, a point is reached where it is about time to perform investigations with the aim of making actual predictions for upcoming experiments. One way towards a more realistic description of the Schwinger effect is to consider more *realistic electric fields*. Indeed, there have been studies which considered the electric field of a short pulse with sub-cycle structure recently [34, 36, 37, 38, 39, 40]:

$$\mathbf{E}(t) = \mathbf{E}_0 \cos(\omega t - \varphi) \exp\left(-\frac{t^2}{2\tau^2}\right) . \tag{2.24}$$

Here, $\omega$ denotes the laser frequency, $\tau$ is a measure for the total pulse length and $\varphi$ is the carrier-envelope absolute phase. Recently, there has been an investigation which included the effect of a linearly varying frequency $\omega(t) = \omega + b\,t$ as well [39].

It has been shown that the momentum distribution of created particles strongly depends on the various parameters, most notably on the carrier phase $\varphi$. It has also been pointed out that these distinct signatures in the momentum distribution might facilitate the experimental observation of the Schwinger effect because of the clear discrimination between particles originating from vacuum pair creation and background events. Further investigations could help to optimize the various laser parameters for upcoming experiments. My own contributions to this research topic [36, 37] are summarized in Section 4.4.

### 2.3.2 Enhancement of the Schwinger effect

Another topic which raised much interest very recently is the issue of dynamical enhancement of the pair creation rate. As the Schwinger effect is exponentially suppressed, it might happen that the reachable intensities in upcoming high-intensity laser facilities might still be too small so that no clear signals from vacuum pair creation are detected. Therefore, there have been various investigations on the issue of how to enhance the pair creation rate by means of dynamical effects [35, 98, 99, 100].

The main idea of the *dynamically assisted Schwinger effect* is rather simple: Superimposing a strong and slowly varying electric field (non-perturbative pair creation) with a weak and rapidly varying electric field (dynamical pair creation) should result in a decrease of the tunneling barrier and, consequently, to a drastic enhancement of the pair creation rate [35]. Indeed, it has been shown that the combination of the different scales results in a subtle change in the pair creation process so that an enhancement of the Schwinger effect can be expected [35, 99].

Recent investigations confirmed this picture of dynamic enhancement by means of quantum kinetic methods [101, 102]. Additionally, these studies gave valuable insight in the momentum distribution of created pairs and showed that the achievable enhancement strongly depends on the choice of the various parameters. Accordingly, a focus of future research should also be laid on pulse shaping analysis in order to further maximize the attainable pair creation rate.

### 2.3.3 Depletion of the laser field

Most investigations of the Schwinger effect considered the electric field as classical background. They either totally neglected the self-induced electric field due to the pair creation process or treated it in mean field approximation. These investigations indicated that the backreaction mechanism can be well ignored in the sub-critical field strength regime [65, 71]. Lately, there have been first attempts to account for the feedback of photons in a quantum kinetic formulation as well [103].

Very recently, the backreaction issue beyond mean field approximation has attracted much interest [104, 105, 106, 107]. These investigations indicate that already a single electron-positron pair might lead to an avalanche-like electromagnetic cascade restricting the attainable laser intensities in upcoming high-intensity laser experiments: The quick acceleration in the laser field could lead to the emission of hard photons which in turn create electron-positron pairs via dynamical pair creation as already observed in the SLAC-E144 pair production experiment [108]. Accordingly, the evolution of a QED cascade might result in a total *destruction of the laser field* even in the sub-critical field strength regime in the last resort. It is surely worth to further investigate this issue because of its great importance for future high-intensity laser experiments.

### 2.3.4 Schwinger effect in space- and time-dependent electric fields

Investigations of the Schwinger effect mainly focused on the temporal dependence of the electric field $\mathbf{E}(t)$ in the past. It is one main result of these studies that short time scales yield an enhancement of the number of created particles [19, 70, 109]. On the other hand, there have been investigations of the Schwinger effect in spatially varying electric fields $\mathbf{E}(\mathbf{x})$ as well. These studies indicated that spatial inhomogeneities tend to decrease the pair creation rate as distinguished from temporal inhomogeneities [22, 23, 109].

This diminishment due to spatial variations can be easily understood: Vacuum fluctuations have to delocalize in order to gain energy and turn into real particles, with the Compton wavelength $\lambda_C$ being the characteristic delocalization scale. As soon as the work done by the electric field over its spatial extent is smaller than twice the electron mass:

$$e \int \mathbf{E}(\mathbf{x}) \cdot \mathrm{d}\mathbf{x} < 2m \; , \tag{2.25}$$

the vacuum fluctuations cannot acquire enough energy from the electric field and the pair creation process terminates.

Accordingly, there might be two opposing effects in a *space- and time-dependent electric field* $\mathbf{E}(\mathbf{x}, t)$: Enhancement of the pair creation rate by temporal variations on the one hand and decrease of the pair creation rate by spatial variations on the other hand. In fact, to the best of my knowledge there have not been any first principle investigations on the influence of space- and time-dependent electric fields on the Schwinger effect so far. Hence, the overall effect of the interplay between the different scales is by no means clear.

## 2.4 Objectives

I tried to highlight some current research topics which seem of utmost importance in the previous section. Some of these issues will in fact be touched upon in this thesis, which is dedicated to further investigate the Schwinger effect in inhomogeneous electric fields. In the first part, I focus on the Schwinger effect in time-dependent electric fields $\mathbf{E}(t)$:

- I present the close similarity between the equal-time Wigner formalism and the quantum Vlasov equation for the case of a time-dependent electric field $\mathbf{E}(t)$. Accordingly, it has to be emphasized that the equal-time Wigner formalism is the natural generalization of the quantum Vlasov equation in the presence of space- and time-dependent electric fields $\mathbf{E}(\mathbf{x}, t)$.

- I derive an analytic expression for the equal-time Wigner function for the exactly solvable cases of a static as well as a pulsed electric field. Additionally, I discuss some features of the Schwinger effect in a pulsed electric field with sub-cycle structure. It has to be stressed that this investigation represents an important step towards a more realistic description of vacuum pair creation in upcoming high-intensity laser experiments.

As already mentioned, little is known about the Schwinger effect in space- and time-dependent electric fields $\mathbf{E}(\mathbf{x}, t)$. Therefore, it is about time to shed some light on this issue. Accordingly, the following topics are covered in the second part:

- I perform an ab initio simulation of the Schwinger effect in a simple space- and time-dependent electric field $\mathbf{E}(\mathbf{x}, t)$ in order to better understand the interplay between spatial and temporal variations. Most notably, I will show in which way observable quantities such as the number of created particles or the momentum distribution change. For computational reasons, this study is performed on the basis of QED in $1 + 1$ dimensions.

- I adopt various approximations (local density approximation, derivative expansion) and compare their outcome with the results of the full numerical simulation. These investigation will give an estimate on the range of validity of the different approximations.

Chapter 3

# Schwinger effect in phase space

In this chapter I give a brief review on the Wigner formalism upon which the subsequent investigations are based. In Section 3.1, I focus on the *covariant Wigner formalism* and derive the equation of motion for the covariant Wigner function [81, 83, 85]. In Section 3.2, I present the connection to the *equal-time Wigner formalism* and discuss some of its properties in more detail [31, 33, 85, 86, 87, 92, 93].

## 3.1 Covariant Wigner formalism

The quantum field theory of electromagnetism is based on the QED Lagrangian:[1]

$$\mathcal{L}(\Psi,\bar{\Psi},\mathsf{A}) = \tfrac{1}{2}\left[\bar{\Psi}\gamma^{\mu}[i\partial_{\mu}-eA_{\mu}]\Psi - ([i\partial_{\mu}+eA_{\mu}]\bar{\Psi})\gamma^{\mu}\Psi\right] - m\bar{\Psi}\Psi - \tfrac{1}{4}F^{\mu\nu}F_{\mu\nu}\ . \tag{3.1}$$

The equations of motion for the Dirac field $\Psi(\mathsf{x})$, the adjoint field $\bar{\Psi}(\mathsf{x})$ as well as the photon field $A^{\mu}(\mathsf{x})$ are given by:

$$\gamma^{\mu}\left[i\partial_{\mu}-eA_{\mu}(\mathsf{x})\right]\Psi(\mathsf{x}) - m\Psi(\mathsf{x}) = 0\ , \tag{3.2}$$

$$\left[i\partial_{\mu}+eA_{\mu}(\mathsf{x})\right]\bar{\Psi}(\mathsf{x})\gamma^{\mu} + m\bar{\Psi}(\mathsf{x}) = 0\ , \tag{3.3}$$

and

$$\partial_{\mu}F^{\mu\nu}(\mathsf{x}) = \tfrac{e}{2}\left[\bar{\Psi}(\mathsf{x}),\gamma^{\nu}\Psi(\mathsf{x})\right]\ . \tag{3.4}$$

Note that these equations have to be understood as operator equations in the Heisenberg picture at this point.

Due to the fact that I focus on pair creation in an external electromagnetic field, the following simplification is made: The field strength tensor $F^{\mu\nu}(\mathsf{x})$ is not treated as an operator but rather as a c-number. It is clear that this *mean field approximation* is only justified in the strong-field regime as it suffers from several shortcomings such as the neglect of radiative corrections or final state interactions. On the other hand, this mean field treatment results in the truncation of the BBGKY hierarchy for the Wigner function at the one-body level without further approximations.

[1] Note, that I use the *hermitian* Dirac Lagrangian which treats the oppositely charged Dirac fields symmetrically. As usual, the field strength tensor is defined as $F^{\mu\nu}(\mathsf{x}) \equiv \partial^{\mu}A^{\nu}(\mathsf{x}) - \partial^{\nu}A^{\mu}(\mathsf{x})$.

### 3.1.1 Covariant Wigner operator

The commutator of two Dirac fields $\frac{1}{2}\left[\bar{\Psi}_\beta(\mathsf{x}_1), \Psi_\alpha(\mathsf{x}_2)\right]$ is the starting point for the following considerations. As this density matrix is *not* gauge invariant under local $U(1)$ gauge transformations:

$$A'_\mu(\mathsf{x}) = A_\mu(\mathsf{x}) - \partial_\mu \Lambda(\mathsf{x}) \;\longrightarrow\; \Psi'(\mathsf{x}) = \Psi(\mathsf{x}) e^{ie\Lambda(\mathsf{x})} \,, \tag{3.5}$$

one has to introduce a Wilson-line factor $\mathcal{U}(\mathsf{A};\mathsf{x},\mathsf{y})$:

$$\mathcal{U}(\mathsf{A};\mathsf{x},\mathsf{y}) = \exp\left( ie \int\limits_{\mathsf{x}-\mathsf{y}/2}^{\mathsf{x}+\mathsf{y}/2} \mathrm{d}\mathsf{z}\cdot \mathsf{A}(\mathsf{z}) \right) \,, \tag{3.6}$$

in order to allow for a gauge invariant definition of the *covariant Wigner operator*:[2]

$$\hat{\mathcal{W}}^{(4)}_{\alpha\beta}(\mathsf{x},\mathsf{p}) \equiv \frac{1}{2}\int \mathrm{d}^4 y e^{i\mathsf{p}\cdot\mathsf{y}} \mathcal{U}(\mathsf{A};\mathsf{x},\mathsf{y}) \left[\bar{\Psi}_\beta(\mathsf{x}-\tfrac{\mathsf{y}}{2}), \Psi_\alpha(\mathsf{x}+\tfrac{\mathsf{y}}{2})\right] \,, \tag{3.7}$$

with $\mathsf{x} = \frac{1}{2}(\mathsf{x}_1 + \mathsf{x}_2)$ being the center of mass coordinate and $\mathsf{y} = \mathsf{x}_2 - \mathsf{x}_1$ the relative coordinate.[3] As already mentioned, the necessity of a gauge invariant definition is rather important for an interpretational reason: The Wigner function should coincide with the corresponding distribution function in its classical limit. Accordingly, it should be an observable quantity and as such it must not be gauge dependent.

Note that the path-ordering in $\mathcal{U}(\mathsf{A};\mathsf{x},\mathsf{y})$ has been dropped in comparison to Eq. (2.20) as $A^\mu(\mathsf{x})$ is treated as a classical mean field now. Additionally, it has to be stressed once more that the integration path from $\mathsf{x}-\mathsf{y}/2$ to $\mathsf{x}+\mathsf{y}/2$ in the Wilson-line factor is not unique. However, requiring a proper physical interpretation of $\mathbf{p}$ as kinetic momentum forces the integration path to be chosen along the straight line between the end points. Accordingly, $\hat{\mathcal{W}}^{(4)}(\mathsf{x},\mathsf{p})$ can be written as:

$$\hat{\mathcal{W}}^{(4)}(\mathsf{x},\mathsf{p}) = \frac{1}{2}\int \mathrm{d}^4 y e^{i\mathsf{p}\cdot\mathsf{y}} e^{ie\int_{-1/2}^{1/2}\mathrm{d}\xi \mathsf{A}(\mathsf{x}+\xi\mathsf{y})\cdot\mathsf{y}} \left[\bar{\Psi}(\mathsf{x}-\tfrac{\mathsf{y}}{2}), \Psi(\mathsf{x}+\tfrac{\mathsf{y}}{2})\right] \,. \tag{3.8}$$

It can be seen explicitly that the covariant Wigner operator transforms like an ordinary Dirac gamma matrix under Hermitian conjugation:

$$[\hat{\mathcal{W}}^{(4)}(\mathsf{x},\mathsf{p})]^\dagger = \gamma^0 \hat{\mathcal{W}}^{(4)}(\mathsf{x},\mathsf{p})\gamma^0 \,. \tag{3.9}$$

[2] Note that this definition differs from Eq. (2.19): Taking the commutator $\frac{1}{2}\left[\bar{\Psi}_\beta(\mathsf{x}_1), \Psi_\alpha(\mathsf{x}_2)\right]$ instead of $\bar{\Psi}_\beta(\mathsf{x}_1)\Psi_\alpha(\mathsf{x}_2)$, the symmetrized electromagnetic current will be obtained later on.

[3] Spinor indices will be suppressed in the following to simplify notation.

### 3.1.2 Equation of motion

The equation of motion for the covariant Wigner operator can be calculated by taking advantage of the Dirac equation Eq. (3.2) – (3.3). Similar to the Dirac equation, one obtains in fact two equations which are the adjoints of each other:[4]

$$\left[\tfrac{1}{2}D_\mu(\mathsf{x},\mathsf{p}) - i\Pi_\mu(\mathsf{x},\mathsf{p})\right]\gamma^\mu \hat{\mathcal{W}}^{(4)}(\mathsf{x},\mathsf{p}) = -im\hat{\mathcal{W}}^{(4)}(\mathsf{x},\mathsf{p}) \,, \tag{3.10}$$

$$\left[\tfrac{1}{2}D_\mu(\mathsf{x},\mathsf{p}) + i\Pi_\mu(\mathsf{x},\mathsf{p})\right]\hat{\mathcal{W}}^{(4)}(\mathsf{x},\mathsf{p})\gamma^\mu = im\hat{\mathcal{W}}^{(4)}(\mathsf{x},\mathsf{p}) \,, \tag{3.11}$$

with $D_\mu(\mathsf{x},\mathsf{p})$ and $\Pi_\mu(\mathsf{x},\mathsf{p})$ denoting the following self-adjoint operators:[5]

$$D_\mu(\mathsf{x},\mathsf{p}) \equiv \partial^{\mathsf{x}}_\mu - e\int_{-\frac{1}{2}}^{\frac{1}{2}} \mathrm{d}\xi\, F_{\mu\nu}(\mathsf{x} - i\xi\partial_{\mathsf{p}})\partial^\nu_{\mathsf{p}} \,, \tag{3.12}$$

$$\Pi_\mu(\mathsf{x},\mathsf{p}) \equiv p_\mu - ie\int_{-\frac{1}{2}}^{\frac{1}{2}} \mathrm{d}\xi\xi F_{\mu\nu}(\mathsf{x} - i\xi\partial_{\mathsf{p}})\partial^\nu_{\mathsf{p}} \,. \tag{3.13}$$

In order to switch from the covariant Wigner operator $\hat{\mathcal{W}}^{(4)}(\mathsf{x},\mathsf{p})$ to the *covariant Wigner function* $\mathcal{W}^{(4)}(\mathsf{x},\mathsf{p})$, one takes the vacuum expectation value:[6]

$$\mathcal{W}^{(4)}(\mathsf{x},\mathsf{p}) = \langle\Omega|\hat{\mathcal{W}}^{(4)}(\mathsf{x},\mathsf{p})|\Omega\rangle \,. \tag{3.14}$$

The corresponding equation of motion for the covariant Wigner function is accordingly found by taking the vacuum expectation value of Eq. (3.10) – (3.11).

Note that the mean field approximation, treating $F^{\mu\nu}(\mathsf{x})$ as a c-number instead of an operator, becomes apparent at this point: If the field strength tensor was still considered as an operator, the equation of motion for the covariant Wigner function $\langle\Omega|\hat{W}^{(4)}|\Omega\rangle$ would couple to a two-body term $\langle\Omega|F^{\mu\nu}\hat{W}^{(4)}|\Omega\rangle$. As the equation of motion for the two-body term depends on a three-body term and so forth, this would generate the infinite BBGKY in the end. However, as the field strength tensor is in fact taken to be a c-number, the BBGKY hierarchy is already truncated at the one-body level.

It seems advantageous to either add or subtract Eq. (3.10) from Eq. (3.11) for later use. Upon taking the vacuum expectation value, this yields two new self-adjoint

[4]The detailed calculation can be found in Appendix A.1.

[5]The differential operators with respect to x and p are defined so that:

$$\partial^{\mathsf{x}}_\mu \equiv \tfrac{\partial}{\partial x^\mu} \quad \text{and} \quad \partial^\mu_{\mathsf{x}} \equiv \tfrac{\partial}{\partial x_\mu} \qquad \text{as well as} \qquad \partial^{\mathsf{p}}_\mu \equiv \tfrac{\partial}{\partial p^\mu} \quad \text{and} \quad \partial^\mu_{\mathsf{p}} \equiv \tfrac{\partial}{\partial p_\mu} \,.$$

[6]As the field operators are considered as Heisenberg operators throughout, $|\Omega\rangle$ has to be understood as Heisenberg vacuum as well.

equations of motion for the covariant Wigner function:[7]

$$\tfrac{1}{2}D_\mu\big[\gamma^\mu\,,\mathcal{W}^{(4)}\big] - i\Pi_\mu\big\{\gamma^\mu\,,\mathcal{W}^{(4)}\big\} \;=\; -2im\mathcal{W}^{(4)}\;, \tag{3.15}$$

$$\tfrac{1}{2}D_\mu\big\{\gamma^\mu\,,\mathcal{W}^{(4)}\big\} - i\Pi_\mu\left[\gamma^\mu\,,\,\mathcal{W}^{(4)}\right] \;=\; 0\;. \tag{3.16}$$

Along with the equation of motion for the mean electromagnetic field which results from Eq. (3.4), a self-consistent differential equation system is finally obtained:

$$\partial_\mu^{\mathsf{x}} F^{\mu\nu}(\mathsf{x}) = \langle\Omega|\tfrac{e}{2}\left[\bar{\Psi}(\mathsf{x}),\gamma^\nu\Psi(\mathsf{x})\right]|\Omega\rangle = e\int\frac{\mathrm{d}^4p}{(2\pi)^4}\,\mathrm{tr}[\gamma^\nu\mathcal{W}^{(4)}(\mathsf{x},\mathsf{p})]\;. \tag{3.17}$$

### 3.1.3 Spinor decomposition

As the covariant Wigner function transforms as a Dirac gamma matrix, one may decompose $\mathcal{W}^{(4)}(\mathsf{x},\mathsf{p})$ in terms of the Dirac bilinears. As a matter of fact, this defines 16 *covariant Wigner components* $\mathbb{W}(\mathsf{x},\mathsf{p})$ with specific transformation behavior:[8]

$$\mathcal{W}^{(4)}(\mathsf{x},\mathsf{p}) = \tfrac{1}{4}\left[\mathbb{S} + i\gamma_5\mathbb{P} + \gamma^\mu\mathbb{V}_\mu + \gamma^\mu\gamma_5\mathbb{A}_\mu + \sigma^{\mu\nu}\mathbb{T}_{\mu\nu}\right]\;. \tag{3.18}$$

Note that all $\mathbb{W}(\mathsf{x},\mathsf{p})$ can be chosen to be real because of Eq. (3.9). Under orthochronous Lorentz transformations they transform as scalar $\mathbb{S}(\mathsf{x},\mathsf{p})$, pseudoscalar $\mathbb{P}(\mathsf{x},\mathsf{p})$, vector $\mathbb{V}_\mu(\mathsf{x},\mathsf{p})$, axialvector $\mathbb{A}_\mu(\mathsf{x},\mathsf{p})$ and tensor $\mathbb{T}_{\mu\nu}(\mathsf{x},\mathsf{p})$, respectively.

In order to decompose Eq. (3.15) – (3.16) in the basis of Dirac bilinears, one has to calculate both the commutator and the anticommutator between $\gamma^\mu$ and the Dirac bilinears. This gives:[9]

| | $\{\gamma^\mu,\cdot\}$ | $[\gamma^\mu,\cdot]$ |
|---|---|---|
| $\mathrm{I}$ | $2\gamma^\mu$ | $0$ |
| $\gamma_5$ | $0$ | $2\gamma^\mu\gamma_5$ |
| $\gamma^\nu$ | $2g^{\mu\nu}$ | $-2i\sigma^{\mu\nu}$ |
| $\gamma^\nu\gamma_5$ | $\epsilon^{\mu\nu\alpha\beta}\sigma_{\alpha\beta}$ | $2g^{\mu\nu}\gamma_5$ |
| $\sigma^{\nu\rho}\equiv\frac{i}{2}[\gamma^\nu,\gamma^\rho]$ | $-2\epsilon^{\mu\nu\rho\alpha}\gamma_\alpha\gamma_5$ | $2i(g^{\mu\nu}\gamma^\rho - g^{\mu\rho}\gamma^\nu)$ |

(3.19)

[7]Alternatively, it is also possible to first multiply Eq. (3.10) from the left and the right side by a factor $\gamma^0$. In this case, the resulting equation of motion becomes symmetric under the exchange of commutators $[\cdot,\cdot]_-$ with anticommutators $[\cdot,\cdot]_+$:

$$\tfrac{1}{2}D_\mu\left[\gamma^0\gamma^\mu,\mathcal{W}^{(4)}\gamma^0\right]_\pm - i\Pi_\mu\left[\gamma^0\gamma^\mu,\mathcal{W}^{(4)}\gamma^0\right]_\mp = -im\left[\gamma^0,\mathcal{W}^{(4)}\gamma^0\right]_\mp\;.$$

[8]The components of the covariant Wigner function will be denoted by upper-case fonts $\mathbb{W}(\mathsf{x},\mathsf{p})$, whereas the components of the equal-time Wigner function will be denoted by lower-case fonts $\mathbb{w}(\mathbf{x},\mathbf{p},t)$ in the following.

[9]Here, the usual convention $\epsilon^{0123}=1$ for the totally antisymmetric tensor is used.

Accordingly, the terms appearing in Eq. (3.15) – (3.16) read:

$$\{\gamma^\mu, \mathcal{W}^{(4)}\} = \tfrac{1}{2}\left[\gamma^\mu \mathbb{S} + \mathbb{V}^\mu + \tfrac{1}{2}\epsilon^{\mu\nu\alpha\beta}\sigma_{\alpha\beta}\mathbb{A}_\nu - \epsilon^{\mu\nu\rho\alpha}\gamma_\alpha\gamma_5\mathbb{T}_{\nu\rho}\right], \quad (3.20)$$
$$\left[\gamma^\mu, \mathcal{W}^{(4)}\right] = \tfrac{1}{2}\left[i\gamma^\mu\gamma_5\mathbb{P} - i\sigma^{\mu\nu}\mathbb{V}_\nu + \gamma_5\mathbb{A}^\mu + 2ig^{\mu\nu}\gamma^\rho\mathbb{T}_{\nu\rho}\right]. \quad (3.21)$$

The equations of motion for the covariant Wigner components are then found by equating the coefficients of the individual Dirac bilinears. On the one hand, this results in a set of inhomogeneous equations originating from Eq. (3.15):

$$\Pi^\mu \mathbb{V}_\mu = m\mathbb{S}, \quad (3.22)$$
$$D^\mu \mathbb{A}_\mu = 2m\mathbb{P}, \quad (3.23)$$
$$\Pi^\mu \mathbb{S} - D_\nu \mathbb{T}^{\nu\mu} = m\mathbb{V}^\mu, \quad (3.24)$$
$$D^\mu \mathbb{P} - 2\epsilon^{\mu\nu\rho\sigma}\Pi_\nu \mathbb{T}_{\rho\sigma} = -2m\mathbb{A}^\mu, \quad (3.25)$$
$$(D^\mu \mathbb{V}^\nu - D^\nu \mathbb{V}^\mu) + 2\epsilon^{\mu\nu\rho\sigma}\Pi_\rho \mathbb{A}_\sigma = 4m\mathbb{T}^{\mu\nu}. \quad (3.26)$$

On the other hand, a set of homogeneous equations is derived from Eq. (3.16):[10]

$$D^\mu \mathbb{V}_\mu = 0, \quad (3.27)$$
$$\Pi^\mu \mathbb{A}_\mu = 0, \quad (3.28)$$
$$D^\mu \mathbb{S} + 4\Pi_\nu \mathbb{T}^{\nu\mu} = 0, \quad (3.29)$$
$$\Pi^\mu \mathbb{P} + \tfrac{1}{2}\epsilon^{\mu\nu\rho\sigma} D_\nu \mathbb{T}_{\rho\sigma} = 0, \quad (3.30)$$
$$(\Pi^\mu \mathbb{V}^\nu - \Pi^\nu \mathbb{V}^\mu) - \tfrac{1}{2}\epsilon^{\mu\nu\rho\sigma} D_\rho \mathbb{A}_\sigma = 0. \quad (3.31)$$

## 3.2 Equal-time Wigner formalism

The covariant Wigner formalism provides a clear theoretical description, however, the experimental situation rather suggests a non-covariant formulation in terms of an initial value problem: Accordingly, covariance should be given up in favor of a clearer and simpler interpretation. In fact, there are two different but equivalent ways to obtain an equal-time Wigner formalism: One can either derive the equation of motion for the *equal-time Wigner function* $\mathcal{W}^{(3)}(\mathbf{x}, \mathbf{p}, t)$ in a non-covariant way from the very beginning or start with the covariant formulation and perform an energy average afterwards.[11]

[10]The set of equations Eq. (3.22) – (3.31) is sometimes called VGE equations in the literature, named after D. Vasak, M. Gyulassy and H. T. Elze [83].

[11]The equal-time Wigner formalism has first been derived in a non-covariant way from the very beginning [31] whereas the connection between the covariant and the non-covariant approach has been shown only later [85, 86, 87]. Note that the second approach is adopted in the following.

### 3.2.1 Energy average

The equal-time Wigner function is defined as the energy average of the covariant Wigner function:

$$\mathcal{W}^{(3)}(\mathbf{x},\mathbf{p},t) \equiv \int \frac{\mathrm{d}p_0}{(2\pi)} \mathcal{W}^{(4)}(\mathsf{x},\mathsf{p}) \ . \tag{3.32}$$

Plugging in the explicit expression for the covariant Wigner function, this reads:

$$\mathcal{W}^{(3)}(\mathbf{x},\mathbf{p},t) = \frac{1}{2}\int \mathrm{d}^3y e^{-i\mathbf{p}\cdot\mathbf{y}} e^{-ie\int_{-1/2}^{1/2}\mathrm{d}\xi \mathbf{A}(\mathbf{x}+\xi\mathbf{y},t)\cdot\mathbf{y}} \langle\Omega|\left[\bar{\Psi}(\mathbf{x}-\tfrac{\mathbf{y}}{2},t),\Psi(\mathbf{x}+\tfrac{\mathbf{y}}{2},t)\right]|\Omega\rangle \tag{3.33}$$

with $t = x^0$. The equal-time Wigner function can then be decomposed into its Dirac bilinears in complete analogy to Eq. (3.18):[12]

$$\mathcal{W}^{(3)}(\mathbf{x},\mathbf{p},t) = \tfrac{1}{4}\left[\mathbb{s} + i\gamma_5\mathbb{p} + \gamma^\mu\mathbb{v}_\mu + \gamma^\mu\gamma_5\mathbb{a}_\mu + \sigma^{\mu\nu}\mathbb{t}_{\mu\nu}\right] \ . \tag{3.34}$$

Note that the equations of motion for the covariant Wigner components cannot be set up as an initial value problem by fixing an asymptotic field configuration at $x^0 \to -\infty$. This is in contrast to the equal-time approach which will indeed allow for a formulation in terms of an initial value problem.[13] It has to be emphasized, however, that dynamical information on $\mathcal{W}^{(4)}(\mathsf{x},\mathsf{p})$ is lost by taking the energy average. As a consequence, in order to recover all the information which is thrown away by averaging out the off-shell behavior of $\mathcal{W}^{(4)}(\mathsf{x},\mathsf{p})$, one would have to consider not only the equations of motion for $\mathbb{w}(\mathbf{x},\mathbf{p},t)$ but also for all higher energy moments:

$$\mathbb{w}^{[n]}(\mathbf{x},\mathbf{p},t) = \int \frac{\mathrm{d}p_0}{(2\pi)}\, p_0^n \mathbb{W}(\mathsf{x},\mathsf{p}) \ . \tag{3.35}$$

Accordingly, the equal-time Wigner formalism would have to involve the equations of motion for *all* energy moments $\mathbb{w}^{[n]}(\mathbf{x},\mathbf{p},t)$ in order to provide a complete description of the system. This would, however, result in an infinite hierarchy of equations again. This hierarchy is in fact truncated at the level of the equal-time Wigner components $\mathbb{w}(\mathbf{x},\mathbf{p},t)$ for an electromagnetic field which is treated as classical mean field. Accordingly, the equal-time Wigner formalism in mean field approximation is complete in the sense that knowledge of $\mathbb{w}(\mathbf{x},\mathbf{p},t)$ suffices to recover all information contained in the covariant Wigner function $\mathcal{W}^{(4)}(\mathsf{x},\mathsf{p})$ [92, 93].

[12] The *equal-time Wigner components* are defined by the energy average as well:

$$\mathbb{w}(\mathbf{x},\mathbf{p},t) = \int \frac{\mathrm{d}p_0}{(2\pi)} \mathbb{W}(\mathsf{x},\mathsf{p})\,.$$

[13] The simplest choice of initial condition, which will be used later in this thesis as well, is to take the vacuum value of the equal-time Wigner components $\mathbb{w}_{\mathrm{vac}}(\mathbf{x},\mathbf{p},t_{\mathrm{vac}})$, with $t_{\mathrm{vac}} \to -\infty$.

### 3.2.2 Equation of motion

The derivation of the equation of motion within the equal-time Wigner formalism is based on the equations of motion for the covariant Wigner components. Taking the energy average of Eq. (3.22) – (3.31), it turns out that the subset of equations which does not contain any explicit dependence on $p_0$ via $\Pi_0(\mathsf{x},\mathsf{p})$ trivially translates into *time evolution equations*[14] for the equal-time Wigner components:[15]

$$D_t\, \mathbb{s} \qquad - 2\mathbf{\Pi}\cdot \mathbb{t}_1 = 0 \;, \tag{3.36}$$
$$D_t\, \mathbb{p} \qquad + 2\mathbf{\Pi}\cdot \mathbb{t}_2 = -2m\, \mathbb{a}_0 \;, \tag{3.37}$$
$$D_t\, \mathbb{v}_0 + \mathbf{D}\cdot \mathbb{v} = 0 \;, \tag{3.38}$$
$$D_t\, \mathbb{a}_0 + \mathbf{D}\cdot \mathbb{a} = 2m\, \mathbb{p} \;, \tag{3.39}$$
$$D_t\, \mathbb{v} + \mathbf{D}\, \mathbb{v}_0 + 2\mathbf{\Pi}\times \mathbb{a} = -2m\, \mathbb{t}_1 \;, \tag{3.40}$$
$$D_t\, \mathbb{a} + \mathbf{D}\, \mathbb{a}_0 + 2\mathbf{\Pi}\times \mathbb{v} = 0 \;, \tag{3.41}$$
$$D_t\, \mathbb{t}_1 + \mathbf{D}\times \mathbb{t}_2 + 2\mathbf{\Pi}\, \mathbb{s} = 2m\, \mathbb{v} \;, \tag{3.42}$$
$$D_t\, \mathbb{t}_2 - \mathbf{D}\times \mathbb{t}_1 - 2\mathbf{\Pi}\, \mathbb{p} = 0 \;, \tag{3.43}$$

with the vectors $\mathbb{t}_1(\mathbf{x},\mathbf{p},t)$ and $\mathbb{t}_2(\mathbf{x},\mathbf{p},t)$ being defined in terms of the antisymmetric tensor $\mathbb{t}_{\mu\nu}$:[16]

$$\mathbb{t}_1 = 2\mathbb{t}^{i0}\mathbf{e}_i \qquad \text{and} \qquad \mathbb{t}_2 = \epsilon_{ijk}\mathbb{t}^{jk}\mathbf{e}_i \;. \tag{3.44}$$

The pseudo-differential operators $D_t(\mathbf{x},\mathbf{p},t)$, $\mathbf{D}(\mathbf{x},\mathbf{p},t)$ and $\mathbf{\Pi}(\mathbf{x},\mathbf{p},t)$, which are obtained by taking the energy average of Eq. (3.12) – (3.13), are given by:

$$D_t(\mathbf{x},\mathbf{p},t) \equiv \frac{\partial}{\partial t} + e\int_{-\frac{1}{2}}^{\frac{1}{2}} \mathrm{d}\xi \mathbf{E}(\mathbf{x}+i\xi\nabla_{\mathbf{p}},t)\cdot\nabla_{\mathbf{p}} \;, \tag{3.45}$$

$$\mathbf{D}(\mathbf{x},\mathbf{p},t) \equiv \nabla_{\mathbf{x}} + e\int_{-\frac{1}{2}}^{\frac{1}{2}} \mathrm{d}\xi \mathbf{B}(\mathbf{x}+i\xi\nabla_{\mathbf{p}},t)\times\nabla_{\mathbf{p}} \;, \tag{3.46}$$

$$\mathbf{\Pi}(\mathbf{x},\mathbf{p},t) \equiv \mathbf{p} - ie\int_{-\frac{1}{2}}^{\frac{1}{2}} \mathrm{d}\xi \xi \mathbf{B}(\mathbf{x}+i\xi\nabla_{\mathbf{p}},t)\times\nabla_{\mathbf{p}} \;. \tag{3.47}$$

As a matter of fact, the equations of motion for the covariant Wigner components Eq. (3.22) – (3.31) give rise to still another set of *constraint equations*, resulting from the energy average of the subset of equations which does show an explicit dependence on $p_0$ via $\Pi_0(\mathsf{x},\mathsf{p})$:

$$\mathbb{s}^{[1]} + \Pi_t\,\mathbb{s} - \tfrac{1}{2}\mathbf{D}\cdot\mathbb{t}_1 = m\,\mathbb{v}_0\ , \tag{3.48}$$

$$\mathbb{p}^{[1]} + \Pi_t\,\mathbb{p} + \tfrac{1}{2}\mathbf{D}\cdot\mathbb{t}_2 = 0\ , \tag{3.49}$$

$$\mathbb{v}_0^{[1]} + \Pi_t\,\mathbb{v}_0 - \mathbf{\Pi}\cdot\mathbb{v} = m\,\mathbb{s}\ , \tag{3.50}$$

$$\mathbb{a}_0^{[1]} + \Pi_t\,\mathbb{a}_0 - \mathbf{\Pi}\cdot\mathbb{a} = 0\ , \tag{3.51}$$

$$\mathbb{v}^{[1]} + \Pi_t\,\mathbb{v} + \tfrac{1}{2}\mathbf{D}\times\mathbb{a} - \mathbf{\Pi}\,\mathbb{v}_0 = 0\ , \tag{3.52}$$

$$\mathbb{a}^{[1]} + \Pi_t\,\mathbb{a} + \tfrac{1}{2}\mathbf{D}\times\mathbb{v} - \mathbf{\Pi}\,\mathbb{a}_0 = -\,m\,\mathbb{t}_2\ , \tag{3.53}$$

$$\mathbb{t}_1^{[1]} + \Pi_t\,\mathbb{t}_1 + \tfrac{1}{2}\mathbf{D}\,\mathbb{s} - \mathbf{\Pi}\times\mathbb{t}_2 = 0\ , \tag{3.54}$$

$$\mathbb{t}_2^{[1]} + \Pi_t\,\mathbb{t}_2 - \tfrac{1}{2}\mathbf{D}\,\mathbb{p} + \mathbf{\Pi}\times\mathbb{t}_1 = -\,m\,\mathbb{a}\ . \tag{3.55}$$

[14] The set of equations Eq. (3.36) – (3.43) is also called BGR equations in the literature, named after I. Bialynicki-Birula, P. Gornicki and J. Rafelski [31].

[15] One must assume that the electromagnetic field can be expanded in a Taylor series with respect to the temporal coordinate and that the covariant Wigner components and all its derivatives with respect to $p_0$ vanish for $|p_0| \to \infty$. Details can be found in Appendix A.2.

[16] This is in analogy to the definition of the electric and magnetic field in terms of $F^{\mu\nu}(\mathsf{x})$.

This system of equations describes the relation between the first energy moments $\mathbb{w}^{[1]}(\mathbf{x},\mathbf{p},t)$ and the equal-time Wigner components $\mathbb{w}(\mathbf{x},\mathbf{p},t)$. Moreover, the operator $\Pi_t(\mathbf{x},\mathbf{p},t)$ is given by:

$$\Pi_t(\mathbf{x},\mathbf{p},t) = ie\int_{-\frac{1}{2}}^{\frac{1}{2}} \mathrm{d}\xi\,\xi\,\mathbf{E}(\mathbf{x}+i\xi\nabla_{\mathbf{p}},t)\cdot\nabla_{\mathbf{p}}\ . \tag{3.56}$$

The separation into transport and constraint equations continues for all higher energy moments as well: Multiplying the equations of motion for the covariant Wigner components Eq. (3.22) – (3.31) by a factor $p_0^n$ and subsequently taking the energy average results in a set of transport equations for $\mathbb{w}^{[n]}(\mathbf{x},\mathbf{p},t)$ and a set of constraint equations. As mentioned in the previous section, this finally gives an infinite hierarchy of equations for all energy moments.

In mean field approximation, however, this infinite hierarchy is truncated at the level of the equal-time Wigner components $\mathbb{w}(\mathbf{x},\mathbf{p},t)$. This means that the transport equations for $\mathbb{w}^{[n]}(\mathbf{x},\mathbf{p},t)$ are identically fulfilled once $\mathbb{w}(\mathbf{x},\mathbf{p},t)$ obey the equations of motion Eq. (3.36) – (3.43).[17] Accordingly, in order to calculate the higher energy moments $\mathbb{w}^{[n]}(\mathbf{x},\mathbf{p},t)$ one does not have to solve the infinite hierarchy but it suffices to deduce them from $\mathbb{w}(\mathbf{x},\mathbf{p},t)$ via the constraint equations.

Due to the fact that the equal-time Wigner components $\mathbb{w}(\mathbf{x},\mathbf{p},t)$ encode the information of all higher energy moments $\mathbb{w}^{[n]}(\mathbf{x},\mathbf{p},t)$, the equal-time Wigner formalism based on the equations of motion Eq. (3.36) – (3.43) is called complete.

[17] For the first energy moments $\mathbb{w}^{[1]}(\mathbf{x},\mathbf{p},t)$ this is demonstrated in Appendix A.3. The general proof can be found in [93].

### 3.2.3 Observables and conservation laws

It is one objective of this thesis to calculate observable quantities such as the momentum distribution or the total number of created particles for the Schwinger effect in space- and time-dependent electric fields $\mathbf{E}(\mathbf{x}, t)$ later on.

As a matter of fact, the *observables* in this theory are defined via Noether's theorem and therefore correspond to conserved quantities, most notably the total charge $\mathcal{Q}$, the total energy $\mathcal{E}$, the total linear momentum $\mathbf{P}$, the total angular momentum $\mathbf{M}$ as well as the Lorentz boost operator $\mathbf{K}$.

#### 3.2.3.1 Electric charge

The invariance of the QED action under global $U(1)$ transformations:

$$\Psi'(\mathsf{x}) = \Psi(\mathsf{x}) e^{i\alpha} \,, \tag{3.57}$$

results in the electromagnetic current:

$$j^\mu(\mathsf{x}) = \tfrac{e}{2} \left[ \bar{\Psi}(\mathsf{x}), \gamma^\mu \Psi(\mathsf{x}) \right] \,, \tag{3.58}$$

with:

$$\partial_\mu^{\mathsf{x}} j^\mu(\mathsf{x}) = 0 \,. \tag{3.59}$$

The vacuum expectation value of this electromagnetic current can in fact be expressed in terms of the covariant Wigner function $\mathcal{W}^{(4)}(\mathsf{x}, \mathsf{p})$:

$$\langle \Omega | j^\mu(\mathsf{x}) | \Omega \rangle = e \int \frac{\mathrm{d}^4 p}{(2\pi)^4} \, \mathrm{tr} \left[ \gamma^\mu \mathcal{W}^{(4)}(\mathsf{x}, \mathsf{p}) \right] \,, \tag{3.60}$$

so that the space integral of its 0-component, which corresponds to the total electric charge $\mathcal{Q}$, is constant in time. Expressing it in terms of the equal-time Wigner components, it is given by:

$$\mathcal{Q} = \int \mathrm{d}^3 x \, \langle \Omega | j^0(\mathsf{x}) | \Omega \rangle = e \int \mathrm{d}\Gamma \, \mathbb{v}_0(\mathbf{x}, \mathbf{p}, t) \,, \tag{3.61}$$

with $\mathrm{d}\Gamma = \mathrm{d}^3 x \mathrm{d}^3 p / (2\pi)^3$ being the phase-space volume element.

#### 3.2.3.2 Energy and linear momentum

Invariance of the QED action under space-time translations results in the canonical energy-momentum tensor $\tilde{T}^{\mu\nu}(\mathsf{x})$, with:[18]

$$\partial_\mu^{\mathsf{x}} \tilde{T}^{\mu\nu}(\mathsf{x}) = 0 \ . \tag{3.62}$$

The energy-momentum tensor $T^{\mu\nu}(\mathsf{x})$ is composed of two parts. The electromagnetic part, which is in fact symmetric and gauge-invariant, reads:

$$T^{\mu\nu}_{\text{em}}(\mathsf{x}) = \tfrac{1}{4} g^{\mu\nu} F^{\alpha\beta}(\mathsf{x}) F_{\alpha\beta}(\mathsf{x}) - g^{\nu\alpha} F^{\mu\beta}(\mathsf{x}) F_{\alpha\beta}(\mathsf{x}) \ , \tag{3.63}$$

whereas the gauge-invariant fermionic part, which treats the oppositely charged Dirac fields symmetrically, is given by [112]:[19]

$$T^{\mu\nu}_{\text{D}}(\mathsf{x}) = \tfrac{1}{4}\left[\bar{\Psi}(\mathsf{x}), \gamma^\mu [i\partial^\nu_{\mathsf{x}} - eA^\nu(\mathsf{x})]\Psi(\mathsf{x})\right] - \tfrac{1}{4}\left[[i\partial^\nu_{\mathsf{x}} + eA^\nu(\mathsf{x})]\bar{\Psi}(\mathsf{x}), \gamma^\mu \Psi(\mathsf{x})\right] \ . \tag{3.64}$$

It is again possible to express the vacuum expectation value of the fermionic part in terms of the covariant Wigner function $\mathcal{W}^{(4)}(\mathsf{x}, \mathsf{p})$:

$$\langle \Omega | T^{\mu\nu}_{\text{D}}(\mathsf{x}) | \Omega \rangle = \int \frac{\mathrm{d}^4 p}{(2\pi)^4}\, p^\nu \, \mathrm{tr}\left[\gamma^\mu \mathcal{W}^{(4)}(\mathsf{x}, \mathsf{p})\right] . \tag{3.65}$$

As the energy-momentum tensor is conserved, the space integral of the 0-components are again constant in time. In fact, this defines the total energy $\mathcal{E}$ and the total linear momentum $\mathbf{P}$:

$$\mathcal{E} = \int \mathrm{d}^3 x \, \langle \Omega | T^{00}(\mathsf{x}) | \Omega \rangle \quad \text{and} \quad \mathbf{P} = \int \mathrm{d}^3 x \, \langle \Omega | T^{0i}(\mathsf{x}) \mathbf{e}_i | \Omega \rangle \ . \tag{3.66}$$

[18] The canonical energy-momentum tensor $\tilde{T}^{\mu\nu}(\mathsf{x})$ obtained from Noether's theorem is neither symmetric nor gauge invariant. It is, however, always possible to find a remedy by adding a total derivative term which is antisymmetric under the exchange of $\mu$ and $\lambda$ [110, 111]:

$$T^{\mu\nu}(\mathsf{x}) = \tilde{T}^{\mu\nu}(\mathsf{x}) + \partial^{\mathsf{x}}_\lambda \Sigma^{\mu\nu\lambda}(\mathsf{x}) = \tilde{T}^{\mu\nu}(\mathsf{x}) + \partial^{\mathsf{x}}_\lambda [F^{\mu\lambda}(\mathsf{x}) A^\nu(\mathsf{x})] \ .$$

[19] The fermionic part of the energy-momentum tensor $T^{\mu\nu}_{\text{D}}(\mathsf{x})$ is still not symmetric. A symmetric definition would be given by the linear combination:

$$T^{\mu\nu}_{\text{D,sym}}(\mathsf{x}) = \tfrac{1}{2}[T^{\mu\nu}_{\text{D}}(\mathsf{x}) + T^{\nu\mu}_{\text{D}}(\mathsf{x})] \ .$$

Note, however, that a symmetric definition is not absolutely necessary for the purpose of defining the total energy $\mathcal{E}$ and the total linear momentum $\mathbf{P}$.

Expressing $\mathcal{E}$ in terms of the equal-time Wigner components, this gives:[20]

$$\mathcal{E} = \int \mathrm{d}\Gamma \big[ m\mathbb{s}(\mathbf{x},\mathbf{p},t) + \mathbf{p}\cdot\mathbb{v}(\mathbf{x},\mathbf{p},t)\big] + \tfrac{1}{2}\int \mathrm{d}^3x \left[ |\mathbf{E}(\mathbf{x},t)|^2 + |\mathbf{B}(\mathbf{x},t)|^2 \right] , \quad (3.67)$$

whereas $\mathbf{P}$ reads:

$$\mathbf{P} = \int \mathrm{d}\Gamma\, \mathbf{p}\, \mathbb{v}_0(\mathbf{x},\mathbf{p},t) + \int \mathrm{d}^3x\, \mathbf{E}(\mathbf{x},t)\times\mathbf{B}(\mathbf{x},t) \ . \quad (3.68)$$

#### 3.2.3.3 Angular momentum and Lorentz boost

Invariance of the QED action under homogeneous Lorentz transformations results in the generalized angular momentum tensor $J^{\mu\nu\lambda}(\mathsf{x})$, with:

$$\partial^{\mathsf{x}}_{\mu} J^{\mu\nu\lambda}(\mathsf{x}) = 0 \ . \quad (3.69)$$

Treating the oppositely charged Dirac fields symmetrically, it is given by [110]:[21]

$$J^{\mu\nu\rho}(\mathsf{x}) = x^{\nu}T^{\mu\rho}_{\mathrm{sym}}(\mathsf{x}) - x^{\rho}T^{\mu\nu}_{\mathrm{sym}}(\mathsf{x}) + \tfrac{1}{4}\left[\bar{\Psi}(\mathsf{x}), \left\{\gamma^{\mu}, \tfrac{\sigma^{\nu\rho}}{2}\right\}\Psi(\mathsf{x})\right] \ . \quad (3.70)$$

The first two terms describe the orbital part whereas the last term originates from the spin of the Dirac fields. The vacuum expectation value of the orbital part can be taken from the previous section, whereas the spin contribution is given by:[22]

$$\langle\Omega| J^{\mu\nu\rho}_{\mathrm{spin}}(\mathsf{x})|\Omega\rangle = \tfrac{1}{2}\int \frac{\mathrm{d}^4p}{(2\pi)^4}\, \mathrm{tr}\left[\left\{\gamma^{\mu}, \tfrac{\sigma^{\nu\rho}}{2}\right\}\mathcal{W}^{(4)}(\mathsf{x},\mathsf{p})\right] \ . \quad (3.71)$$

[20] In order to calculate the fermionic part of the total energy, the expression for the first energy moment $\mathbb{v}_0^{[1]}(\mathbf{x},\mathbf{p},t)$ in Eq. (3.50) is needed:

$$\int \mathrm{d}^3x\, \langle\Omega|T^{00}_{\mathrm{D}}(\mathsf{x})|\Omega\rangle = \int \mathrm{d}\Gamma\, \mathbb{v}_0^{[1]}(\mathbf{x},\mathbf{p},t) \ .$$

As an integration over the whole phase space is performed, this can be simplified to give:

$$\int \mathrm{d}^3x\, \langle\Omega|T^{00}_{\mathrm{D}}(\mathsf{x})|\Omega\rangle = \int \mathrm{d}\Gamma [m\mathbb{s} + \boldsymbol{\Pi}\cdot\mathbb{v} - \Pi_t \mathbb{v}_0] = \int \mathrm{d}\Gamma [m\mathbb{s} + \mathbf{p}\cdot\mathbb{v}] \ .$$

[21] Note, that the symmetric energy-momentum tensor $T^{\mu\nu}_{\mathrm{sym}}(\mathsf{x})$ has to be used in this expression.

[22] As one is forced to use the symmetric energy-momentum tensor, the fermionic part is given by:

$$\langle\Omega|T^{\mu\nu}_{\mathrm{D,sym}}(\mathsf{x})|\Omega\rangle = \tfrac{1}{2}\int \tfrac{\mathrm{d}^4p}{(2\pi)^4}\, \mathrm{tr}\left[(p^{\nu}\gamma^{\mu} + p^{\mu}\gamma^{\nu})\mathcal{W}^{(4)}(\mathsf{x},\mathsf{p})\right] .$$

Even though this expression differs from Eq. (3.65), it turns out that the integration over the whole space gives the same result:

$$\int \mathrm{d}^3x \langle\Omega|T^{0\nu}_{\mathrm{D}}(\mathsf{x})|\Omega\rangle = \int \mathrm{d}^3x \langle\Omega|T^{0\nu}_{\mathrm{D,sym}}(\mathsf{x})|\Omega\rangle \ ,$$

such that the observable quantities Eq. (3.67) and Eq. (3.68) remain unchanged.

As the generalized angular momentum tensor is conserved, the space integral of the 0-components are again constant in time. In fact, this defines the total angular momentum **M** and the Lorentz boost operator **K**:

$$\mathbf{M} = \int \mathrm{d}^3x\, \langle\Omega|\tfrac{1}{2}\epsilon_{ijk}J^{0jk}(\mathsf{x})\mathbf{e}_i|\Omega\rangle \quad \text{and} \quad \mathbf{K} = \int \mathrm{d}^3x\, \langle\Omega|J^{00i}(\mathsf{x})\mathbf{e}_i|\Omega\rangle \ . \tag{3.72}$$

Expressing these quantities in terms of the equal-time Wigner components, **M** is given by:

$$\mathbf{M} = \int \mathrm{d}\Gamma \left[\mathbf{x}\times\mathbf{p}\,\mathbb{v}_0(\mathbf{x},\mathbf{p},t) - \tfrac{1}{2}\mathbb{a}(\mathbf{x},\mathbf{p},t)\right] + \int \mathrm{d}^3x\,\mathbf{x}\times\mathbf{E}(\mathbf{x},t)\times\mathbf{B}(\mathbf{x},t) \ , \tag{3.73}$$

whereas **K** reads:

$$\mathbf{K} = t\,\mathbf{P} - \int \mathrm{d}\Gamma\,\mathbf{x}\big[m\mathbb{s}(\mathbf{x},\mathbf{p},t) + \mathbf{p}\cdot\mathbb{v}(\mathbf{x},\mathbf{p},t)\big] - \tfrac{1}{2}\int \mathrm{d}^3x\,\mathbf{x}\big[|\mathbf{E}(\mathbf{x},t)|^2 + |\mathbf{B}(\mathbf{x},t)|^2\big] \ . \tag{3.74}$$

Chapter 4

# Schwinger effect for $\mathbf{E}(t)$

This chapter is dedicated to the quantum kinetic formulation of the Schwinger effect in spatially homogeneous, time-dependent electric fields $\mathbf{E}(t)$. In Section 4.1, I briefly review the first principle derivation of the quantum Vlasov equation in its nowadays widely used form [29, 30, 65]. In Section 4.2, I show that the equal-time Wigner formalism for spatially homogeneous, time-dependent electric fields is in fact totally equivalent to the quantum Vlasov equation. In the subsequent Section 4.3, I derive analytic expressions for both the one-particle distribution function and the equal-time Wigner components for the exactly solvable cases of a static as well as a pulsed electric field [33]. In Section 4.4, I eventually discuss the Schwinger effect in a pulsed electric field with sub-cycle structure on the basis of a numerical simulation of the quantum Vlasov equation [36, 37].

## 4.1 Quantum Vlasov equation

As already mentioned, it was only in the mid 1990s that a first principle derivation of a quantum kinetic equation accounting for electron-positron pair creation in the presence of a spatially homogeneous, time-dependent electric field $\mathbf{E}(t)$ has been given. Since then a huge number of studies for various types of time dependencies has been performed and achieved in fact significant progress [36, 37, 57, 66, 67, 68, 69, 70, 102, 103].

This formalism is based upon the QED Lagrangian Eq. (3.1), with the field strength tensor treated as as c-number again.[1] Accordingly, the equation of motion

[1]In this mean field approximation, it is assumed that the magnetic field vanishes whereas the electric field is time dependent and points along a given direction:

$$\mathbf{B}(t) = 0 \quad \text{and} \quad \mathbf{E}(t) = E(t)\mathbf{e}_3 \ .$$

It is convenient to represent this type of electromagnetic field in temporal gauge $A_0(\mathrm{x}) = 0$. Accordingly, the electric field is derived from the vector potential:

$$\mathbf{E}(t) = -\tfrac{\mathrm{d}}{\mathrm{d}t}\mathbf{A}(t) \quad \text{with} \quad \mathbf{A}(t) = A(t)\mathbf{e}_3 \ .$$

for the Dirac field Eq. (3.2) reads:

$$\left[i\gamma^0 \tfrac{\partial}{\partial t} + i\boldsymbol{\gamma} \cdot [\nabla_{\mathbf{x}} - ie\mathbf{A}(t)] - m\right] \Psi(\mathbf{x}, t) = 0 \; . \tag{4.1}$$

A decomposition of the Dirac field into its Fourier modes is indicated as the vector potential is spatially homogeneous:

$$\Psi(\mathbf{x}, t) = \int \frac{\mathrm{d}^3 q}{(2\pi)^3}\, e^{i\mathbf{q}\cdot\mathbf{x}} \psi(\mathbf{q}, t) \; , \tag{4.2}$$

so that the corresponding equation of motion for the Fourier modes reads:[2]

$$\left[i\gamma^0 \tfrac{\partial}{\partial t} - \boldsymbol{\gamma} \cdot \boldsymbol{\pi}(\mathbf{q}, t) - m\right] \psi(\mathbf{q}, t) = 0 \; . \tag{4.3}$$

This equation is solved by introducing the following ansatz:

$$\psi(\mathbf{q}, t) = \left[i\gamma^0 \tfrac{\partial}{\partial t} - \boldsymbol{\gamma} \cdot \boldsymbol{\pi}(\mathbf{q}, t) + m\right] \phi(\mathbf{q}, t) \; , \tag{4.4}$$

so that the first order differential equation for $\psi(\mathbf{q}, t)$ is transformed into a second order differential equation for $\phi(\mathbf{q}, t)$:[3]

$$\left[\tfrac{\partial^2}{\partial t^2} + \omega^2(\mathbf{q}, t) + ieE(t)\gamma^0\gamma^3\right] \phi(\mathbf{q}, t) = 0 \; . \tag{4.5}$$

It is convenient to expand $\phi(\mathbf{q}, t)$ in a basis[4] spanned by the eigenvectors of $\gamma^0\gamma^3$:

$$\phi(\mathbf{q}, t) = \sum_{r=1}^{4} g_r(\mathbf{q}, t) R_r \; . \tag{4.6}$$

[2]The following notation is used throughout: The variable $\mathbf{q}$ is a canonical momentum, whereas $\boldsymbol{\pi}(\mathbf{q}, t) = \mathbf{q} - e\mathbf{A}(t)$ denotes the time-dependent kinetic momentum on a trajectory. Moreover, this has to be clearly distinguished from the kinetic momentum in phase space $\mathbf{p}$.

[3]It is convenient to introduce the following notation for later use: The transverse energy squared is defined as $\epsilon_\perp^2 = m^2 + \mathbf{q}_\perp^2$ so that the total energy squared is given by:

$$\omega^2(\mathbf{q}, t) = m^2 + \boldsymbol{\pi}^2(\mathbf{q}, t) = \epsilon_\perp^2 + \pi_3^2(q_3, t) \; .$$

[4]The Dirac bilinear $\gamma^0\gamma^3 = -i\sigma^{03}$ is proportional to the Lorentz boost operator along the electric field direction $K_3 = \frac{1}{2}\sigma^{03}$. The four corresponding eigenvectors $R_r$:

$$K_3 R_{r=\{1,2\}} = +\tfrac{i}{2} R_{r=\{1,2\}} \quad \text{and} \quad K_3 R_{r=\{3,4\}} = -\tfrac{i}{2} R_{r=\{3,4\}}$$

form a complete basis. As the commutator $[K_3, M_3] = 0$ vanishes, where $M_3 = \frac{1}{2}\sigma^{12}$ denotes the spin operator along the electric field direction, one may further characterize the eigenvectors according to their intrinsic spin:

$$M_3 R_{r=\{1,3\}} = +\tfrac{1}{2} R_{r=\{1,3\}} \quad \text{and} \quad M_3 R_{r=\{2,4\}} = -\tfrac{1}{2} R_{r=\{2,4\}} \; .$$

Inserting this ansatz into Eq. (4.3), it turns out that each $g_r(\mathbf{q},t)$ obeys the equation of motion of a time-dependent harmonic oscillator with an additional complex term:

$$\left[\frac{\partial^2}{\partial t^2} + \omega^2(\mathbf{q},t) + ieE(t)\right] g_r(\mathbf{q},t) = 0 \quad \text{with} \quad r = \{1,2\} \,, \tag{4.7}$$

$$\left[\frac{\partial^2}{\partial t^2} + \omega^2(\mathbf{q},t) - ieE(t)\right] g_r(\mathbf{q},t) = 0 \quad \text{with} \quad r = \{3,4\} \,, \tag{4.8}$$

which are in general not exactly solvable; exceptions are the constant electric field $E(t) = E_0$ and the pulsed electric field $E(t) = E_0 \operatorname{sech}^2(\frac{t}{\tau})$ to be discussed in more detail later on. Note that Eq. (4.7) – (4.8) are second order ODEs, each of them possessing two linearly independent solutions $g_r^{(\pm)}(\mathbf{q},t)$, so that $\phi(\mathbf{q},t)$ is composed of eight independent contributions in total.[5] In contrast, the Dirac equation Eq. (4.3) allows merely for four linearly independent solutions. This redundancy is removed by choosing only one complete set of solutions, either the one with $r = \{1,2\}$ or the one with $r = \{3,4\}$ [28]. Choosing the first set, for instance, Eq. (4.4) can be written as a linear combination:

$$\psi(\mathbf{q},t) = \sum_{r=1}^{2} \left[u_r(\mathbf{q},t)a_r^{(+)}(\mathbf{q}) + v_r(-\mathbf{q},t)a_r^{(-)}(-\mathbf{q})\right] \,, \tag{4.9}$$

with $a_r^{(\pm)}(\mathbf{q})$ being the coefficients of the corresponding terms. Moreover, the spinors are given by:[6]

$$u_r(\mathbf{q},t) = \left[i\gamma^0 \frac{\partial}{\partial t} - \boldsymbol{\gamma}\cdot\boldsymbol{\pi}(\mathbf{q},t) + m\right] g^{(+)}(\mathbf{q},t)R_r \,, \tag{4.10}$$

$$v_r(-\mathbf{q},t) = \left[i\gamma^0 \frac{\partial}{\partial t} - \boldsymbol{\gamma}\cdot\boldsymbol{\pi}(\mathbf{q},t) + m\right] g^{(-)}(\mathbf{q},t)R_r \,. \tag{4.11}$$

The coefficients are interpreted as anticommuting creation/annihilation operators for particles and antiparticles upon canonical quantization:

$$a_r^{(+)}(\mathbf{q}) \to a_r(\mathbf{q}) \qquad \text{and} \qquad a_r^{(-)}(-\mathbf{q}) \to b_r^\dagger(-\mathbf{q}) \,, \tag{4.12}$$

with the only non-vanishing anticommutators being given by:

$$\left\{a_r(\mathbf{q}), a_s^\dagger(\mathbf{q}')\right\} = \left\{b_r(\mathbf{q})\,,\, b_s^\dagger(\mathbf{q}')\right\} = (2\pi)^3\delta_{rs}\delta(\mathbf{q}-\mathbf{q}') \,. \tag{4.13}$$

[5] In the limit of a vanishing electric field, they turn into the positive and negative energy plane wave solutions:

$$g_{r,\mathrm{vac}}^{(\pm)}(\mathbf{q},t) \sim e^{\mp i\omega(\mathbf{q})t} \quad \text{with} \quad \omega^2(\mathbf{q}) = m^2 + \mathbf{q}^2 \,.$$

[6] The linearly independent solutions $g_1^{(\pm)}(\mathbf{q},t)$ and $g_2^{(\pm)}(\mathbf{q},t)$ are identical due to the uniqueness theorem for ODEs:

$$g_1^{(\pm)}(\mathbf{q},t) = g_2^{(\pm)}(\mathbf{q},t) \equiv g^{(\pm)}(\mathbf{q},t) \,.$$

It has to be emphasized that a particle interpretation of the field quanta is only possible in the case of plane wave solutions.[7] The actual mode functions $g^{(\pm)}(\mathbf{q},t)$, however, are no plane waves once an electric field is present so that a particle interpretation is not feasible. A further consequence of a non-vanishing electric field is that the Hamiltonian operator achieves off-diagonal elements, accounting for the possibility of pair creation as well as pair annihilation.

The off-diagonal Hamiltonian operator can be diagonalized by performing a unitary non-equivalent change of basis to a *quasi-particle representation* via a time-dependent Bogoliubov transformation:

$$A_r(\mathbf{q},t) \quad = \alpha(\mathbf{q},t)\, a_r(\mathbf{q}) - \beta^*(\mathbf{q},t)\, b_r^\dagger(-\mathbf{q}) \;, \tag{4.14}$$

$$B_r^\dagger(-\mathbf{q},t) = \beta(\mathbf{q},t)\, a_r(\mathbf{q}) + \alpha^*(\mathbf{q},t)\, b_r^\dagger(-\mathbf{q}) \;. \tag{4.15}$$

In order to be a canonical transformation, the Bogoliubov coefficients $\alpha(\mathbf{q},t)$ and $\beta(\mathbf{q},t)$ must fulfill:

$$|\alpha(\mathbf{q},t)|^2 + |\beta(\mathbf{q},t)|^2 = 1 \;. \tag{4.16}$$

The creation/annihilation operators, even though they are now considered as being time-dependent, still fulfill the equal-time anticommutation relations:

$$\left\{A_r(\mathbf{q},t), A_s^\dagger(\mathbf{q}',t)\right\} = \left\{B_r(\mathbf{q},t)\,,\, B_s^\dagger(\mathbf{q}',t)\right\} = (2\pi)^3\delta_{rs}\delta(\mathbf{q}-\mathbf{q}') \;. \tag{4.17}$$

In the vacuum, when there is no electric field present, the Bogoliubov transformation is trivial and the two operator bases coincide:

$$\alpha(\mathbf{q},t_{\text{vac}}) = 1 \quad \text{and} \quad \beta(\mathbf{q},t_{\text{vac}}) = 0 \qquad \text{with} \qquad t_{\text{vac}} \to -\infty \;. \tag{4.18}$$

In fact, the time-dependent Bogoliubov transformation is equivalent to expanding the Fourier modes $\psi(\mathbf{q},t)$ in a different basis:

$$\psi(\mathbf{q},t) = \sum_{r=1}^{2} \Big[U_r(\mathbf{q},t)A_r(\mathbf{q},t) + V_r(-\mathbf{q},t)B_r^\dagger(-\mathbf{q},t)\Big] \;. \tag{4.19}$$

As a matter of fact, any complete basis of spinors $U_r(\mathbf{q},t)$ and $V_r(\mathbf{q},t)$ would do this job. The *adiabatic basis*, however, which is chosen in close analogy to the basis

[7]The spinor normalization is chosen according to $u_r^\dagger(\mathbf{q},t)u_s(\mathbf{q},t) = v_r^\dagger(\mathbf{q},t)v_s(\mathbf{q},t) = \delta_{rs}$, so that the properly normalized positive and negative energy plane wave solutions are given by:

$$g^{(\pm)}_{\text{vac}}(\mathbf{q},t) = \frac{e^{\mp i\omega(\mathbf{q})t}}{\sqrt{2\omega(\mathbf{q})[\omega(\mathbf{q}) \mp q_3]}} \;.$$

of spinors $u_r(\mathbf{q},t)$ and $v_r(\mathbf{q},t)$, respectively, fits especially well:

$$U_r(\mathbf{q},t) = \left[\ \gamma^0\omega(\mathbf{q},t) - \boldsymbol{\gamma}\cdot\boldsymbol{\pi}(\mathbf{q},t) + m\right] G^{(+)}(\mathbf{q},t)R_r \ , \tag{4.20}$$

$$V_r(-\mathbf{q},t) = \left[-\gamma^0\omega(\mathbf{q},t) - \boldsymbol{\gamma}\cdot\boldsymbol{\pi}(\mathbf{q},t) + m\right] G^{(-)}(\mathbf{q},t)R_r \ , \tag{4.21}$$

with the adiabatic mode functions $G^{(\pm)}(\mathbf{q},t)$ being defined as:[8]

$$G^{(\pm)}(\mathbf{q},t) = \frac{e^{\mp i\Theta(\mathbf{q},t_0,t)}}{\sqrt{2\omega(\mathbf{q},t)[\omega(\mathbf{q},t) \mp \pi_3(q_3,t)]}} \qquad \text{with} \qquad \Theta(\mathbf{q},t_0,t) = \int_{t_0}^{t} \omega(\mathbf{q},t')dt' \ . \tag{4.22}$$

The lower bound $t_0$ of the dynamical phase $\Theta(\mathbf{q},t;t_0)$ is not fully determined as it only fixes an arbitrary phase. Note that the adiabatic mode functions $G^{(\pm)}(\mathbf{q},t)$ behave like plane waves in the limit of a vanishing electric field.[9]

The *one-particle distribution function* $\mathcal{F}(\mathbf{q},t)$ is defined as the instantaneous quasi-particle number density for a given canonical momentum $\mathbf{q}$. As quasi-particles with different values of $r$ behave identically when there is no magnetic field present, it is in fact convenient to include the sum over both values $r = \{1,2\}$ in the definition:[10]

$$\mathcal{F}(\mathbf{q},t) \equiv \lim_{V\to\infty} \sum_{r=1}^{2} \frac{\langle\Omega|A_r^\dagger(\mathbf{q},t)A_r(\mathbf{q},t)|\Omega\rangle}{V} = 2|\beta(\mathbf{q},t)|^2 \ , \tag{4.23}$$

with vacuum initial conditions for $t_{\text{vac}} \to -\infty$ being assumed:

$$\langle\Omega|a_r^\dagger(\mathbf{q})a_r(\mathbf{q})|\Omega\rangle = \langle\Omega|b_r^\dagger(\mathbf{q})b_r(\mathbf{q})|\Omega\rangle = 0 \ . \tag{4.24}$$

Accordingly, the knowledge of $\beta(\mathbf{q},t)$ allows for the calculation of $\mathcal{F}(\mathbf{q},t)$. Note that the different representations Eq. (4.10) – (4.11) and Eq. (4.20) – (4.21), respectively, translate into a relation between the Bogoliubov coefficients:

$$\alpha(\mathbf{q},t) = \ \ i\epsilon_\perp G^{(-)}(\mathbf{q},t)\left[\tfrac{\partial}{\partial t} - i\omega(\mathbf{q},t)\right] g^{(+)}(\mathbf{q},t) \ , \tag{4.25}$$

$$\beta(\mathbf{q},t) = -i\epsilon_\perp G^{(+)}(\mathbf{q},t)\left[\tfrac{\partial}{\partial t} + i\omega(\mathbf{q},t)\right] g^{(+)}(\mathbf{q},t) \ , \tag{4.26}$$

[8] In fact, this choice is very convenient as $U_r(\mathbf{q},t)$ and $V_r(\mathbf{q},t)$ coincide with $u_r(\mathbf{q},t)$ and $v_r(\mathbf{q},t)$, respectively, in the limit of a vanishing electric field.

[9] This is the deeper reason why the one-particle distribution function $\mathcal{F}(\mathbf{q},t)$ can be interpreted as the momentum distribution of real particles at asymptotic times when the electric field vanishes.

[10] Here, $V$ denotes the configuration space volume, which has to be taken to infinity. This factor is necessary to cancel the divergence which appears in the anticommutators:

$$\left\{a_r(\mathbf{q}), a_r^\dagger(\mathbf{q})\right\} = \left\{b_r(\mathbf{q})\, ,\, b_r^\dagger(\mathbf{q})\right\} = (2\pi)^3\delta(\mathbf{0}) = V \ ,$$

Alternatively, one could derive the quantum Vlasov equation in a finite box from the very beginning so that $\mathbf{q}$ can take only discrete values. In this case, the limit $V \to \infty$ would be taken at the very end of the derivation.

so that their time derivatives form an ODE system:

$$\dot{\alpha}(\mathbf{q},t) = \frac{1}{2}Q(\mathbf{q},t)\beta(\mathbf{q},t)e^{2i\Theta(\mathbf{q},t_0,t)} \quad , \tag{4.27}$$

$$\dot{\beta}(\mathbf{q},t) = -\frac{1}{2}Q(\mathbf{q},t)\alpha(\mathbf{q},t)e^{-2i\Theta(\mathbf{q},t_0,t)} \ , \tag{4.28}$$

with $Q(\mathbf{q},t)$ being defined as:

$$Q(\mathbf{q},t) = \frac{eE(t)\epsilon_\perp}{\omega^2(\mathbf{q},t)} \ . \tag{4.29}$$

Introducing the auxiliary quantity $\mathcal{C}(\mathbf{q},t)$ which describes the density of quasi-particle pair creation:[11]

$$\mathcal{C}(\mathbf{q},t) \equiv \lim_{V\to\infty}\sum_{r=1}^{2}\frac{\langle\Omega|B_r^\dagger(\mathbf{q},t)A_r^\dagger(\mathbf{q},t)|\Omega\rangle}{V} = 2\alpha^*(\mathbf{q},t)\beta(\mathbf{q},t) \ , \tag{4.30}$$

the ODE system Eq. (4.27) – (4.28) can be rewritten in terms of the functions $\mathcal{C}(\mathbf{q},t)$ and $\mathcal{F}(\mathbf{q},t)$:[12]

$$\dot{\mathcal{C}}(\mathbf{q},t) = -Q(\mathbf{q},t)\left[1-\mathcal{F}(\mathbf{q},t)\right]e^{-2i\Theta(\mathbf{q},t_0,t)} \ , \tag{4.31}$$

$$\dot{\mathcal{F}}(\mathbf{q},t) = -Q(\mathbf{q},t)\,\mathrm{Re}\left[\mathcal{C}(\mathbf{q},t)e^{2i\Theta(\mathbf{q},t_0,t)}\right] \quad . \tag{4.32}$$

Integrating the first equation from $t_{\mathrm{vac}}$ to $t$ yields the quantum Vlasov equation for the one-particle distribution function $\mathcal{F}(\mathbf{q},t)$ in its integro-differential form [30]:

$$\dot{\mathcal{F}}(\mathbf{q},t) = Q(\mathbf{q},t)\int_{t_{\mathrm{vac}}}^{t} Q(\mathbf{q},t')[1-\mathcal{F}(\mathbf{q},t')]\cos\left[2\Theta(\mathbf{q},t',t)\right] \ , \tag{4.33}$$

along with the vacuum initial condition $\mathcal{F}(\mathbf{q},t_{\mathrm{vac}}) = 0$. Accordingly, the pair creation process shows non-Markovian behavior owing to memory effects and non-locality in time. Additionally, $[1-\mathcal{F}(\mathbf{q},t)]$ might be interpreted as a Pauli-blocking factor which accounts for the correct quantum statistics.

It turned out to be advantageous to rewrite this integro-differential equation in terms of a first order ODE system as the numerical solution of integro-differential equations is usually quite challenging [65]. To this end, two auxiliary quantities are

[11]The sum over $r = \{1,2\}$ is taken again as one is only interest in the total density of quasi-particle pair creation. Additionally, the infinite volume limit $V \to \infty$ has to be taken in order to cancel the divergence originating from the anticommutators.

[12]In order to derive these equations, the relation $\frac{\partial}{\partial t}|\beta(\mathbf{q},t)|^2 = 2\,\mathrm{Re}[\beta(\mathbf{q},t)\dot{\beta}^*(\mathbf{q},t)]$ is used.

introduced:

$$\mathcal{G}(\mathbf{q},t) = \int_{t_{\text{vac}}}^{t} Q(\mathbf{q},t')[1-\mathcal{F}(\mathbf{q},t')]\cos\left[2\Theta(\mathbf{q},t',t)\right] , \tag{4.34}$$

$$\mathcal{H}(\mathbf{q},t) = \int_{t_{\text{vac}}}^{t} Q(\mathbf{q},t')[1-\mathcal{F}(\mathbf{q},t')]\sin\left[2\Theta(\mathbf{q},t',t)\right] , \tag{4.35}$$

so that:

$$\dot{\mathcal{F}}(\mathbf{q},t) = Q(\mathbf{q},t)\mathcal{G}(\mathbf{q},t) , \tag{4.36}$$

$$\dot{\mathcal{G}}(\mathbf{q},t) = Q(\mathbf{q},t)[1-\mathcal{F}(\mathbf{q},t)] - 2\omega(\mathbf{q},t)\mathcal{H}(\mathbf{q},t) , \tag{4.37}$$

$$\dot{\mathcal{H}}(\mathbf{q},t) = 2\omega(\mathbf{q},t)\mathcal{G}(\mathbf{q},t) , \tag{4.38}$$

along with vacuum initial conditions $\mathcal{F}(\mathbf{q},t_{\text{vac}}) = \mathcal{G}(\mathbf{q},t_{\text{vac}}) = \mathcal{H}(\mathbf{q},t_{\text{vac}}) = 0$.

As mentioned previously, a consistent mean field description of the Schwinger effect requires to include the self-induced electric field due to the pair creation process:[13] The charged particles, which are created by the applied electric field $\mathbf{E}_{\text{ext}}(t)$ serve as a source of an internal electric field $\mathbf{E}_{\text{int}}(t)$. Accordingly, one is faced with the problem of describing the pair creation process in the presence of the total electric field $\mathbf{E}(t) = \mathbf{E}_{\text{ext}}(t) + \mathbf{E}_{\text{int}}(t)$ in a self-consistent way.

The internal electric field in mean field approximation can in fact be calculated from the vacuum expectation value of the electromagnetic current:[14]

$$\dot{E}_{\text{int}}(t) = -\langle\Omega|j^3(t)|\Omega\rangle = -\tfrac{e}{2}\int \mathrm{d}^3x \langle\Omega|\left[\bar{\Psi}(\mathbf{x},t), \gamma^3\Psi(\mathbf{x},t)\right]|\Omega\rangle . \tag{4.39}$$

As a matter of fact, this expression exhibits a logarithmic UV-divergence, which has to be regularized and renormalized. This finally yields [65]:

$$\dot{E}_{\text{int}}(t) = -2e\int \frac{\mathrm{d}^3q}{(2\pi)^3}\left[\frac{\pi_3(q_3,t)}{\omega(\mathbf{q},t)}\mathcal{F}(\mathbf{q},t) + \frac{\omega(\mathbf{q},t)}{eE(t)}\dot{\mathcal{F}}(\mathbf{q},t) - \frac{e\dot{E}(t)\epsilon_\perp^2}{4\omega^5(\mathbf{q},t)}\right] . \tag{4.40}$$

The first two terms represent the conduction and polarization current, respectively, whereas the last term is introduced as a counterterm in the course of charge renormalization. Together with the quantum Vlasov equation, this provides a consistent mean field description of the Schwinger effect in time-dependent electric fields.

[13] As I will not consider the backreaction issue in the following, I only give the final result for completeness' sake. Details can be found in [53, 65, 71].

[14] Note, that $\mathbf{E}_{\text{int}}(t)$ points along the same direction as $\mathbf{E}_{\text{ext}}(t)$ by virtue of cylindrical symmetry.

## 4.2 Equal-time Wigner formalism

There are now two at first sight rather different formalisms at hand: The equation of motion for the equal-time Wigner function has been derived by Wigner transforming the vacuum expectation value of the fermionic density matrix whereas the quantum Vlasov equation results from canonical quantization of the Dirac field and subsequent Bogoliubov transformation to a quasi-particle representation.

It has to be emphasized that the equal-time Wigner formalism allows for an appropriate description of the pair creation process in the presence of a space- and time-dependent electromagnetic field $F^{\mu\nu}(\mathrm{x})$ whereas the quantum Vlasov equation is only valid for spatially homogeneous, time-dependent electric fields $\mathbf{E}(t) = E(t)\mathbf{e}_3$. Accordingly, the question arises whether there is a deeper connection between them so that the equal-time Wigner formalism can in fact be related to the quantum Vlasov equation.[15]

As already mentioned, it is an advantage of the equal-time Wigner formalism that it is formulated in terms of an initial value problem. The appropriate initial conditions for describing the Schwinger effect are provided by the vacuum value of the equal-time Wigner function, which can be easily calculated by explicitly evaluating Eq. (3.33) for the free Dirac field. The only non-vanishing equal-time Wigner components are then given by:[16]

$$\mathbb{s}_{\mathrm{vac}}(\mathbf{p}) = -\frac{2m}{\omega(\mathbf{p})} \qquad \text{and} \qquad \mathbb{v}_{\mathrm{vac}}(\mathbf{p}) = -\frac{2\mathbf{p}}{\omega(\mathbf{p})} \ . \tag{4.41}$$

They are in fact spatially homogeneous and time-independent as it should be.

Considering the equations of motion for the equal-time Wigner components Eq. (3.36) – (3.43) in the presence of a time-dependent electric field $\mathbf{E}(t) = E(t)\mathbf{e}_3$, the pseudo-differential operators Eq. (3.45) – (3.47) reduce to local ones:[17]

$$D_t(\mathbf{x}, \mathbf{p}, t) \longrightarrow \frac{\partial}{\partial t} + eE(t)\frac{\partial}{\partial p_3} \ , \tag{4.42}$$

$$\mathbf{D}(\mathbf{x}, \mathbf{p}, t) \longrightarrow \nabla_{\mathbf{x}} \ , \tag{4.43}$$

$$\mathbf{\Pi}(\mathbf{x}, \mathbf{p}, t) \longrightarrow \mathbf{p} \ . \tag{4.44}$$

[15] The equal-time Wigner formalism in the limit of a time-dependent electric field has in fact been considered already at the time when the formalism was invented [31]. However, as the nowadays widely used form of the quantum Vlasov equation has not been derived by then [30, 65], the deep connection between them has been pointed out only recently in the context of this thesis [33].

[16] The detailed calculation can be found in Appendix A.4.

[17] For a more general time-dependent electromagnetic field $F^{\mu\nu}(t)$, these operators are given by:

$$D_t \to \frac{\partial}{\partial t} + e\mathbf{E}(t)\cdot\nabla_{\mathbf{p}} \quad \text{and} \quad \mathbf{D} \to \nabla_{\mathbf{x}} + e\mathbf{B}\times\nabla_{\mathbf{p}} \quad \text{and} \quad \mathbf{\Pi} \to \mathbf{p} \ .$$

Moreover, all spatial derivatives vanish owing to the spatial homogeneity of both the electric field $E(t)$ and the vacuum value of the equal-time Wigner components $\mathbb{w}_{\text{vac}}(\mathbf{p})$. Accordingly, the equations of motion for the *spatially homogeneous* equal-time Wigner components $\mathbb{w}(\mathbf{p},t)$ are given by:

$$\begin{aligned}
D_t\,\mathbb{s} \quad - 2\mathbf{p}\cdot\mathbb{t}_1 &= 0 \ , && (4.45)\\
D_t\,\mathbb{p} \quad + 2\mathbf{p}\cdot\mathbb{t}_2 &= -2m\ \mathbb{a}_0 \ , && (4.46)\\
D_t\,\mathbb{v}_0 &= 0 \ , && (4.47)\\
D_t\,\mathbb{a}_0 &= 2m\ \mathbb{p} \ , && (4.48)\\
D_t\,\mathbb{v} \quad + 2\mathbf{p}\times\mathbb{a} &= -2m\ \mathbb{t}_1 \ , && (4.49)\\
D_t\,\mathbb{a} \quad + 2\mathbf{p}\times\mathbb{v} &= 0 \ , && (4.50)\\
D_t\,\mathbb{t}_1 \quad + 2\mathbf{p}\,\mathbb{s} &= 2m\ \mathbb{v} \ , && (4.51)\\
D_t\,\mathbb{t}_2 \quad - 2\mathbf{p}\,\mathbb{p} &= 0 \ . && (4.52)
\end{aligned}$$

First of all, the equal-time Wigner component $\mathbb{v}_0(\mathbf{p},t)$ decouples from this first order, partial differential equation (PDE) system. Due to the fact that $\mathbb{v}_{0,\text{vac}}(\mathbf{p})=0$, this yields:[18]

$$\mathbb{v}_0(\mathbf{p},t)=0\ . \tag{4.53}$$

Moreover, the equal-time Wigner components $\mathbb{p}(\mathbf{p},t)$, $\mathbb{a}_0(\mathbf{p},t)$ and $\mathbb{t}_2(\mathbf{p},t)$ form a closed set. As their vacuum values vanish as well, one finds:

$$\mathbb{p}(\mathbf{p},t)=\mathbb{a}_0(\mathbf{p},t)=\mathbb{t}_2(\mathbf{p},t)=0\ . \tag{4.54}$$

As a consequence, one is left with the PDE system:

$$\left[\frac{\partial}{\partial t}+eE(t)\frac{\partial}{\partial p_3}\right]\mathbb{z}(\mathbf{p},t)=\mathbb{M}(\mathbf{p})\mathbb{z}(\mathbf{p},t)\ , \tag{4.55}$$

where $\mathbb{z}(\mathbf{p},t)\equiv\{\mathbb{s},\mathbb{v},\mathbb{a},\mathbb{t}_1\}(\mathbf{p},t)$ denotes the vector consisting of the remaining ten equal-time Wigner components. Moreover, $\mathbb{M}(\mathbf{p})$ is a shorthand notation for the $10\times10$ matrix in accordance with Eq. (4.45) and Eq. (4.49) – (4.51). As Eq. (4.55) is in fact a hyperbolic PDE system, it can be solved by adopting the method of characteristics.[19] Formally, this is done by replacing the kinetic momentum in phase space $\mathbf{p}$ by the time-dependent kinetic momentum on a trajectory $\boldsymbol{\pi}(\mathbf{q},t)$ so that

[18] According to the interpretation of $\mathbb{v}_0(\mathbf{p},t)$ as the charge density, this result is interpreted as local charge neutrality in the spatially homogeneous situation.

[19] The differential operator in Eq. (4.55), acting on any differentiable function of the independent variables $\mathbf{p}$ and $t$, might be viewed as a total derivative with respect to a new parameter $\alpha$, assuming

the PDE system is transformed into an ODE system:[20]

$$\dot{\mathbb{z}}(\mathbf{q},t) = \mathbb{M}(\mathbf{q},t)\mathbb{z}(\mathbf{q},t) \ . \tag{4.56}$$

As $\mathbb{z}(\mathbf{q},t)$ is a 10-component vector, it is suggested to search for an orthonormal basis of the corresponding 10-dimensional space in such a way that:

$$\mathbb{z}(\mathbf{q},t) = -2\sum_{i=1}^{10} \mathcal{E}_i(\mathbf{q},t)\mathbb{e}_i(\mathbf{q},t) \ . \tag{4.57}$$

Here, $\mathbb{e}_i(\mathbf{q},t)$ are the basis vectors whereas $\mathcal{E}_i(\mathbf{q},t)$ denote the expansion coefficients. The factor $-2$ is chosen for later convenience. Exploiting the vacuum initial conditions, it is convenient to choose the first basis vector $\mathbb{e}_1(\mathbf{q},t)$ in such a way that $\mathcal{E}_1(\mathbf{q},t_{\text{vac}}) = 1$ whereas all other $\mathcal{E}_i(\mathbf{q},t_{\text{vac}}) = 0$. As a consequence, one finds that a closed subset of only three basis vectors couples to the vacuum initial conditions:[21]

$$\mathbb{M}(\mathbf{q},t)\mathbb{e}_1(\mathbf{q},t) = 0 \qquad \dot{\mathbb{e}}_1(\mathbf{q},t) = -Q(\mathbf{q},t)\mathbb{e}_2(\mathbf{q},t)\,, \tag{4.58}$$

$$\mathbb{M}(\mathbf{q},t)\mathbb{e}_2(\mathbf{q},t) = 2\omega(\mathbf{q},t)\mathbb{e}_3(\mathbf{q},t) \qquad \dot{\mathbb{e}}_2(\mathbf{q},t) = Q(\mathbf{q},t)\mathbb{e}_1(\mathbf{q},t)\,, \tag{4.59}$$

$$\mathbb{M}(\mathbf{q},t)\mathbb{e}_3(\mathbf{q},t) = -2\omega(\mathbf{q},t)\mathbb{e}_2(\mathbf{q},t) \qquad \dot{\mathbb{e}}_3(\mathbf{q},t) = 0\ , \tag{4.60}$$

so that the system is fully characterized by means of three expansion coefficients $\mathcal{E}_i(\mathbf{q},t)$ with $i = \{1,2,3\}$. According to Eq. (4.56), one finds:

$$\dot{\mathcal{E}}_1(\mathbf{q},t) = -Q(\mathbf{q},t)\mathcal{E}_2(\mathbf{q},t) \ , \tag{4.61}$$

$$\dot{\mathcal{E}}_2(\mathbf{q},t) = Q(\mathbf{q},t)\mathcal{E}_1(\mathbf{q},t) - 2\omega(\mathbf{q},t)\mathcal{E}_3(\mathbf{q},t) \ , \tag{4.62}$$

$$\dot{\mathcal{E}}_3(\mathbf{q},t) = 2\omega(\mathbf{q},t)\mathcal{E}_2(\mathbf{q},t) \ , \tag{4.63}$$

along with vacuum initial conditions $\mathcal{E}_1(\mathbf{q},t_{\text{vac}}) = 1$ and $\mathcal{E}_2(\mathbf{q},t_{\text{vac}}) = \mathcal{E}_3(\mathbf{q},t_{\text{vac}}) = 0$.

---

that the former independent variables depend on $\alpha$ now:

$$[\tfrac{\partial}{\partial t} + eE(t)\tfrac{\partial}{\partial p_3}]f(\mathbf{p},t) \equiv \tfrac{d}{d\alpha}f(\mathbf{p}(\alpha),t(\alpha)) \ ,$$

with $t(\alpha) \equiv t$ and $\mathbf{p}(\alpha) \equiv \boldsymbol{\pi}(\mathbf{q},t)$. Accordingly, the *kinetic momentum in phase space* $\mathbf{p}$ is traded for the *time-dependent kinetic momentum on a trajectory* $\boldsymbol{\pi}(\mathbf{q},t) = \mathbf{q} - e\mathbf{A}(t)$, with $\mathbf{q}$ serving as an integration constant and corresponding to the canonical momentum.

[20]The following notation will be used throughout: Any function depending on $\mathbf{p}$ and $t$ is understood as being defined in phase space, whereas any function depending on $\mathbf{q}$ and $t$ is considered on a trajectroy. They are related to each other via:

$$f(\mathbf{q},t) \equiv f(\mathbf{p},t)|_{\mathbf{p}\to\mathbf{q}-e\mathbf{A}(t)} \quad \text{and} \quad f(\mathbf{p},t) \equiv f(\mathbf{q},t)|_{\mathbf{q}\to\mathbf{p}+e\mathbf{A}(t)} \ .$$

Note, that $f(\mathbf{p},t)$ possesses only an explicit time dependence whereas $f(\mathbf{q},t)$ shows both an explicit and an implicit time dependence via $\boldsymbol{\pi}(\mathbf{q},t)$.

[21]The detailed calculation can be found in Appendix A.5.

Introducing the notation:

$$\mathcal{E}_1(\mathbf{q},t) \equiv 1-\mathcal{F}(\mathbf{q},t) \ , \tag{4.64}$$

$$\mathcal{E}_2(\mathbf{q},t) \equiv \mathcal{G}(\mathbf{q},t) \ , \tag{4.65}$$

$$\mathcal{E}_3(\mathbf{q},t) \equiv \mathcal{H}(\mathbf{q},t) \ , \tag{4.66}$$

so that $\mathcal{F}(\mathbf{q},t)$ measures the deviation from the vacuum state, one exactly recovers the quantum Vlasov equation Eq. (4.36) – (4.38) in its differential form along with identical initial conditions. Accordingly, one can conclude that the equal-time Wigner formalism in the limit of a time-dependent electric field $\mathbf{E}(t) = E(t)\mathbf{e}_3$ is totally equivalent to the corresponding quantum Vlasov equation. Most notably, the one-particle distribution function can be expressed in terms of the equal-time Wigner components:[22]

$$\mathcal{F}(\mathbf{p},t) = \frac{m\big[\mathbb{s}(\mathbf{p},t) - \mathbb{s}_{\mathrm{vac}}(\mathbf{p})\big] + \mathbf{p}\cdot\big[\mathbb{v}(\mathbf{p},t) - \mathbb{v}_{\mathrm{vac}}(\mathbf{p})\big]}{2\omega(\mathbf{p})} \ . \tag{4.67}$$

Alternatively, it is possible to express the equal-time Wigner components $\mathbb{w}(\mathbf{p},t)$ in terms of $\mathcal{F}(\mathbf{p},t)$, $\mathcal{G}(\mathbf{p},t)$ and $\mathcal{H}(\mathbf{p},t)$ as well. As a matter of fact, it turns out that only 7 out of possible 16 equal-time Wigner components $\mathbb{w}(\mathbf{p},t)$ take non-vanishing values:

$$\mathbb{s}(\mathbf{p},t) = -\frac{2m}{\omega(\mathbf{p})}\big[1-\mathcal{F}(\mathbf{p},t) + \frac{p_3}{\epsilon_\perp}\mathcal{G}(\mathbf{p},t)\big] \ , \tag{4.68}$$

$$\mathbb{v}_1(\mathbf{p},t) = -\frac{2p_1}{\omega(\mathbf{p})}\big[1-\mathcal{F}(\mathbf{p},t) + \frac{p_3}{\epsilon_\perp}\mathcal{G}(\mathbf{p},t)\big] \ , \tag{4.69}$$

$$\mathbb{v}_2(\mathbf{p},t) = -\frac{2p_2}{\omega(\mathbf{p})}\big[1-\mathcal{F}(\mathbf{p},t) + \frac{p_3}{\epsilon_\perp}\mathcal{G}(\mathbf{p},t)\big] \ , \tag{4.70}$$

$$\mathbb{v}_3(\mathbf{p},t) = -\frac{2p_3}{\omega(\mathbf{p})}\big[1-\mathcal{F}(\mathbf{p},t) - \frac{\epsilon_\perp}{p_3}\mathcal{G}(\mathbf{p},t)\big] \ , \tag{4.71}$$

$$\mathbb{a}_1(\mathbf{p},t) = -\frac{2p_2}{\epsilon_\perp}\mathcal{H}(\mathbf{p},t) \ , \tag{4.72}$$

$$\mathbb{a}_2(\mathbf{p},t) = \frac{2p_1}{\epsilon_\perp}\mathcal{H}(\mathbf{p},t) \ , \tag{4.73}$$

$$\mathbb{t}_{1,3}(\mathbf{p},t) = \frac{2m}{\epsilon_\perp}\mathcal{H}(\mathbf{p},t) \ . \tag{4.74}$$

[22]Note again, that any function which is considered on a trajectroy $f(\mathbf{q},t)$ can be expressed in phase space $f(\mathbf{p},t)$ by replacing the time-dependent kinetic momentum on a trajectory by the phase space kinetic momentum, $\boldsymbol{\pi}(\mathbf{q},t) \to \mathbf{p}$:

$$f(\mathbf{p},t) = f(\mathbf{q},t)|_{\mathbf{q}\to\mathbf{p}+e\mathbf{A}(t)} \ .$$

## 4.3 Analytically solvable electric fields

It is in fact possible to derive an analytic expression for the one-particle distribution function $\mathcal{F}(\mathbf{q},t)$ for some specific spatially homogeneous, time-dependent electric fields $\mathbf{E}(t) = E(t)\mathbf{e}_3$. This in turn means that an analytic expression for the equal-time Wigner components $\mathbb{w}(\mathbf{p},t)$ can be derived as well due to the fact that the equal-time Wigner formalism is equivalent to the quantum Vlasov equation in that case.

It has been shown previously that the one-particle distribution function $\mathcal{F}(\mathbf{q},t)$ is defined via the Bogoliubov coefficient $\beta(\mathbf{q},t)$:

$$\mathcal{F}(\mathbf{q},t) = 2|\beta(\mathbf{q},t)|^2 \ , \tag{4.75}$$

with:

$$\beta(\mathbf{q},t) = -i\epsilon_\perp G^{(+)}(\mathbf{q},t) \left[\tfrac{\partial}{\partial t} + i\omega(\mathbf{q},t)\right] g^{(+)}(\mathbf{q},t) \ . \tag{4.76}$$

Accordingly, an analytic expression for $\mathcal{F}(\mathbf{q},t)$ can be derived once the mode functions $g^{(\pm)}(\mathbf{q},t)$ as well as the adiabatic mode functions $G^{(\pm)}(\mathbf{q},t)$ are known. As a matter of fact, the fundamental system of the equation of motion of a time-dependent harmonic oscillator:

$$\left[\tfrac{\partial^2}{\partial t^2} + \omega^2(\mathbf{q},t) + ieE(t)\right] g(\mathbf{q},t) = 0 \ , \tag{4.77}$$

consists of the mode functions $g^{(\pm)}(\mathbf{q},t)$. Note, however, that this differential equation is only for a restricted number of electric fields exactly solvable, most notably the static electric field $E(t) = E_0$ as well as the pulsed electric field $E(t) = E_0 \operatorname{sech}^2(\frac{t}{\tau})$. The adiabatic mode functions $G^{(\pm)}(\mathbf{q},t)$, on the other hand, can be calculated once the dynamical phase $\Theta(\mathbf{q},t_0,t)$ is analytically computable:

$$G^{(\pm)}(\mathbf{q},t) = \frac{e^{\mp i\Theta(\mathbf{q},t_0,t)}}{\sqrt{2\omega(\mathbf{q},t)[\omega(\mathbf{q},t) \mp \pi_3(q_3,t)]}} \qquad \text{with} \qquad \Theta(\mathbf{q},t_0,t) = \int_{t_0}^{t} \omega(\mathbf{q},t')dt' \ . \tag{4.78}$$

Given that the one-particle distribution function $\mathcal{F}(\mathbf{q},t)$ is derivable according to this recipe, the auxiliary function $\mathcal{G}(\mathbf{q},t)$ and $\mathcal{H}(\mathbf{q},t)$ can be calculated as well:

$$\mathcal{G}(\mathbf{q},t) = \frac{1}{Q(\mathbf{q},t)}\dot{\mathcal{F}}(\mathbf{q},t) \ , \tag{4.79}$$

$$\mathcal{H}(\mathbf{q},t) = \frac{Q(\mathbf{q},t)}{2\omega(\mathbf{q},t)}\left[1 - \mathcal{F}(\mathbf{q},t) + \frac{\dot{Q}(\mathbf{q},t)}{Q^3(\mathbf{q},t)}\dot{\mathcal{F}}(\mathbf{q},t) - \frac{1}{Q^2(\mathbf{q},t)}\ddot{\mathcal{F}}(\mathbf{q},t)\right] \ , \tag{4.80}$$

so that the equal-time Wigner components $\mathbb{w}(\mathbf{p},t)$ result from Eq. (4.68) – (4.74).

### 4.3.1 Static electric field

The static electric field $E(t) = E_0$ can be represented in terms of the vector potential:

$$A(t) = -E_0 t \ , \tag{4.81}$$

so that the equation of motion for the mode functions reads:[23]

$$\left[\frac{\partial^2}{\partial t^2} + \epsilon_\perp^2 + [q_3 + eE_0 t]^2 + ieE_0\right] g(\mathbf{q}, t) = 0 \ . \tag{4.82}$$

This equation turns into the parabolic cylinder differential equation:[24]

$$\left[\frac{\partial^2}{\partial u^2} + \frac{1}{4}u^2 + \frac{1}{2}(1+\eta)\right] g(u) = 0 \ , \tag{4.83}$$

when performing the variable transformation:[25]

$$u \equiv \sqrt{\frac{2}{eE_0}}\left[q_3 + eE_0 t\right] \qquad \text{with} \qquad u \in (-\infty, \infty) \ , \tag{4.84}$$

as well as introducing the following shorthand notations:

$$\eta \equiv \frac{\epsilon_\perp^2}{eE_0} = \frac{1}{\epsilon} + \frac{\mathbf{q}_\perp^2}{eE_0} \qquad \text{with} \qquad \epsilon = \frac{E_0}{E_{\rm cr}} \ . \tag{4.85}$$

The fundamental system consists of parabolic cylinder functions:

$$g^{(+)}(u) = N^{(+)} D_{-1+\frac{i\eta}{2}}\left(-ue^{-\frac{i\pi}{4}}\right) \ , \tag{4.86}$$

$$g^{(-)}(u) = N^{(-)} D_{-\frac{i\eta}{2}}\left(-ue^{\frac{i\pi}{4}}\right) \ , \tag{4.87}$$

with $N^{(\pm)}$ being normalization factors.[26] The asymptotic behavior of the mode functions for $u \to -\infty$ is given by:[27]

$$g^{(+)}(u) \overset{u\to-\infty}{\longrightarrow} N^{(+)} \frac{1}{|u|} e^{\frac{i}{4}\left[u^2 + 2\eta \ln(|u|) + \pi\right]} e^{\frac{\pi\eta}{8}} \ , \tag{4.88}$$

$$g^{(-)}(u) \overset{u\to-\infty}{\longrightarrow} N^{(-)} e^{-\frac{i}{4}\left[u^2 + 2\eta \ln(|u|)\right]} e^{\frac{\pi\eta}{8}} \ . \tag{4.89}$$

[23] The dependence on the orthogonal momentum $\mathbf{q}_\perp$ will not be explicitly indicated in the following as the differential equation depends on it only parametrically.

[24] For details on parabolic cylinder functions see [113], chapter 19.

[25] Note that the mode functions $g^{(\pm)}(u)$ will depend only on $u$ due to the linear relation between $q_3$ and $t$.

[26] The normalization factors will be chosen in such a way that the mode functions and the adiabatic mode functions coincide for $u \to -\infty$:

$$g^{(\pm)}(u) = G^{(\pm)}(u) \quad \text{for} \quad u \to -\infty \ .$$

[27] Here, without loss of generality $E_0 > 0$ so that $u \to -\infty$ corresponds to $t \to -\infty$.

The adiabatic mode functions, on the other hand, are defined as:

$$G^{(\pm)}(u) = \frac{e^{\mp i\Theta(u_0,u)}}{\sqrt{m^2\epsilon\sqrt{2\eta+u^2}[\sqrt{2\eta+u^2}\mp u]}} \ , \tag{4.90}$$

with the dynamical phase being given by:

$$\Theta(u_0,u) = \frac{1}{2}\int_{u_0}^{u} \mathrm{d}v\sqrt{2\eta+v^2} \ . \tag{4.91}$$

As mentioned previously, the actual value of $u_0$ does not really matter as it only fixes an arbitrary phase at a certain instant of time. Choosing the symmetric point $u_0 = 0$, for instance, one finds:[28]

$$\Theta(0,u) = \tfrac{1}{4}\big[u\sqrt{2\eta+u^2} + 2\eta\ln(u+\sqrt{2\eta+u^2}) - \eta\ln(2\eta)\big] \ . \tag{4.92}$$

The asymptotic behavior of the adiabatic mode functions for $u \to -\infty$ is given by:

$$G^{(+)}(u) \overset{u\to-\infty}{\longrightarrow} \frac{1}{\sqrt{2|u|m^2\epsilon}}e^{\frac{i}{4}\left[u^2+2\eta\ln(|u|)+\eta+\eta\ln(\frac{2}{\eta})\right]} \ , \tag{4.93}$$

$$G^{(-)}(u) \overset{u\to-\infty}{\longrightarrow} \frac{1}{\sqrt{\eta\, m^2\epsilon}}e^{-\frac{i}{4}\left[u^2+2\eta\ln(|u|)+\eta+\eta\ln(\frac{2}{\eta})\right]} \ . \tag{4.94}$$

Accordingly, the properly normalized mode functions read:

$$g^{(+)}(u) = \frac{1}{\sqrt{2m^2\epsilon}}e^{\frac{i}{4}\left[\eta+\eta\ln(\frac{2}{\eta})-\pi\right]}e^{-\frac{\pi\eta}{8}}D_{-1+\frac{i\eta}{2}}\big(-ue^{-\frac{i\pi}{4}}\big) \ , \tag{4.95}$$

$$g^{(-)}(u) = \frac{1}{\sqrt{\eta\, m^2\epsilon}}e^{-\frac{i}{4}\left[\eta+\eta\ln(\frac{2}{\eta})\right]}e^{-\frac{\pi\eta}{8}}D_{-\frac{i\eta}{2}}\big(-ue^{i\frac{\pi}{4}}\big) \ . \tag{4.96}$$

Consequently, it is possible to derive an analytic expression for $\mathcal{F}(u) = 2|\beta(u)|^2$ in terms of parabolic cylinder functions,[29] which is displayed in Fig. 4.1:

$$\mathcal{F}(u) = \tfrac{1}{4}\big[1+\tfrac{u}{\sqrt{2\eta+u^2}}\big]e^{-\frac{\pi\eta}{4}}\Big|\big[\sqrt{2\eta+u^2}-u\big]D_{-1+\frac{i\eta}{2}}\big(-ue^{-\frac{i\pi}{4}}\big)-2e^{\frac{i\pi}{4}}D_{\frac{i\eta}{2}}\big(-ue^{-\frac{i\pi}{4}}\big)\Big|^2 . \tag{4.97}$$

[28]The integral is given by:

$$\int \mathrm{d}u\sqrt{2\eta+u^2} = \tfrac{1}{2}u\sqrt{2\eta+u^2} + \eta\ln(u+\sqrt{2\eta+u^2}) + \eta\ln(2) \ .$$

[29]The derivative of the parabolic cylinder functions with respect to $u$, which appears in the definition of $\beta(u)$, is given by:

$$\tfrac{\partial}{\partial u}D_{-1+\frac{i\eta}{2}}\big(-ue^{-\frac{i\pi}{4}}\big) = -\tfrac{i}{2}uD_{-1+\frac{i\eta}{2}}\big(-ue^{-\frac{i\pi}{4}}\big) + e^{-\frac{i\pi}{4}}D_{\frac{i\eta}{2}}\big(-ue^{-\frac{i\pi}{4}}\big) \ .$$

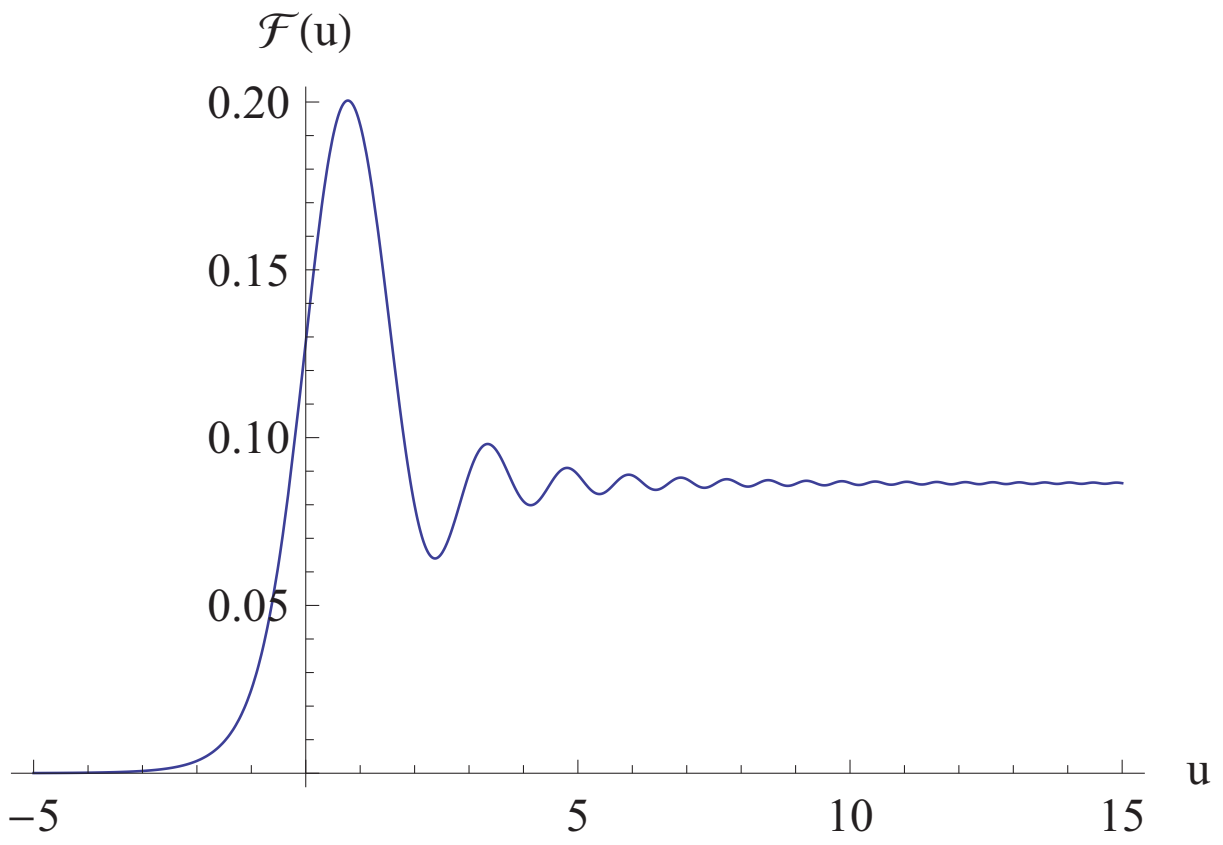

Figure 4.1: One-particle distribution function $\mathcal{F}(u)$ for $\epsilon = 1$ and $\mathbf{q}_\perp = 0$. Starting off with $\mathcal{F}(u \to -\infty) = 0$, the asymptotic value $\mathcal{F}(u \to \infty) = 2e^{-\pi\eta}$ is reached after some transient oscillations.

#### 4.3.1.1 Pair creation rate

It has already been mentioned that the first term in the expression for the *vacuum decay rate*:

$$\mathcal{P}[\text{vac.}] = \frac{(eE_0)^2}{4\pi^3} \sum_{n=1}^{\infty} \frac{1}{n^2} \exp\left(-\frac{n\pi m^2}{eE_0}\right) , \tag{4.98}$$

can be identified with the *pair creation rate* $\dot{\mathcal{N}}[e^+e^-]$:[30]

$$\dot{\mathcal{N}}[e^+e^-] = \frac{(eE_0)^2}{8\pi^2} \int_{-\infty}^{\infty} \mathrm{d}u \int_{-1/\epsilon}^{\infty} \mathrm{d}\eta \mathcal{F}'(u) = \lim_{u\to\infty} \frac{(eE_0)^2}{8\pi^2} \int_{-1/\epsilon}^{\infty} \mathrm{d}\eta \mathcal{F}(u) . \tag{4.99}$$

Here, the prime denotes the derivative with respect to $u$. Accordingly, the asymptotic behavior of the parabolic cylinder functions for $u \to \infty$ is needed:

$$D_{-1+i\frac{\eta}{2}}(-ue^{-\frac{i\pi}{4}}) \overset{u\to\infty}{\longrightarrow} \frac{\sqrt{2\pi}}{\Gamma(1-\frac{i\eta}{2})} e^{-\frac{i}{4}\left[u^2+2\eta\ln(|u|)\right]} e^{-\frac{\pi\eta}{8}} , \tag{4.100}$$

$$D_{i\frac{\eta}{2}}(-ue^{-\frac{i\pi}{4}}) \overset{u\to\infty}{\longrightarrow} e^{\frac{i}{4}\left[u^2+2\eta\ln(|u|)\right]} e^{-\frac{3\pi\eta}{8}} , \tag{4.101}$$

according to which:

$$\lim_{u\to\infty} \mathcal{F}(u) = 2e^{-\pi\eta} . \tag{4.102}$$

[30]The pair creation rate, i.e. the change in the number of created particles per volume and time, is defined as:

$$\dot{\mathcal{N}}[e^+e^-] = \int \frac{\mathrm{d}^3q}{(2\pi)^3} \dot{\mathcal{F}}(\mathbf{q}, t) .$$

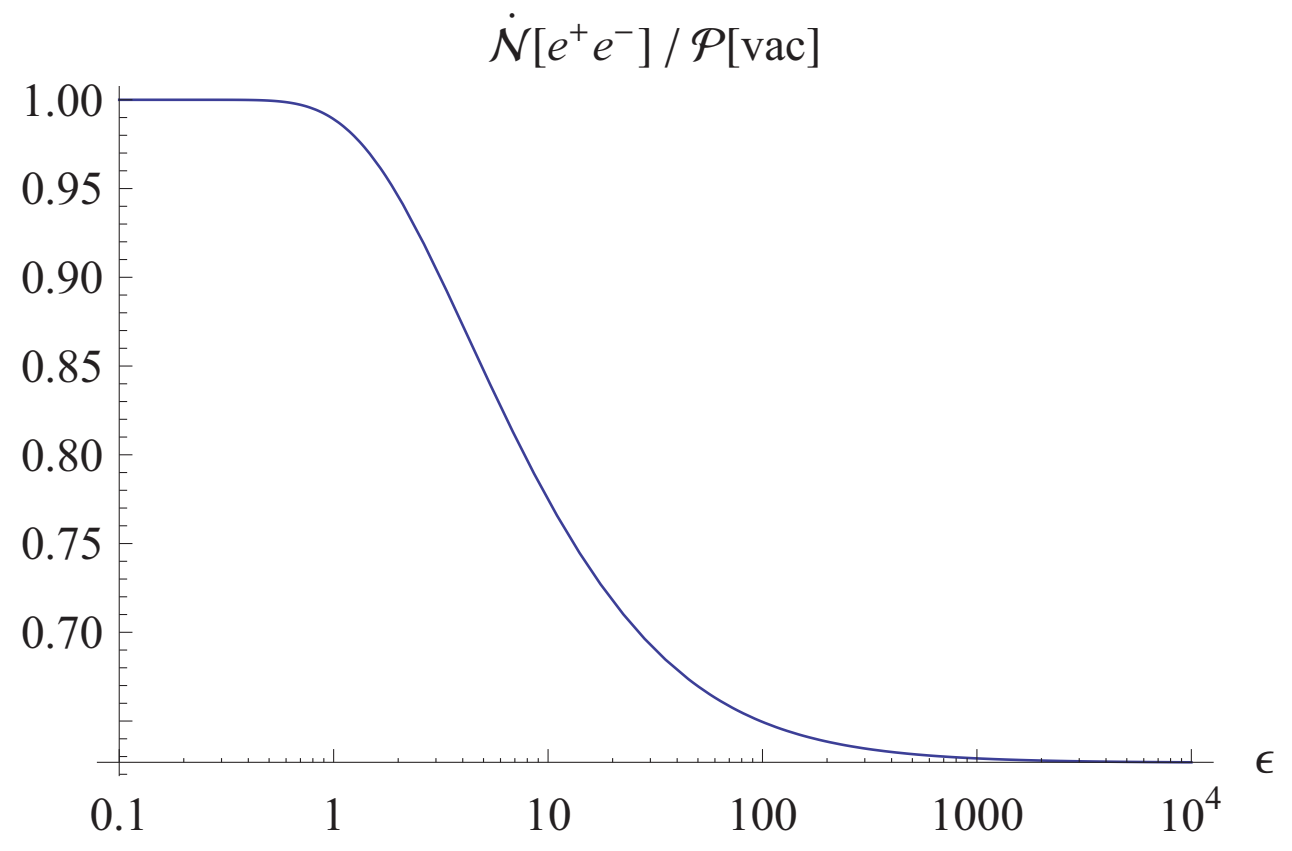

Figure 4.2: Ratio between the pair creation rate $\dot{\mathcal{N}}[e^+e^-]$ and the vacuum decay rate $\mathcal{P}[\text{vac.}]$ as function of $\epsilon$.

Accordingly, one finally obtains for the pair creation rate:

$$\dot{\mathcal{N}}[e^+e^-] = \frac{(eE_0)^2}{4\pi^3} \exp\left(-\frac{\pi m^2}{eE_0}\right) . \tag{4.103}$$

In Fig. 4.2, the ratio between the pair creation rate $\dot{\mathcal{N}}[e^+e^-]$ and the vacuum decay rate $\mathcal{P}[\text{vac.}]$ is displayed. The difference between those two is rather minor in the sub-critical field strength regime $\epsilon < 1$ whereas it becomes sizeable for super-critical field strengths $\epsilon > 1$.

#### 4.3.1.2 Equal-time Wigner components

It has been mentioned that knowledge of the one-particle distribution function allows for determining analytic expressions for the equal-time Wigner components as well. In order to simplify notation, the following shorthand notations are introduced:[31]

$$\mathcal{D}_1(u) \equiv |D_{-1+\frac{i\eta}{2}}\big(-ue^{-\frac{i\pi}{4}}\big)|^2 , \tag{4.104}$$

$$\mathcal{D}_2(u) \equiv |D_{\frac{i\eta}{2}}\big(-ue^{-\frac{i\pi}{4}}\big)|^2 , \tag{4.105}$$

$$\mathcal{D}_3(u) \equiv e^{\frac{i\pi}{4}} D_{\frac{i\eta}{2}}\big(-ue^{-\frac{i\pi}{4}}\big) D_{-1-\frac{i\eta}{2}}\big(-ue^{\frac{i\pi}{4}}\big) + c.c. , \tag{4.106}$$

$$\mathcal{D}_4(u) \equiv e^{-\frac{i\pi}{4}} D_{\frac{i\eta}{2}}\big(-ue^{-\frac{i\pi}{4}}\big) D_{-1-\frac{i\eta}{2}}\big(-ue^{\frac{i\pi}{4}}\big) + c.c. , \tag{4.107}$$

so that $\mathcal{F}(u)$ can be written as:

$$\mathcal{F}(u) = e^{-\frac{\pi\eta}{4}} \Big[\frac{\eta}{2}\big[1 - \frac{u}{\sqrt{2\eta+u^2}}\big]\mathcal{D}_1(u) + \big[1 + \frac{u}{\sqrt{2\eta+u^2}}\big]\mathcal{D}_2(u) - \frac{\eta}{\sqrt{2\eta+u^2}}\mathcal{D}_3(u)\Big] . \tag{4.108}$$

[31] Note that $\mathcal{D}_4(u)$ is not yet required but will be used when calculating $\mathcal{G}(u)$ and $\mathcal{H}(u)$.

Due to the fact that:

$$\frac{\partial}{\partial u}\mathcal{D}_1(u) = \mathcal{D}_4(u) \ , \tag{4.109}$$
$$\frac{\partial}{\partial u}\mathcal{D}_2(u) = -\frac{\eta}{2}\mathcal{D}_4(u) \ , \tag{4.110}$$
$$\frac{\partial}{\partial u}\mathcal{D}_3(u) = -u\mathcal{D}_4(u) \ , \tag{4.111}$$

as well as:

$$Q(u) = \frac{2\sqrt{\eta\, m^2 \epsilon}}{2\eta + u^2} \ , \tag{4.112}$$

it is possible to calculate $\mathcal{G}(u)$ and $\mathcal{H}(u)$ according to Eq. (4.79) and Eq. (4.80), respectively:

$$\mathcal{G}(u) = \sqrt{\tfrac{2\eta}{2\eta+u^2}}e^{-\frac{\pi\eta}{4}}\Big[-\tfrac{\eta}{2}\mathcal{D}_1(u) + \mathcal{D}_2(u) + \tfrac{u}{2}\mathcal{D}_3(u)\Big] \ , \tag{4.113}$$
$$\mathcal{H}(u) = \sqrt{\tfrac{2\eta}{(2\eta+u^2)^3}}\Big(1 - e^{-\frac{\pi\eta}{4}}\Big[\tfrac{\eta}{2}\mathcal{D}_1(u) + \mathcal{D}_2(u) - \tfrac{\sqrt{(2\eta+u^2)^3}}{2}\mathcal{D}_4(u)\Big]\Big) \ . \tag{4.114}$$

The transformation back to phase space is required in order to finally calculate the equal-time Wigner components as shown in Eq. (4.68) – (4.74). The corresponding change of variables $\mathbf{q} \to \mathbf{p} + e\mathbf{A}(t)$ reads:

$$u \to \sqrt{\tfrac{2}{eE_0}}p_3 \qquad \text{and} \qquad \mathbf{q}_\perp \to \mathbf{p}_\perp \ , \tag{4.115}$$

so that:[32]

$$\mathcal{F}(\mathbf{p}) = e^{-\frac{\pi\eta}{4}}\Big[\tfrac{\epsilon_\perp^2}{2eE_0}\big[1 - \tfrac{p_3}{\omega(\mathbf{p})}\big]\mathcal{D}_1(\mathbf{p}) + \big[1 + \tfrac{p_3}{\omega(\mathbf{p})}\big]\mathcal{D}_2(\mathbf{p}) - \tfrac{\epsilon_\perp^2}{\sqrt{2eE_0}\omega(\mathbf{p})}\mathcal{D}_3(\mathbf{p})\Big], \tag{4.116}$$
$$\mathcal{G}(\mathbf{p}) = \tfrac{\epsilon_\perp}{\omega(\mathbf{p})}e^{-\frac{\pi\eta}{4}}\Big[-\tfrac{\epsilon_\perp^2}{2eE_0}\mathcal{D}_1(\mathbf{p}) + \mathcal{D}_2(\mathbf{p}) + \tfrac{p_3}{\sqrt{2eE_0}}\mathcal{D}_3(\mathbf{p})\Big] \ , \tag{4.117}$$
$$\mathcal{H}(\mathbf{p}) = \tfrac{eE_0\epsilon_\perp}{2\omega^3(\mathbf{p})}\Big(1 - e^{-\frac{\pi\eta}{4}}\Big[\tfrac{\epsilon_\perp^2}{2eE_0}\mathcal{D}_1(\mathbf{p}) + \mathcal{D}_2(\mathbf{p}) - \tfrac{4\omega^3(\mathbf{p})}{\sqrt{(2eE_0)^3}}\mathcal{D}_4(\mathbf{p})\Big]\Big) \ . \tag{4.118}$$

Accordingly, all equal-time Wigner components only depend on the phase space kinetic momentum $\mathbf{p}$ but not on the time variable $t$. This is due to the fact that the electric field $E(t) = E_0$ is static as well.

[32] Note that $\mathcal{D}_i(\mathbf{p})$ are defined in such a way that they are obtained from $\mathcal{D}_i(u)$ according to:

$$\mathcal{D}_i(\mathbf{p}) \equiv \mathcal{D}_i\Big(u \to \sqrt{\tfrac{2}{eE_0}}p_3\Big) \qquad \text{with} \qquad \mathbf{q}_\perp \to \mathbf{p}_\perp \ .$$

### 4.3.2 Pulsed electric field

The pulsed electric field $E(t) = E_0 \operatorname{sech}^2(\frac{t}{\tau})$ can be represented in terms of the vector potential:

$$A(t) = -E_0 \tau \tanh(\tfrac{t}{\tau}) \; , \tag{4.119}$$

so that the equation of motion for the mode functions reads:

$$\Big[\frac{\partial^2}{\partial t^2} + \epsilon_\perp^2 + [q_3 + eE_0\tau \tanh(\tfrac{t}{\tau})]^2 + ieE_0 \operatorname{sech}^2(\tfrac{t}{\tau})\Big]\, g(\mathbf{q}, t) = 0 \; . \tag{4.120}$$

In order to actually solve this equation it seems advantageous to introduce the dimensionless time variable:[33]

$$u \equiv \tfrac{1}{2}\big[1 + \tanh(\tfrac{t}{\tau})\big] \qquad \text{with} \qquad u \in [0,1] \; , \tag{4.121}$$

so that:[34]

$$\pi_3(q_3, u) \;=\; q_3 + eE_0\tau(2u-1) \; , \tag{4.122}$$

$$\omega^2(\mathbf{q}, u) \;=\; \epsilon_\perp^2 + \pi_3^2(q_3, u) \; . \tag{4.123}$$

Accordingly, the equation of motion can be written as:

$$\big[\tfrac{4}{\tau^2} u(1-u)\tfrac{\partial}{\partial u}\big[u(1-u)\tfrac{\partial}{\partial u}\big] + \omega^2(\mathbf{q}, u) + 4ieE_0 u(1-u)\big]\, g(\mathbf{q}, u) = 0 \; . \tag{4.124}$$

Introducing an ansatz for the mode function:

$$g(\mathbf{q}, u) \equiv u^{-\frac{i\tau\omega(\mathbf{q},0)}{2}} (1-u)^{\frac{i\tau\omega(\mathbf{q},1)}{2}} h(\mathbf{q}, u) \; , \tag{4.125}$$

the equation of motion turns into the hypergeometric differential equation:[35]

$$\Big[u(1-u)\frac{\partial^2}{\partial u^2} + \big(c(\mathbf{q}) - \big[a(\mathbf{q}) + b(\mathbf{q}) + 1\big]u\big)\frac{\partial}{\partial u} - a(\mathbf{q})b(\mathbf{q})\Big]\, h(\mathbf{q}, u) = 0 \; , \tag{4.126}$$

with $u$-independent parameters:

$$a(\mathbf{q}) \;=\; -ieE_0\tau^2 - \frac{i\tau\omega(\mathbf{q},0)}{2} + \frac{i\tau\omega(\mathbf{q},1)}{2} \; , \tag{4.127}$$

$$b(\mathbf{q}) \;=\; 1 + ieE_0\tau^2 - \frac{i\tau\omega(\mathbf{q},0)}{2} + \frac{i\tau\omega(\mathbf{q},1)}{2} \; , \tag{4.128}$$

$$c(\mathbf{q}) \;=\; 1 - i\tau\omega(\mathbf{q},0) \; . \tag{4.129}$$

[33] The asymptotic limit $u \to 0^+$ corresponds to $t \to -\infty$ whereas $u \to 1$ corresponds to $t \to \infty$.

[34] Note that the dependence on the orthogonal momentum $\mathbf{q}_\perp$ will not be indicated again.

[35] For details on hypergeometric functions see [113], chapter 15.

The hypergeometric differential equation is solved in terms of hypergeometric functions so that the mode functions read:[36]

$$g^{(+)}(\mathbf{q},u) = N^{(+)}(\mathbf{q})u^{-\frac{i\tau\omega(\mathbf{q},0)}{2}}(1-u)^{\frac{i\tau\omega(\mathbf{q},1)}{2}}F(a,b,c;u) \ , \tag{4.130}$$

$$g^{(-)}(\mathbf{q},u) = N^{(-)}(\mathbf{q})u^{\frac{i\tau\omega(\mathbf{q},0)}{2}}(1-u)^{-\frac{i\tau\omega(\mathbf{q},1)}{2}}F(1-a,1-b,2-c;u) \ . \tag{4.131}$$

Accordingly, the asymptotic behavior of the mode functions for $u \to 0^+$ is given by:

$$g^{(+)}(\mathbf{q},u) \overset{u\to 0^+}{\longrightarrow} N^{(+)}(\mathbf{q})e^{-\frac{i\tau\omega(\mathbf{q},0)}{2}\ln(u)} \ , \tag{4.132}$$

$$g^{(-)}(\mathbf{q},u) \overset{u\to 0^+}{\longrightarrow} N^{(-)}(\mathbf{q})e^{\frac{i\tau\omega(\mathbf{q},0)}{2}\ln(u)} \ . \tag{4.133}$$

In analogy to the static electric field $E(t) = E_0$, the asymptotic behavior of the adiabatic mode functions:

$$G^{(\pm)}(\mathbf{q},u) = \frac{e^{\mp i\Theta(\mathbf{q},u_0,u)}}{\sqrt{2\omega(\mathbf{q},u)[\omega(\mathbf{q},u) \mp \pi_3(q_3,u)]}} \ , \tag{4.134}$$

with:

$$\Theta(\mathbf{q},u_0,u) = \frac{\tau}{2}\int_{u_0}^{u} \mathrm{d}v \frac{\omega(\mathbf{q},v)}{v(1-v)} \ , \tag{4.135}$$

is determined in order to fix the value of the normalization constants $N^{(\pm)}(\mathbf{q})$. The dynamical phase can in fact be split into a logarithmically divergent part as well as an irrelevant regular part $\Phi(\mathbf{q},u_0,0)$ in the asymptotic limit $u \to 0^+$:[37]

$$\Theta(\mathbf{q},u_0,u) \overset{u\to 0^+}{\longrightarrow} \frac{\tau\omega(\mathbf{q},0)}{2}\ln(u) + \Phi(\mathbf{q},u_0,0) \ , \tag{4.136}$$

so that the asymptotic behavior of the adiabatic mode functions for $u \to 0^+$ is given by:

$$G^{(\pm)}(\mathbf{q},u) \overset{u\to 0^+}{\longrightarrow} \frac{e^{\mp i\Phi(\mathbf{q},u_0,0)}}{\sqrt{2\omega(\mathbf{q},0)[\omega(\mathbf{q},0)\mp\pi_3(q_3,0)]}} e^{\mp\frac{i\tau\omega(\mathbf{q},0)}{2}\ln(u)} \ . \tag{4.137}$$

[36] The fundamental system of the hypergeometric differential equation $\{h^{(+)}(\mathbf{q},u), h^{(-)}(\mathbf{q},u)\}$ in the neighborhood of $u = 0^+$ consists of:

$$h^{(+)}(\mathbf{q},u) = F(a,b,c;u) \quad \text{and} \quad h^{(-)}(\mathbf{q},u) = u^{i\tau\omega(\mathbf{q},0)}(1-u)^{-i\tau\omega(\mathbf{q},1)}F(1-a,1-b,2-c;u) \ .$$

[37] The integral is given by:

$$\begin{aligned}\int \mathrm{d}u \frac{\omega(\mathbf{q},u)}{u(1-u)} = & -i\pi\Big[\omega(\mathbf{q},0)+\omega(\mathbf{q},1)\Big] - 2eE_0\tau\Big[\ln(2) + \ln\Big(\pi_3(\mathbf{q},u)+\omega(\mathbf{q},u)\Big)\Big] \\ & + \omega(\mathbf{q},0)\Big[\ln(u) + \ln\Big(\tfrac{\omega^3(\mathbf{q},0)}{2}\Big) - \ln\Big(\epsilon_\perp^2 + \pi_3(\mathbf{q},0)\pi_3(\mathbf{q},u) + \omega(\mathbf{q},0)\omega(\mathbf{q},u)\Big)\Big] \\ & - \omega(\mathbf{q},1)\Big[\ln(1-u) + \ln\Big(\tfrac{\omega^3(\mathbf{q},1)}{2}\Big) - \ln\Big(\epsilon_\perp^2 + \pi_3(\mathbf{q},1)\pi_3(\mathbf{q},u) + \omega(\mathbf{q},1)\omega(\mathbf{q},u)\Big)\Big]\end{aligned}$$

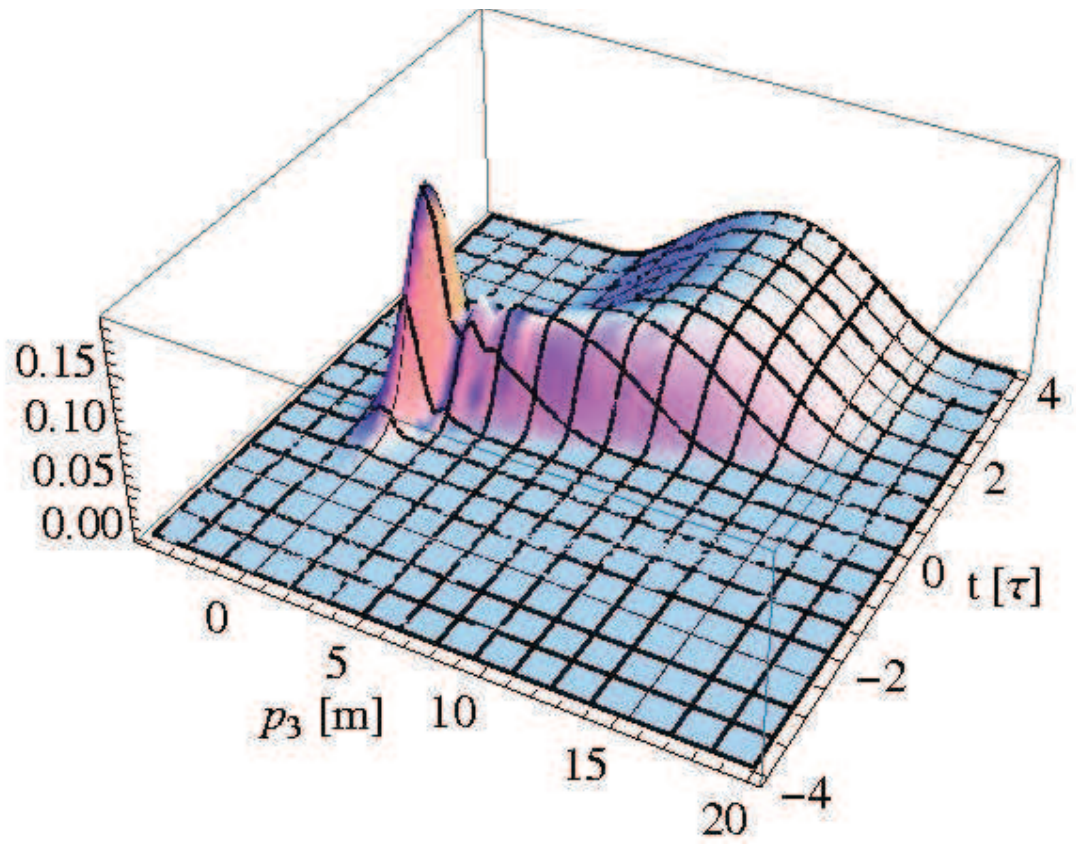

Figure 4.3: Time evolution of the one-particle distribution function $\mathcal{F}(\mathbf{q}, u)$ for $\epsilon = 1$, $\tau = \frac{10}{m}$ and $\mathbf{q}_\perp = 0$. Starting off with $\mathcal{F}(\mathbf{q}, u \to 0^+) = 0$, the asymptotic value $\mathcal{F}(\mathbf{q}, u \to 1)$ is finally obtained. Note that all the remaining figures in this section are displayed in terms of the phase space kinetic momentum $p_3$ instead of the canonical momentum $q_3$.

The normalization constants $N^{(\pm)}(\mathbf{q})$ are accordingly given by:

$$N^{(\pm)}(\mathbf{q}) = \frac{e^{\mp i\Phi(\mathbf{q},u_0,0)}}{\sqrt{2\omega(\mathbf{q},0)[\omega(\mathbf{q},0) \mp \pi_3(q_3,0)]}} \ . \tag{4.138}$$

Consequently, it is possible to derive an analytic expression for $\mathcal{F}(\mathbf{q}, u) = 2|\beta(\mathbf{q}, u)|^2$ in terms of hypergeometric functions,[38] which is displayed in Fig. 4.3:

$$\mathcal{F}(\mathbf{q}, u) \quad = \quad \left|N^{(+)}(\mathbf{q})\right|^2 \left[1 + \tfrac{\pi_3(q_3,u)}{\omega(\mathbf{q},u)}\right] \left|\mathcal{F}_1(\mathbf{q}, u) + i\mathcal{F}_2(\mathbf{q}, u)\right|^2 , \tag{4.139}$$

with:

$$\mathcal{F}_1(\mathbf{q}, u) \quad = \quad \tfrac{2}{\tau} u(1-u) \tfrac{ab}{c} F(1+a, 1+b, 1+c; u) \ , \tag{4.140}$$

$$\mathcal{F}_2(\mathbf{q}, u) \quad = \quad \left[\omega(\mathbf{q}, u) - (1-u)\omega(\mathbf{q}, 0) - u\omega(\mathbf{q}, 1)\right] F(a, b, c; u) \ . \tag{4.141}$$

#### 4.3.2.1 Asymptotic one-particle distribution function

The analytic expression for one-particle distribution function in terms of hypergeometric functions drastically simplifies in the asymptotic limit. The asymptotic

[38] The derivative of the hypergeometric function with respect to the last argument is given by:

$$\tfrac{\partial}{\partial u} F(a, b, c; u) = \tfrac{ab}{c} F(1+a, 1+b, 1+c; u) \ .$$

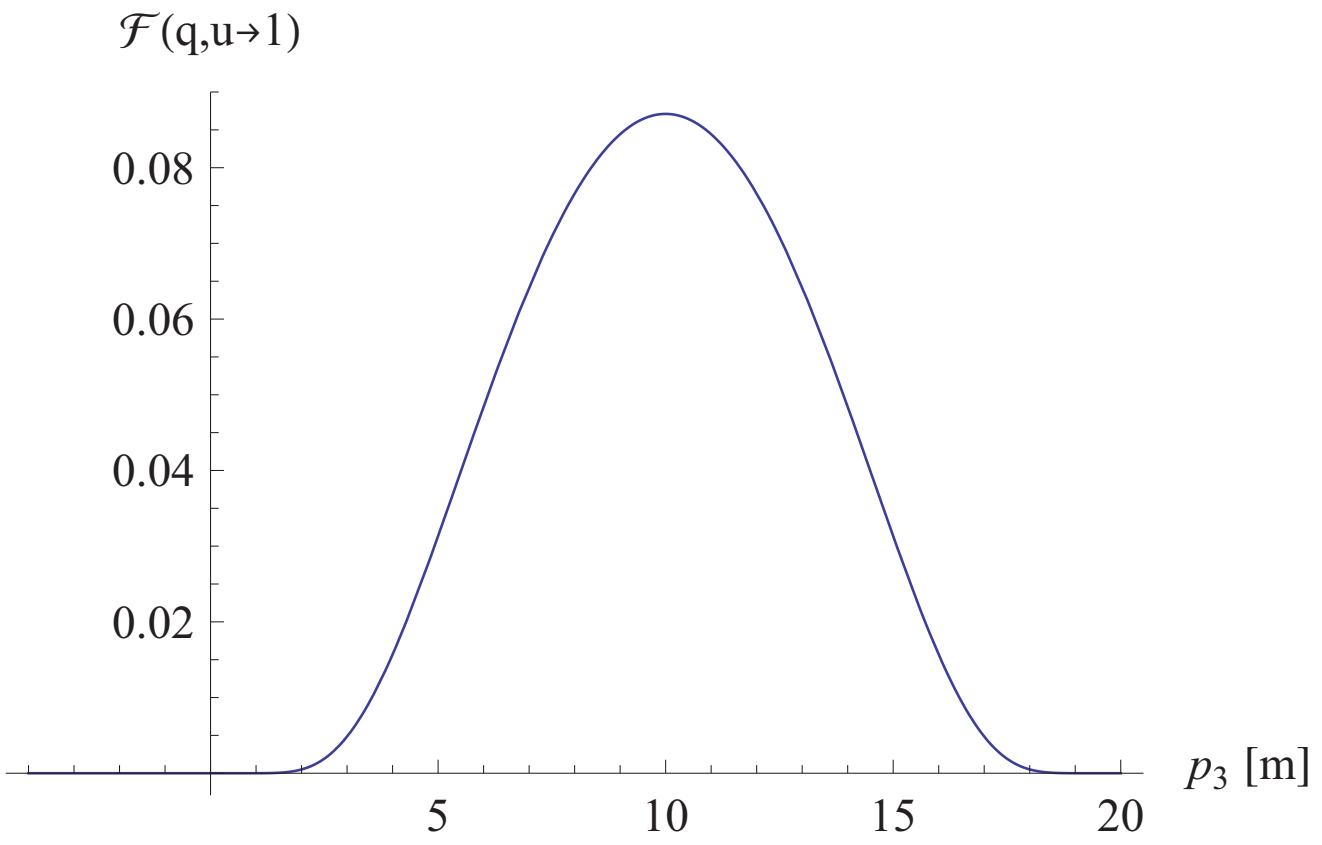

Figure 4.4: Asymptotic one-particle distribution function $\mathcal{F}(\mathbf{q}, u \to 1)$ for $\epsilon = 1$, $\tau = \frac{10}{m}$ and $\mathbf{q}_\perp = 0$. Note again that the interpretation of $\mathcal{F}(\mathbf{q}, u)$ as momentum distribution of real particles is in fact only possible at asymptotic times $u \to 1$.

behavior of the hypergeometric functions for $u \to 1$ is in fact given by:[39]

$$F(a, b, c; u) \quad \overset{u\to 1}{\longrightarrow} \quad (1-u)^{c-a-b} \frac{\Gamma(c)\Gamma(a+b-c)}{\Gamma(a)\Gamma(b)} \ , \tag{4.142}$$

$$F(1+a, 1+b, 1+c; u) \quad \overset{u\to 1}{\longrightarrow} \quad (1-u)^{c-a-b-1} \frac{\Gamma(1+c)\Gamma(1+a+b-c)}{\Gamma(1+a)\Gamma(1+b)} \ , \tag{4.143}$$

so that:

$$\lim_{u\to 1} \mathcal{F}(\mathbf{q}, u) = \frac{2}{\tau^2 \omega(\mathbf{q},0)\omega(\mathbf{q},1)} \frac{\omega(\mathbf{q},1)+\pi_3(q_3,1)}{\omega(\mathbf{q},0)-\pi_3(q_3,0)} \left| \frac{ab}{c} \frac{\Gamma(1+c)\Gamma(1+a+b-c)}{\Gamma(1+a)\Gamma(1+b)} \right|^2 . \tag{4.144}$$

Applying the transformation formula for gamma functions, one finally obtains a simple analytic expression for the asymptotic one-particle distribution function:[40]

$$\mathcal{F}(\mathbf{q}, u \to 1) = \frac{2 \sinh\left(\frac{\pi\tau}{2}[2eE_0\tau + \omega(\mathbf{q},0) - \omega(\mathbf{q},1)]\right) \sinh\left(\frac{\pi\tau}{2}[2eE_0\tau - \omega(\mathbf{q},0) + \omega(\mathbf{q},1)]\right)}{\sinh\left(\pi\tau\omega(\mathbf{q},0)\right) \sinh\left(\pi\tau\omega(\mathbf{q},1)\right)} \ . \tag{4.145}$$

In Fig. 4.4, the asymptotic one-particle distribution function $\mathcal{F}(\mathbf{q}, u \to 1)$ is displayed which is obtained once the electric field vanishes.

[39] Note that the hypergeometric functions have been transformed by means of the linear transformation formula:

$$\begin{aligned} F(\alpha, \beta, \gamma; u) \quad = \quad & \frac{\Gamma(\gamma)\Gamma(\gamma-\alpha-\beta)}{\Gamma(\gamma-\alpha)\Gamma(\gamma-\beta)} F(\alpha, \beta, \alpha+\beta-\gamma+1; 1-u) + \\ & (1-u)^{\gamma-\alpha-\beta} \frac{\Gamma(\gamma)\Gamma(\alpha+\beta-\gamma)}{\Gamma(\alpha)\Gamma(\beta)} F(\gamma-\alpha, \gamma-\beta, \gamma-\alpha-\beta+1; 1-u) \ . \end{aligned}$$

[40] For details on gamma functions see [113], chapter 6. The following relations are used here:

$$\Gamma(1+\alpha) = \alpha\Gamma(\alpha) \qquad \text{and} \qquad \left|\Gamma(1+i\alpha)\right|^2 = \frac{\pi\alpha}{\sinh(\pi\alpha)} \ .$$

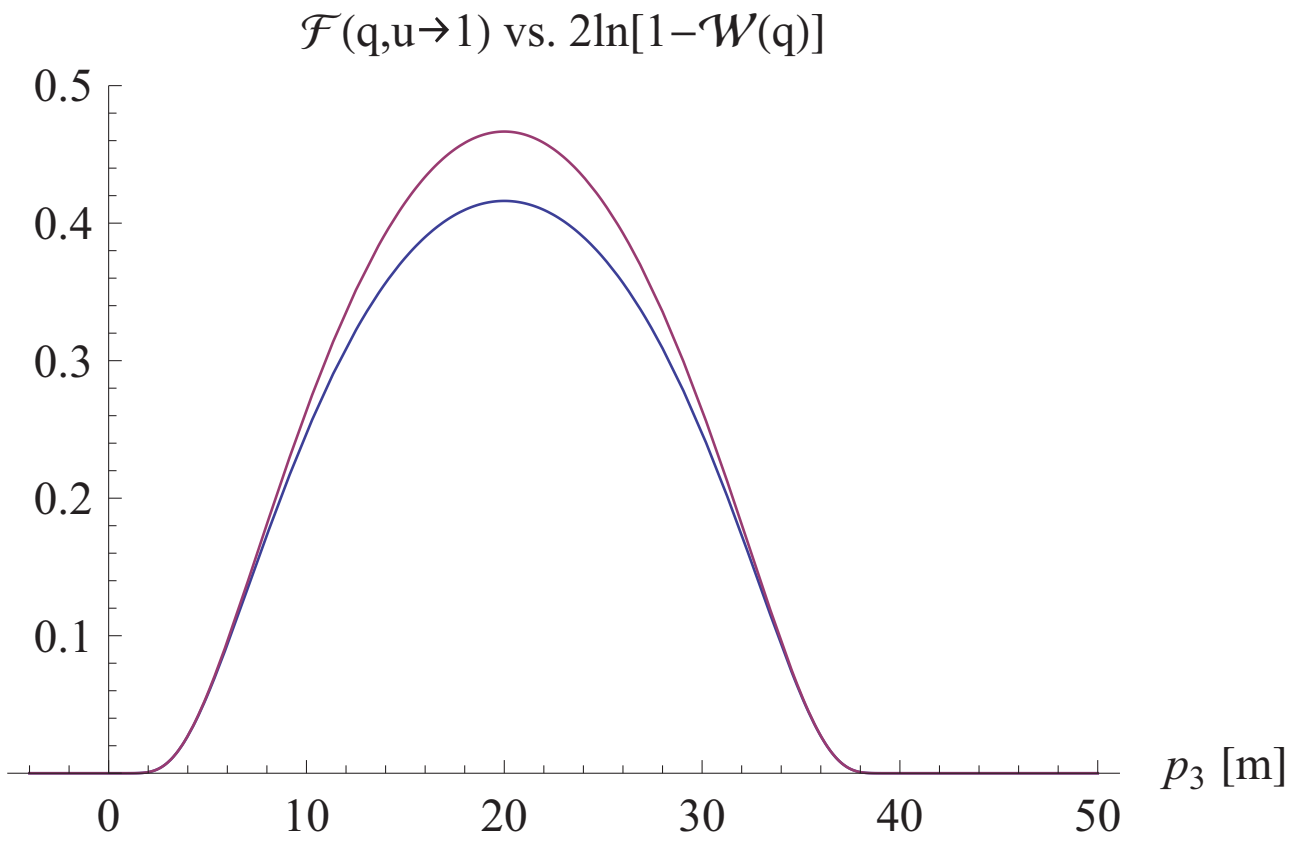

Figure 4.5: Comparison of $\mathcal{F}(\mathbf{q}, u \to 1)$ (blue) with $2\ln[1-\mathcal{W}(\mathbf{q})]$ (purple) for $\epsilon = 2$, $\tau = \frac{10}{m}$ and $\mathbf{q}_\perp = 0$.

#### 4.3.2.2 Particle density vs. imaginary part of the effective action

It is very instructive to compare the result for the *density of created particles* $\mathcal{N}[e^+e^-]$, which is obtained from the quantum Vlasov equation, with the result for the imaginary part of the effective action $2\,\mathrm{Im}\,\mathcal{S}_{\mathrm{eff}}$, which has been obtained from semi-classical methods previously [18].[41] The density of created particles is given by:

$$\mathcal{N}[e^+e^-] = \lim_{u\to 1} \int \frac{\mathrm{d}^3 q}{(2\pi)^3} \mathcal{F}(\mathbf{q}, u) \ , \tag{4.146}$$

whereas the imaginary part of the effective action density reads:

$$2\,\mathrm{Im}\,\mathcal{S}_{\mathrm{eff}} = \int \frac{\mathrm{d}^3 q}{(2\pi)^3}\, 2\ln[1 - \mathcal{W}(\mathbf{q})] \ , \tag{4.147}$$

with:

$$\mathcal{W}(\mathbf{q}) = \frac{\sinh\left(\frac{\pi\tau}{2}[2eE_0\tau + \omega(\mathbf{q},0) - \omega(\mathbf{q},1)]\right)\sinh\left(\frac{\pi\tau}{2}[2eE_0\tau - \omega(\mathbf{q},0) + \omega(\mathbf{q},1)]\right)}{\sinh\left(\frac{\pi\tau}{2}[2eE_0\tau + \omega(\mathbf{q},0) + \omega(\mathbf{q},1)]\right)\sinh\left(\frac{\pi\tau}{2}[2eE_0\tau - \omega(\mathbf{q},0) - \omega(\mathbf{q},1)]\right)} \ . \tag{4.148}$$

In Fig. 4.5, the integrand of Eq. (4.146) is compared with the integrand of Eq. (4.147). One finds that $2\ln[1-\mathcal{W}(\mathbf{q})]$, which is related to vacuum decay probability, is in fact always larger than the asymptotic one-particle distribution function $\mathcal{F}(u \to 1)$,

[41] In contrast to the static electric field $E(t) = E_0$ where it has been only sensible to talk about the *pair creation rate* $\dot{\mathcal{N}}[e^+e^-]$, i.e. the change in the number of created particles per volume and time, it is now appropriate to consider the *density of created particles* $\mathcal{N}[e^+e^-]$, i.e. the number of particles per volume, which are created during an electric pulse of finite duration:

$$\mathcal{N}[e^+e^-] = \int_{t_{\mathrm{vac}}}^{\infty} \mathrm{d}t \int \frac{\mathrm{d}^3 q}{(2\pi)^3} \dot{\mathcal{F}}(\mathbf{q}, t) = \lim_{t\to\infty} \int \frac{\mathrm{d}^3 q}{(2\pi)^3} \mathcal{F}(\mathbf{q}, t) \ .$$

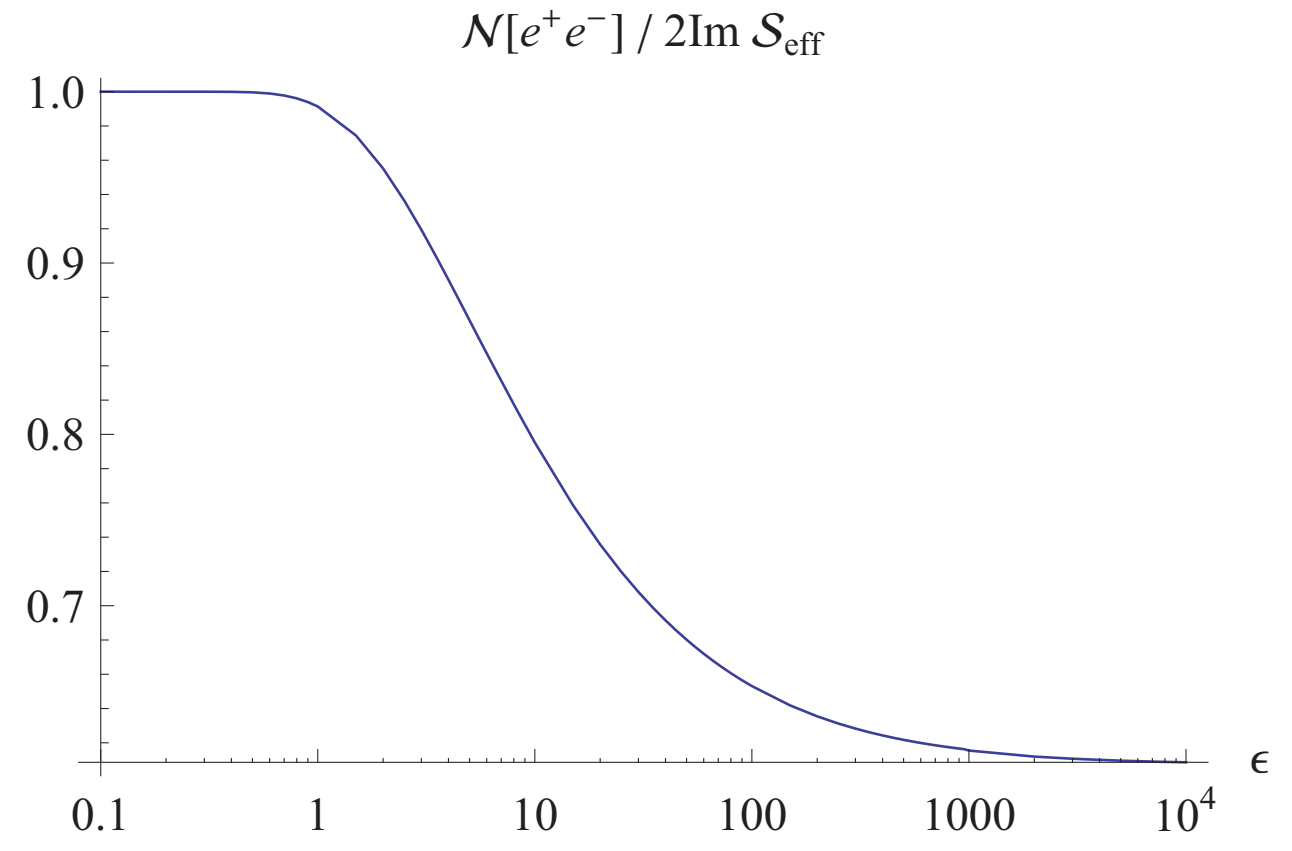

Figure 4.6: Ratio between the density of created particles $\mathcal{N}[e^+e^-]$ and the imaginary part of the effective action density $2\,\mathrm{Im}\,\mathcal{S}_{\mathrm{eff}}$ as function of $\epsilon$ for $\tau = \frac{10}{m}$.

which characterizes the momentum distribution of created particles. This is in accordance with the discussion for the static electric field $E(t) = E_0$ where it has been pointed out that the vacuum decay rate $\mathcal{P}[\text{vac.}]$ is always larger than the pair creation rate $\dot{\mathcal{N}}[e^+e^-]$.

In Fig. 4.6, the ratio between the density of created particles $\mathcal{N}[e^+e^-]$ and the imaginary part of the effective action density $2\,\mathrm{Im}\,\mathcal{S}_{\mathrm{eff}}$ is displayed. Similarly to the static electric field, cf. Fig. 4.2, the difference between those two quantities is again rather minor in the sub-critical field strength regime $\epsilon < 1$ whereas a sizeable deviation occurs for super-critical fields $\epsilon > 1$.

#### 4.3.2.3 Equal-time Wigner components

Given the analytic expression for the one-particle distribution function in Eq. (4.139), it is again possible to derive analytic expressions for the equal-time Wigner components as well. As a matter of fact, one has to calculate $\mathcal{G}(\mathbf{q}, u)$ and $\mathcal{H}(\mathbf{q}, u)$ first, which are given by:

$$\mathcal{G}(\mathbf{q}, u) = \frac{\omega^2(\mathbf{q},u)}{2eE_0\tau\epsilon_\perp}\mathcal{F}'(\mathbf{q}, u) \ , \tag{4.149}$$

$$\mathcal{H}(\mathbf{q}, u) = \frac{2eE_0\epsilon_\perp u(1-u)}{\omega^3(\mathbf{q},u)} \times \left[1 - \mathcal{F}(\mathbf{q}, u) - \frac{\omega^2(\mathbf{q},u)\pi_3(q_3,u)}{eE_0\tau\epsilon_\perp^2}\mathcal{F}'(\mathbf{q}, u) + \left[\frac{\omega^2(\mathbf{q},u)}{2eE_0\tau\epsilon_\perp}\right]^2\mathcal{F}''(\mathbf{q}, u)\right] \ . \tag{4.150}$$

Here, the prime denotes differentiation with respect to $u$ again. Note, however, that both $\mathcal{F}'(\mathbf{q}, u)$ and $\mathcal{F}''(\mathbf{q}, u)$ yield rather cumbersome expressions as there occurs no simplification like for the static electric field $E(t) = E_0$.

The change of variables $\mathbf{q} \to \mathbf{p} + e\mathbf{A}(t)$ then allows for the transformation back to phase space:

$$q_3 \to p_3 - eE_0\tau(2u-1) \qquad \text{and} \qquad \mathbf{q}_\perp \to \mathbf{p}_\perp \ , \tag{4.151}$$

so that:

$$\pi_3(\mathbf{q}, u) \to p_3 \qquad \text{and} \qquad \omega(\mathbf{q}, u) \to \omega(\mathbf{p}) \ . \tag{4.152}$$

Accordingly, any function which depends only on the canonical momentum $q_3$ shows a dependence on both the phase space kinetic momentum $p_3$ and the time variable $u$ after this transformation:[42]

$$\omega(\mathbf{q}, 0) \quad \to \quad \sqrt{\epsilon_\perp^2 + [p_3 - 2ueE_0\tau]^2} \ , \tag{4.153}$$

$$\omega(\mathbf{q}, 1) \quad \to \quad \sqrt{\epsilon_\perp^2 + [p_3 + 2(1-u)eE_0\tau]^2} \ . \tag{4.154}$$

## 4.4 Pulsed electric field with sub-cycle structure

In the previous section, an analytic expression for the one-particle distribution function $\mathcal{F}(\mathbf{q}, t)$ in the presence of a static electric field $E(t) = E_0$ as well as a pulsed electric field $E(t) = E_0 \operatorname{sech}^2(\frac{t}{\tau})$ has been derived. For an arbitrary electric field $E(t)$, however, no such analytic solution can be found so that the investigation of the Schwinger effect has to be based on numerical results.[43]

As a matter of fact, both the static and the pulsed electric field are not very realistic in the sense of representing a field configuration to be realized at high-intensity laser facilities: Due to the fact that the field strengths which are required to create electron-positrons pairs out of the vacuum are most probably produced in the focus of colliding laser pulses, the electric field shows in fact both spatial and temporal variations.[44] However, as the spatial focussing scale is much larger than the Compton wavelength $\lambda_C$, the effect of spatial variations might be ignored in first approximation, resulting in an electric field configuration which is well suited for an investigation by means of the quantum Vlasov equation.

[42] Most notably, the parameters $a(\mathbf{q})$, $b(\mathbf{q})$ and $c(\mathbf{q})$ appearing in the first three arguments of the hypergeometric functions depend on both $p_3$ and $u$ after this transformation. The transformation of the dimensionless time variable $u$ to the usual time variable $t$ is obtained via Eq. (4.121).

[43] A collection of Fortran solvers for the initial value problem for ordinary differential equation systems named *ODEPACK*, which has been invented by the Lawrence Livermore National Laboratory, has been applied in this work, cf.: `http://www.netlib.org/odepack/`

[44] Magnetic fields are neglected as I assume two counter-propagating laser pulses, forming a standing-wave electric field in such a way that the magnetic field vanishes.

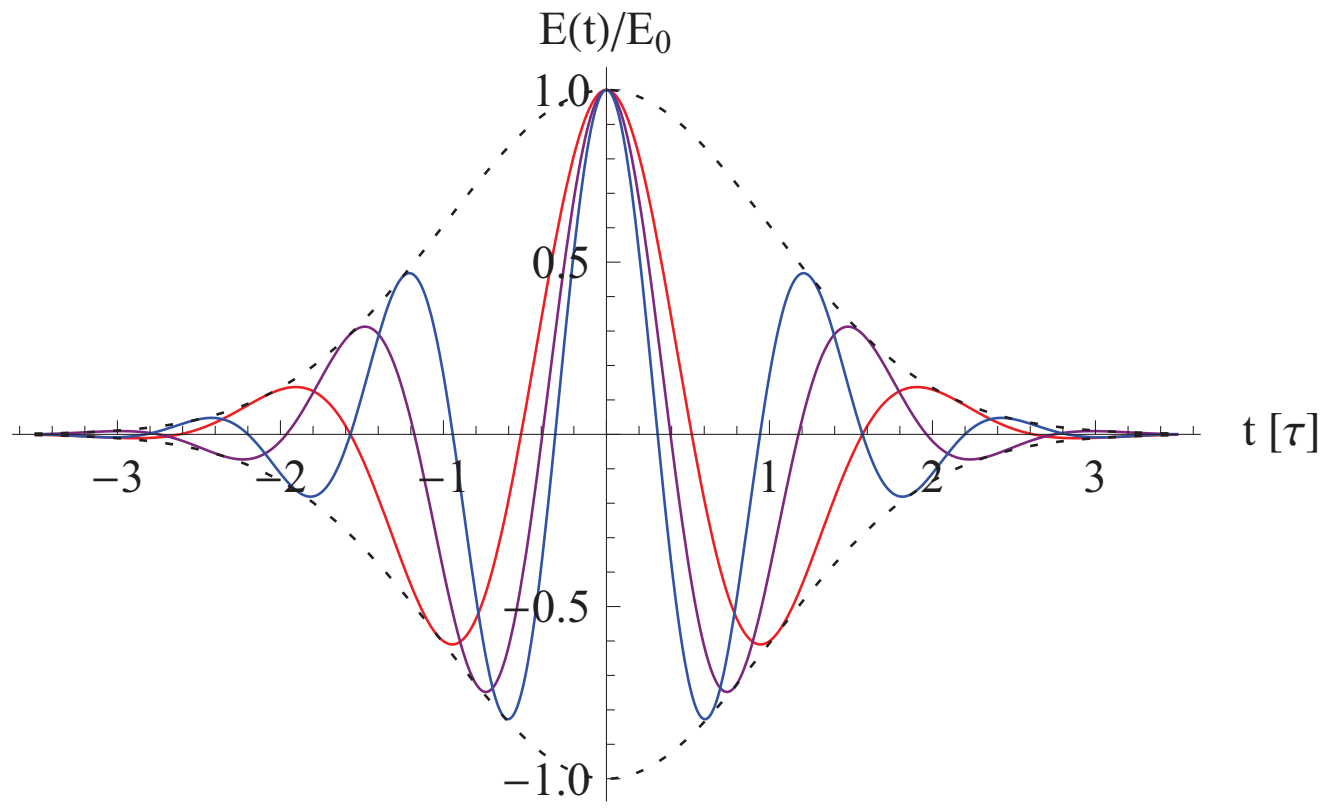

Figure 4.7: Shape of the electric field for $\varphi = 0$ and different values of $\sigma = 3$ (red), $\sigma = 4$ (purple) and $\sigma = 5$ (blue). The Gaussian envelope is given as reference (dotted black).

A simple model of such an electric field in the focus of two colliding laser pulses, which takes into account both the finite pulse duration as well as the sub-cycle structure of a laser pulse, is given by:

$$E(t) = E_0 \cos(\omega t - \varphi) \exp\left(-\frac{t^2}{2\tau^2}\right) \quad . \tag{4.155}$$

Here, $\omega$ denotes the laser frequency, $\tau$ defines the total pulse length and $\varphi$ is the carrier phase which accounts for a possible shift between the maximum of the Gaussian envelope and the maximum of the sub-cycle oscillation.[45] Note that the appearance of a total of three scales makes the investigation of the Schwinger effect in such a type of electric field rather involved.[46] It is convenient to introduce the parameter:

$$\sigma = \omega\tau \ , \tag{4.156}$$

which is a measure of the number of oscillation cycles within the Gaussian envelope. The corresponding vector potential can be expressed in terms of error functions:[47]

$$A(t) = -\sqrt{\frac{\pi}{8}} e^{-i\varphi - \frac{\sigma^2}{2}} E_0 \tau \,\mathrm{Erf}\left(\frac{t}{\sqrt{2}\tau} - \frac{i\sigma}{\sqrt{2}}\right) + c.c. \quad . \tag{4.157}$$

The electric field Eq. (4.155) is displayed for $\varphi = 0$ and different values of $\sigma$ in Fig. 4.7. Additionally, it is shown for $\sigma = 5$ and different values of $\varphi$ in Fig. 4.8.

[45] These studies have in fact been strongly motivated by the sensitive carrier phase dependence of strong-field ionization experiments in atomic, molecular and optical physics [114].

[46] The electric field strength $E_0$, the laser frequency $\omega$ as well as the pulse length $\tau$ have to be considered with respect to the intrinsic scale of QED, which is set by the electron mass $m$.

[47] For details on the error function see [113], chapter 7.

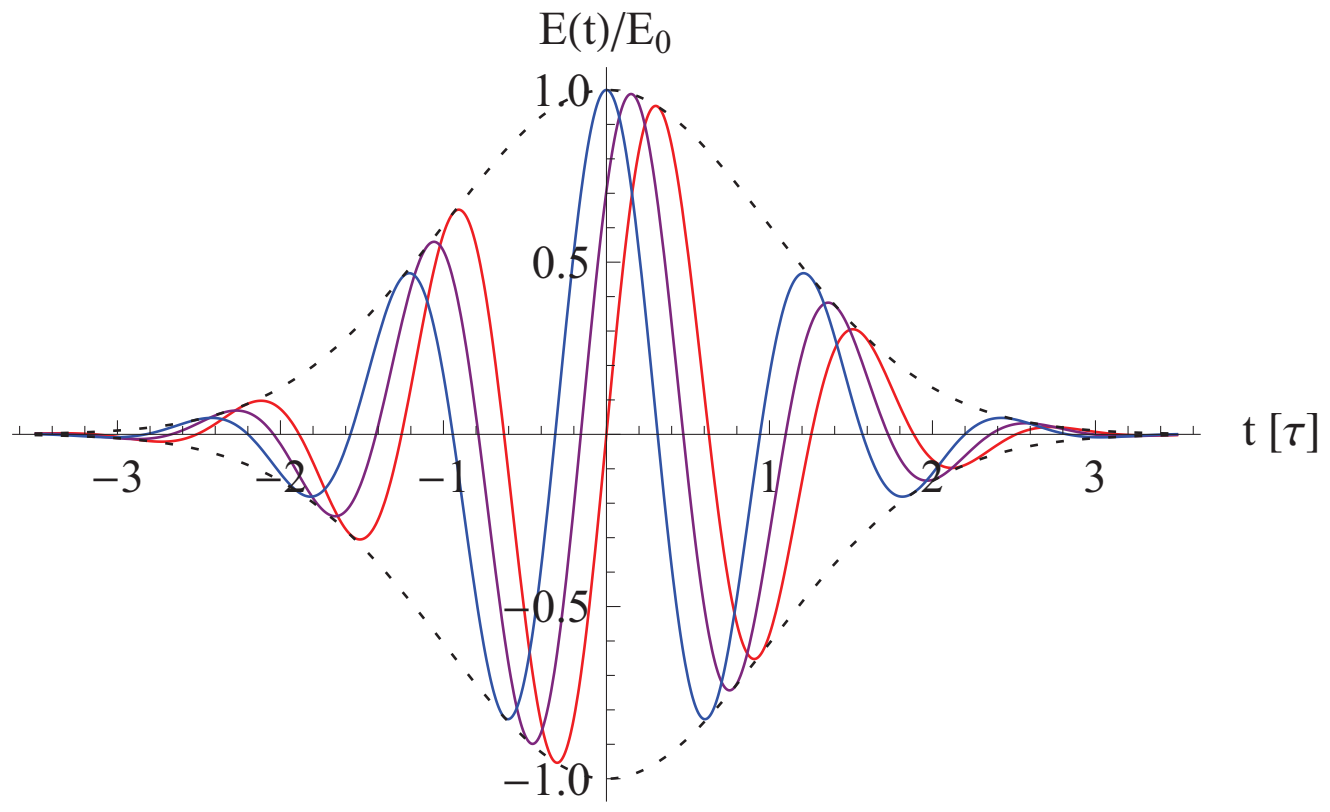

Figure 4.8: Shape of the electric field for $\sigma = 5$ and different values of $\varphi = 0$ (blue), $\varphi = \frac{\pi}{4}$ (purple) and $\varphi = \frac{\pi}{2}$ (red). The Gaussian envelope is given as reference (dotted black).

### 4.4.1 Momentum distribution

As a first issue the momentum distribution of created particles is investigated. To this end, the quantum Vlasov equation in its differential form Eq. (4.36) – (4.38) is numerically solved with the electric field and vector potential being given by Eq. (4.155) and Eq. (4.157), respectively. Note that the backreaction mechanism is neglected as the focus lies on the subcritical field strength regime $\epsilon < 1$.

The parameters are actually chosen in a such a way that $\tau = \frac{100}{m}$ corresponds to a total pulse length of several times $10^{-19}\ s$ which lies in the anticipated range of future XFELs or may become realizable with higher harmonics or secondary beam generation of optical lasers. Moreover, the field strength parameter $\epsilon = 0.1$ and the orthogonal momentum $\mathbf{q}_\perp = 0$.

#### 4.4.1.1 Carrier phase $\varphi = 0$

The Schwinger effect in the presence of the electric field $E(t)$ with vanishing carrier phase $\varphi = 0$ has been investigated by means of a Gaussian WKB approximation previously [34]. It has been shown that the asymptotic momentum distribution in the non-perturbative regime is given by:[48]

$$\mathcal{F}_{\mathrm{WKB}}(\mathbf{q}) \sim \exp\left(-\frac{\pi}{\epsilon}\left[1 - \frac{\tilde{\gamma}^2}{8}\right] - \frac{1}{m^2\epsilon}\left[\tilde{\gamma}^2 q_3^2 + \mathbf{q}_\perp^2\right]\right) , \qquad (4.158)$$

[48]The non-perturbative regime is characterized by a Keldysh parameter $\gamma \ll 1$. Note, however, that the appearance of two different time scales $\frac{1}{\omega}$ and $\tau$ makes the definition of $\gamma$ ambiguous [35]. Considering the laser frequency $\omega$ as dominant time scale, which is justified at least for $\sigma > 1$, suggests the definition:

$$\gamma = \frac{m\omega}{eE_0} .$$

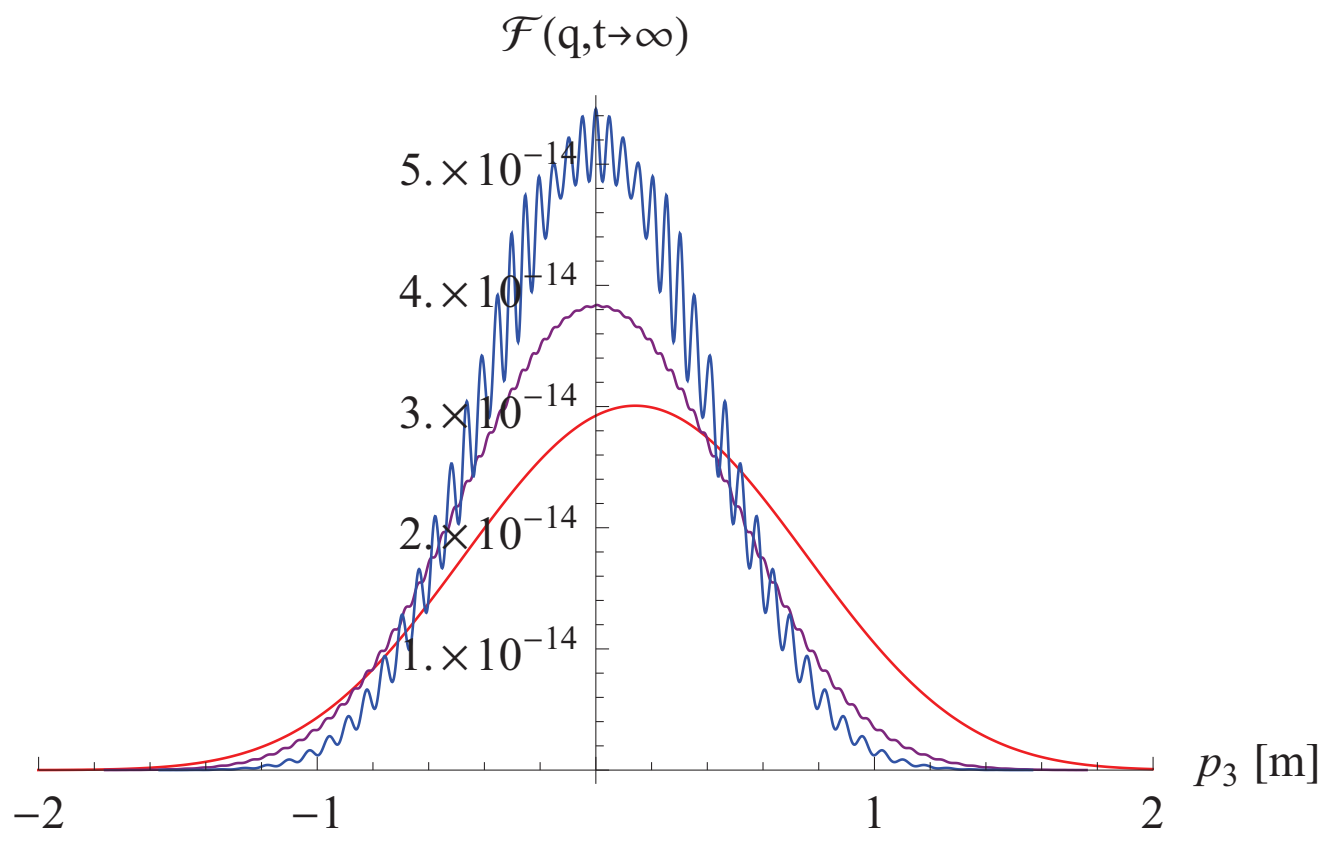

Figure 4.9: Asymptotic one-particle distribution function $\mathcal{F}(\mathbf{q}, t \to \infty)$ when passing from $\sigma = 3$ (red) over $\sigma = 4$ (purple) to $\sigma = 5$ (blue). The main peak shifts to smaller kinetic momenta $p_3$ and distinct oscillations in tune with the laser frequency $\omega$ set in when the number of oscillation cycles is increased. All other parameters are given in the text.

with $\tilde{\gamma}^2 = (1 + \frac{1}{\sigma^2})\gamma^2$. As it turns out, $\mathcal{F}_{\text{WKB}}(\mathbf{q})$ misses in fact several characteristic properties of the momentum distribution. Before actually discussing these shortcomings of the Gaussian WKB approximation, I focus on the asymptotic one-particle distribution function $\mathcal{F}(\mathbf{q}, t \to \infty)$ which is displayed in Fig. 4.9. It is obvious that various effect occur upon passing from $\sigma = 3$ to $\sigma = 5$ which corresponds to increasing the number of oscillation cycles within the Gaussian envelope:

First of all, the main peak of the momentum distribution $p_3^{pk}$ shifts to smaller kinetic momenta and becomes narrower upon increasing the value of $\sigma$: As a matter of fact, the electric field Eq. (4.155) behaves much like a single pulse for small values of $\sigma$ so that all created particles are accelerated into one direction only. By increasing the value of $\sigma$, however, the net acceleration of created particles by the electric field becomes zero, resulting in a momentum distribution which is peaked around:

$$p_3^{pk} = \sqrt{\frac{\pi}{2}} e^{-\frac{\sigma^2}{2}} m^2 \epsilon \tau \overset{\sigma \to \infty}{\longrightarrow} 0 \ . \tag{4.159}$$

The momentum distribution shows in addition distinctive oscillations, with the oscillation scale set by the laser frequency $\omega$. Most notably, these oscillations become even more pronounced for increasing values of $\sigma$. It has to be emphasized that this striking feature is totally missed by the Gaussian WKB approximation as displayed in Fig. 4.10. Additionally note that the width of the momentum distribution is predicted somewhat broader by the Gaussian WKB approximation compared to $\mathcal{F}(\mathbf{q}, t \to \infty)$.

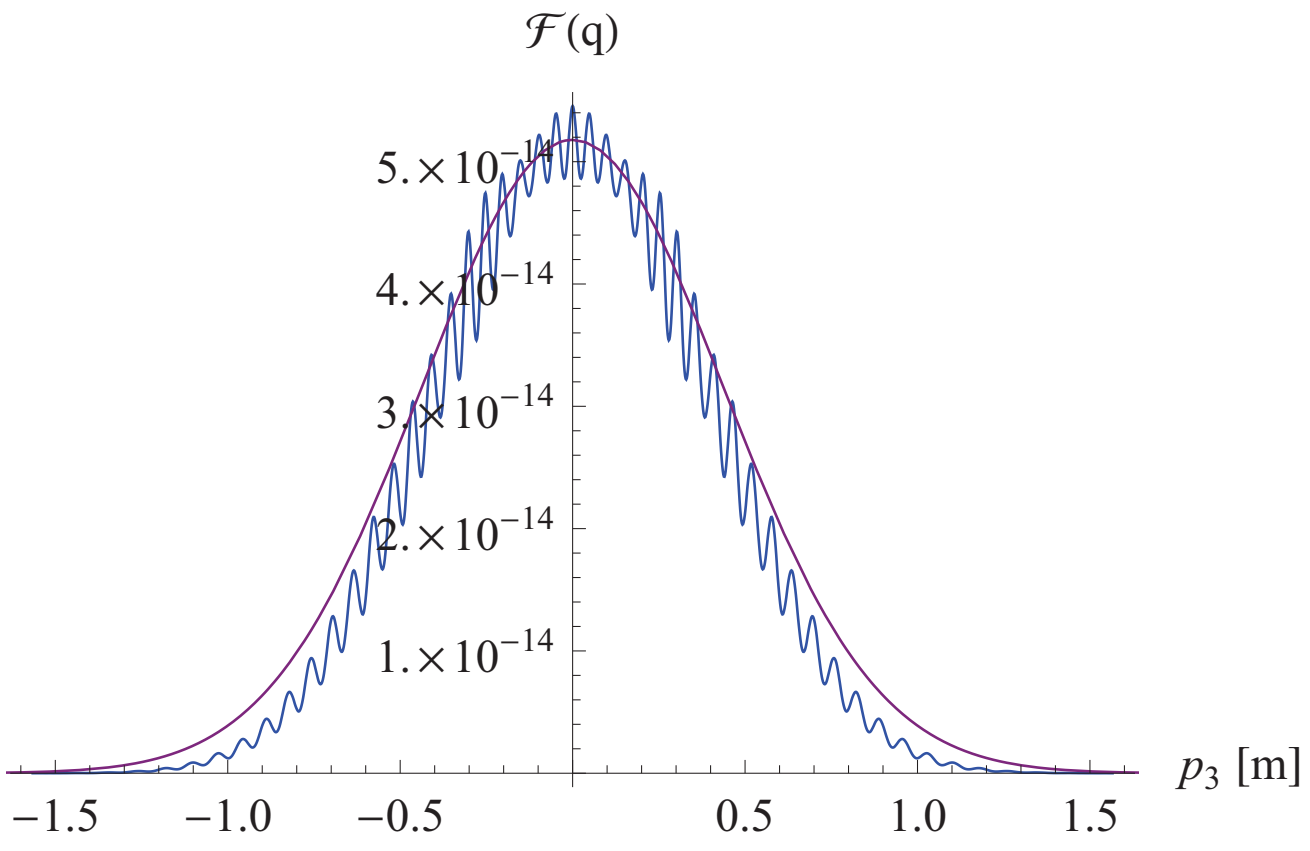

Figure 4.10: Comparison of $\mathcal{F}(\mathbf{q}, t \to \infty)$ (blue) with $\mathcal{F}_{\rm WKB}(\mathbf{q})$ (purple) for $\sigma = 5$. All other parameters are given in the text.

In order to qualitatively understand the oscillatory behavior of the momentum distribution, it is convenient to recall that the Schwinger effect in a spatially homogeneous, time-dependent electric field $E(t)$ can be reformulated as a one-dimensional quantum mechanical over-barrier scattering problem, with the scattering potential given by $-\omega^2(\mathbf{q}, t)$ [17, 19, 115]. According to this approach, the *reflection coefficient* $|\mathcal{R}(\mathbf{q})|^2$ of the scattering problem is related to the asymptotic one-particle distribution function $\mathcal{F}(\mathbf{q}, t \to \infty)$:

$$|\mathcal{R}(\mathbf{q})|^2 = \frac{\mathcal{F}(\mathbf{q}, t \to \infty)}{1 - \mathcal{F}(\mathbf{q}, t \to \infty)} \qquad \longrightarrow \qquad \mathcal{F}(\mathbf{q}, t \to \infty) = \frac{|\mathcal{R}(\mathbf{q})|^2}{1 + |\mathcal{R}(\mathbf{q})|^2} \ . \tag{4.160}$$

Accordingly, the asymptotic one-particle distribution function changes $\mathcal{F}(\mathbf{q}, t \to \infty)$ as the shape of the potential $-\omega^2(\mathbf{q}, t)$ varies for different values of $\mathbf{q}$. Therefore, the oscillatory behavior of the momentum distribution can be interpreted as resonance phenomenon in the equivalent one-dimensional scattering problem. As a matter of fact, this also explains why the spacing between the local maxima is in tune with the laser frequency $\omega$.

As the Gaussian WKB approximation misses some characteristic features of the asymptotic one-particle distribution function $\mathcal{F}(\mathbf{q}, t \to \infty)$, attempts have been made to go beyond this approximation. One possible way is to use the *WKB instanton action* approach, according to which the momentum distribution of created particles is given by [21, 41, 43]:

$$\mathcal{F}^{imp.}_{\rm WKB}(\mathbf{q}) \sim e^{-2\mathcal{S}_{\rm inst.}(\mathbf{q})} \ . \tag{4.161}$$

Here, $\mathcal{S}_{\text{inst.}}(\mathbf{q})$ denotes the instanton action. It has been shown that the leading order instanton action can be defined in the complex $t$-plane as a contour integral:[49]

$$2\mathcal{S}_{\text{inst.}}(\mathbf{q}) = i \oint_\Gamma \mathrm{d}t\, \omega(\mathbf{q}, t) = i \oint_\Gamma \mathrm{d}t\, \sqrt{\epsilon_\perp^2 + [q_3 - eA(t)]^2}\ , \tag{4.162}$$

with the path $\Gamma$ being chosen around the branch cut in the complex $t$-plane. The branch points $t_\pm$ are defined according to:[50]

$$eA(t_\pm) = q_3 \pm i\epsilon_\perp\ . \tag{4.163}$$

Performing a change of variables:

$$T \equiv -\frac{A(t)}{E_0} = \sqrt{\frac{\pi}{8}} e^{-\frac{\sigma^2}{2}} \tau\, \mathrm{Erf}\left(\frac{t}{\sqrt{2}\tau} + \frac{i\sigma}{\sqrt{2}}\right) + c.c.\ , \tag{4.164}$$

the instanton action reads:

$$2\mathcal{S}_{\text{inst.}}(\mathbf{q}) = i \oint_\Gamma \mathrm{d}T\, \frac{\sqrt{\epsilon_\perp^2 + [q_3 + eE_0 T]^2}}{\cos(\omega\, t)} \exp\left(\frac{t^2}{2\tau^2}\right)\ , \tag{4.165}$$

with $t$ being considered as function of $T$.[51] This expression exhibits a number of singularities: The exponential function as well as the square root diverge for large $T$ whereas the cosine term exhibits poles which are determined by:

$$t(T) = \frac{(2n-1)\pi}{2\omega}\ . \tag{4.166}$$

Expanding the integrand of Eq. (4.165) in a Laurent series:[52]

$$eE_0 \sum_{l=0}^{\infty} C_l(\mathbf{q}) T^{-l+1} \sum_{k=0}^{\infty} \frac{t(T)^{2k}}{k! 2^k \tau^{2k}} \left[1 + \frac{\omega^2}{2} t(T)^2 + \frac{5\omega^4}{24} t(T)^4 + \frac{61\omega^6}{720} t(T)^6 + ...\right]\ , \tag{4.167}$$

[49] Actually, Eq. (4.162) gives the instanton action for sQED. It has been shown, however, that the instanton action for sQED can also be applied in QED [41].

[50] For a static electric field with $A(t) = -E_0 t$ this equation has two unique solutions. For the vector potential Eq. (4.157), however, this equation becomes ambiguous. Note that this ambiguity in the definition of the branch points will be disregarded in the following.

[51] In fact, Eq. (4.164) cannot be inverted analytically but only by reversion of the corresponding series. The first few terms in this expansion are given by:

$$t(T) = T + \frac{1+\sigma^2}{6\tau^2} T^3 + \frac{7 + 14\sigma^2 + 9\sigma^4}{120\tau^4} T^5 + \mathcal{O}(T^7)\ .$$

The detailed calculation can be found in Appendix A.6.

[52] Note that only the poles for $n = 0$ and $n = 1$ are taken into account in this expansion whereas the contributions from all the other $n$ are disregarded. As it turns out, the omission of those additional contributions results in a momentum distribution which misses the oscillatory behavior at the end.

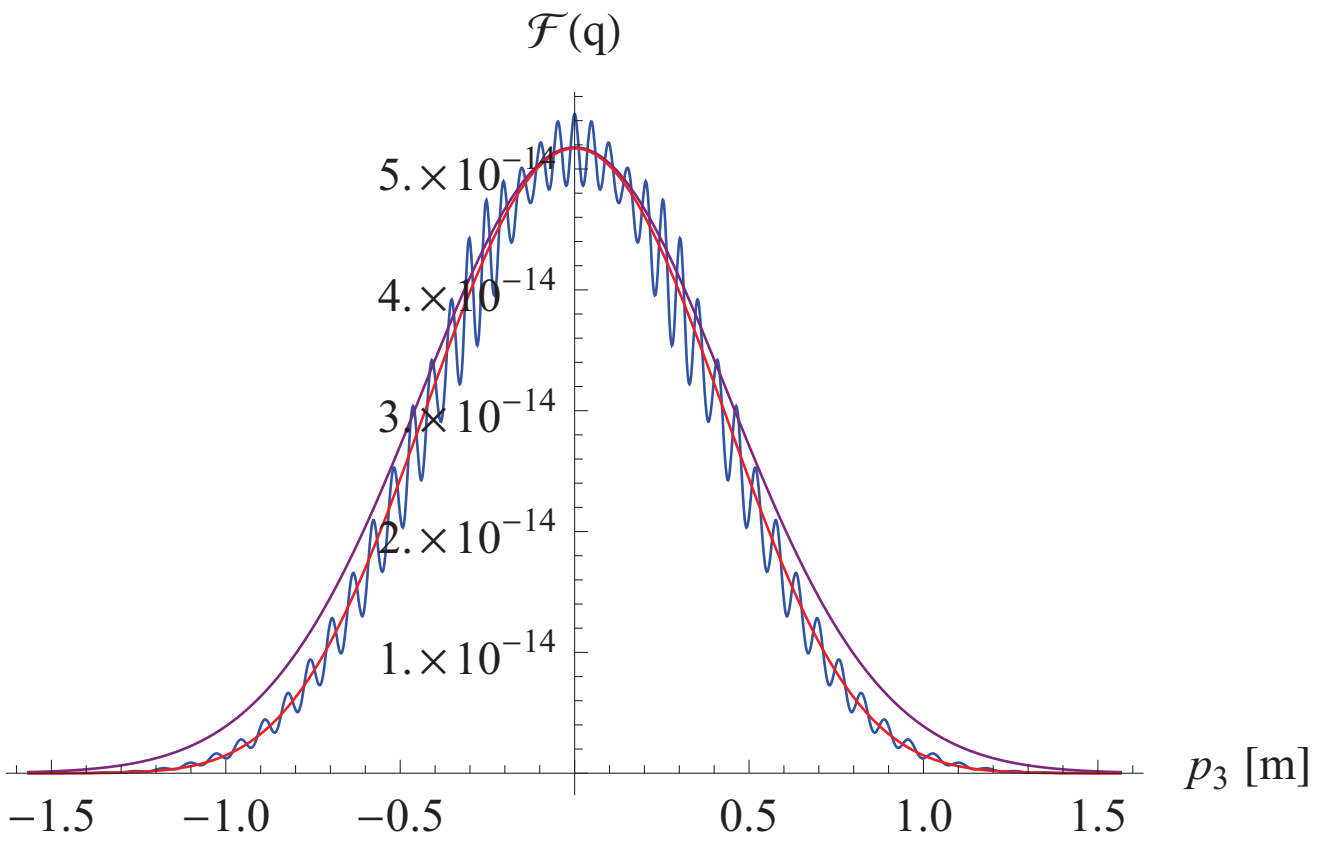

Figure 4.11: Comparison of $\mathcal{F}(\mathbf{q}, t \to \infty)$ (blue), $\mathcal{F}_{\text{WKB}}(\mathbf{q})$ (purple) and $\mathcal{F}^{imp.}_{\text{WKB}}(\mathbf{q})$ (red) for $\sigma = 5$. All other parameters are given in the text.

the instanton action can be calculated by summing up the existing residues. Here, the coefficients $C_l(\mathbf{q})$ are determined in terms of the series expansion:

$$\sqrt{\frac{\epsilon_\perp^2 + [q_3 + eE_0T]^2}{(eE_0T)^2}} = \sum_{l=0}^{\infty} C_l(\mathbf{q}) T^{-l} \ . \tag{4.168}$$

Introducing the following notation for the instanton action:

$$2\mathcal{S}_{\text{inst.}}(\mathbf{q}) = \frac{\pi\epsilon_\perp^2}{eE_0} \sum_{n=0}^{\infty} \frac{\mathcal{S}^{\{2n\}}(\mathbf{q})}{(eE_0\tau)^{2n}} \ , \tag{4.169}$$

the first terms are found to be given by:

$$\mathcal{S}^{\{0\}}(\mathbf{q}) = 1 \ , \tag{4.170}$$

$$\mathcal{S}^{\{2\}}(\mathbf{q}) = \tfrac{1+\sigma^2}{2} \left( q_3^2 - \tfrac{1}{4}\epsilon_\perp^2 \right) \ , \tag{4.171}$$

$$\mathcal{S}^{\{4\}}(\mathbf{q}) = \tfrac{7+14\sigma^2+9\sigma^4}{24} \left( q_3^4 - \tfrac{3}{2} q_3^2 \epsilon_\perp^2 + \tfrac{1}{8}\epsilon_\perp^4 \right) \ , \tag{4.172}$$

$$\mathcal{S}^{\{6\}}(\mathbf{q}) = \tfrac{127+381\sigma^2+463\sigma^4+225\sigma^6}{720} \left( q_3^6 - \tfrac{15}{4} q_3^4 \epsilon_\perp^2 + \tfrac{15}{8} q_3^2 \epsilon_\perp^4 - \tfrac{5}{64}\epsilon_\perp^6 \right) \ . \tag{4.173}$$

In Fig. 4.11, the asymptotic one-particle distribution function $\mathcal{F}(\mathbf{q}, \infty)$ is compared with the Gaussian WKB approximation $\mathcal{F}_{\text{WKB}}(\mathbf{q})$ and the improved WKB approximation $\mathcal{F}^{imp.}_{\text{WKB}}(\mathbf{q})$ for $\sigma = 5$. The improved WKB approximation fits very well the averaged momentum distribution. However, neither the Gaussian nor the improved WKB approximation predict the oscillatory behavior. It was only very recently that this discrepancy has been explained in detail by taking into account interference effects in an extended WKB approximation [38, 40].

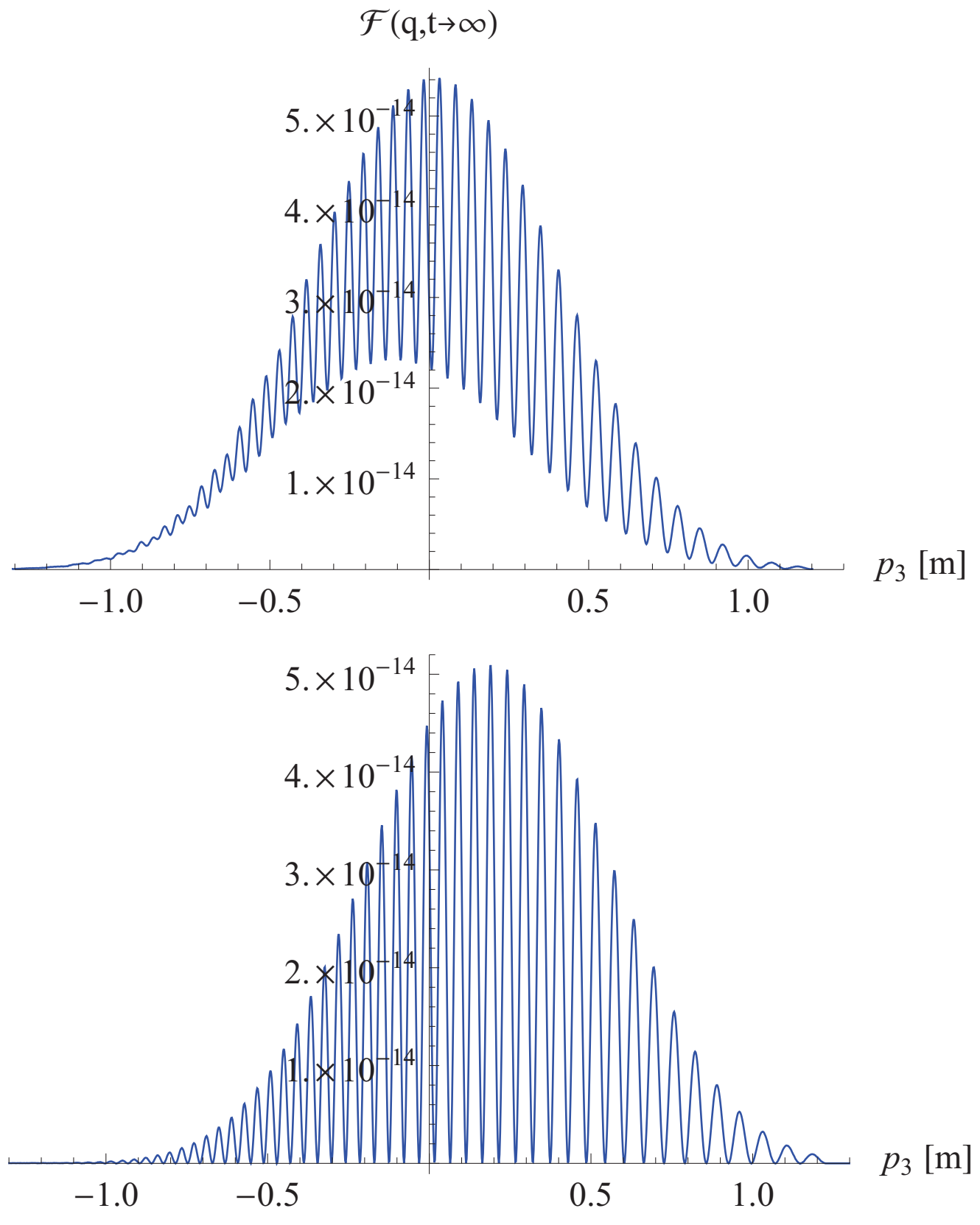

Figure 4.12: Asymptotic one-particle distribution function $\mathcal{F}(\mathbf{q}, t \to \infty)$ for $\varphi = \frac{\pi}{4}$ (*top*) and $\varphi = \frac{\pi}{2}$ (*bottom*). All other parameters are given in the text.

#### 4.4.1.2 Carrier phase $\varphi \neq 0$

I now draw my attention to the Schwinger effect in the presence of the electric field Eq. (4.155) with non-vanishing carrier phase $\varphi \neq 0$. In fact, this field configuration has not been considered before as a non-vanishing carrier phase breaks the symmetry $E(t) = E(-t)$ of the electric field, which in turn makes the imaginary time treatment of the WKB approximation significantly more complicated.

In Fig. 4.12, the asymptotic one-particle distribution function $\mathcal{F}(\mathbf{q}, t \to \infty)$ is displayed for different values of $\varphi$.[53] Apparently, an increasing phase offset makes the oscillatory behavior of the momentum distribution more pronounced. The most distinctive signatures are in fact found for $\varphi = \frac{\pi}{2}$ when $\mathcal{F}(\mathbf{q}, t \to \infty)$ actually vanishes at the minima of the oscillations. Moreover, the main peak shifts to momenta $p_3^{pk} \neq 0$ again, even for large values of $\sigma$.

[53] Equivalently, a negative value of $\varphi < 0$ results in a momentum distribution which is mirrored at $p_3 = 0$.

This characteristic behavior of the asymptotic one-particle distribution function $\mathcal{F}(\mathbf{q}, t \to \infty)$ can be explained in a simple way by considering the equivalent one-dimensional scattering problem again: The electric field is symmetric with respect to time reversal for $\varphi = 0$ whereas it becomes antisymmetric for $\varphi = \frac{\pi}{2}$. Note that the corresponding vector potential $A(t)$ shows exactly the opposite behavior:

$$\varphi = 0: \qquad E(t) = \phantom{-}E(-t) \qquad \text{and} \qquad A(t) = -A(-t) \ , \tag{4.174}$$

$$\varphi = \tfrac{\pi}{2}: \qquad E(t) = -E(-t) \qquad \text{and} \qquad A(t) = \phantom{-}A(-t) \ . \tag{4.175}$$

Owing to the symmetry property of $A(t)$, the scattering potential is in fact symmetric with respect to time reversal for $\varphi = \frac{\pi}{2}$:

$$\varphi = \tfrac{\pi}{2}: \qquad \omega^2(\mathbf{q}, t) = \omega^2(\mathbf{q}, -t) \ . \tag{4.176}$$

As such symmetric scattering potentials allow for perfectly transmitted states:[54]

$$|\mathcal{R}(\mathbf{q})|^2 = 0 \ , \tag{4.177}$$

it is possible to trace the distinctive signatures in the momentum distribution back to a resonance phenomenon: The momentum values at which $\mathcal{F}(\mathbf{q}, t \to \infty)$ vanishes are the same at which perfectly transmitted states in the equivalent one-dimensional scattering problem occur.[55] Note that a more detailed explanation of the oscillatory behavior has recently been given by means of an extended WKB approximation [38, 40].

In analogy to strong-field ionization experiments [114, 116, 119, 120], these distinctive signatures in the momentum distribution may serve as sensitive probe of sub-cycle structure in upcoming high-intensity laser experiments. In addition to the density of created particles, these results suggest a number of new observables such as the main peak position $p_3^{pk}$, the width of the momentum distribution and, most notably, its oscillatory structure. In particular, the characteristics of these oscillations provide precise information about the carrier phase $\varphi$ and the total pulse length parameter $\tau$. As the latter features are hard to control a priori in an absolute manner for a high-intensity laser, these momentum signatures could serve as a tomograph of the laser pulse, providing for a unique means to verify and confirm design goals of future high-intensity laser systems.

[54] This argumentation is guided by intuition from simple one-dimensional scattering problems such as the square well. Unluckily, I am not aware of any rigorous proof of this statement.

[55] Note that there is a close similarity between this effect and the matterless double slit experiment [116, 117, 118].

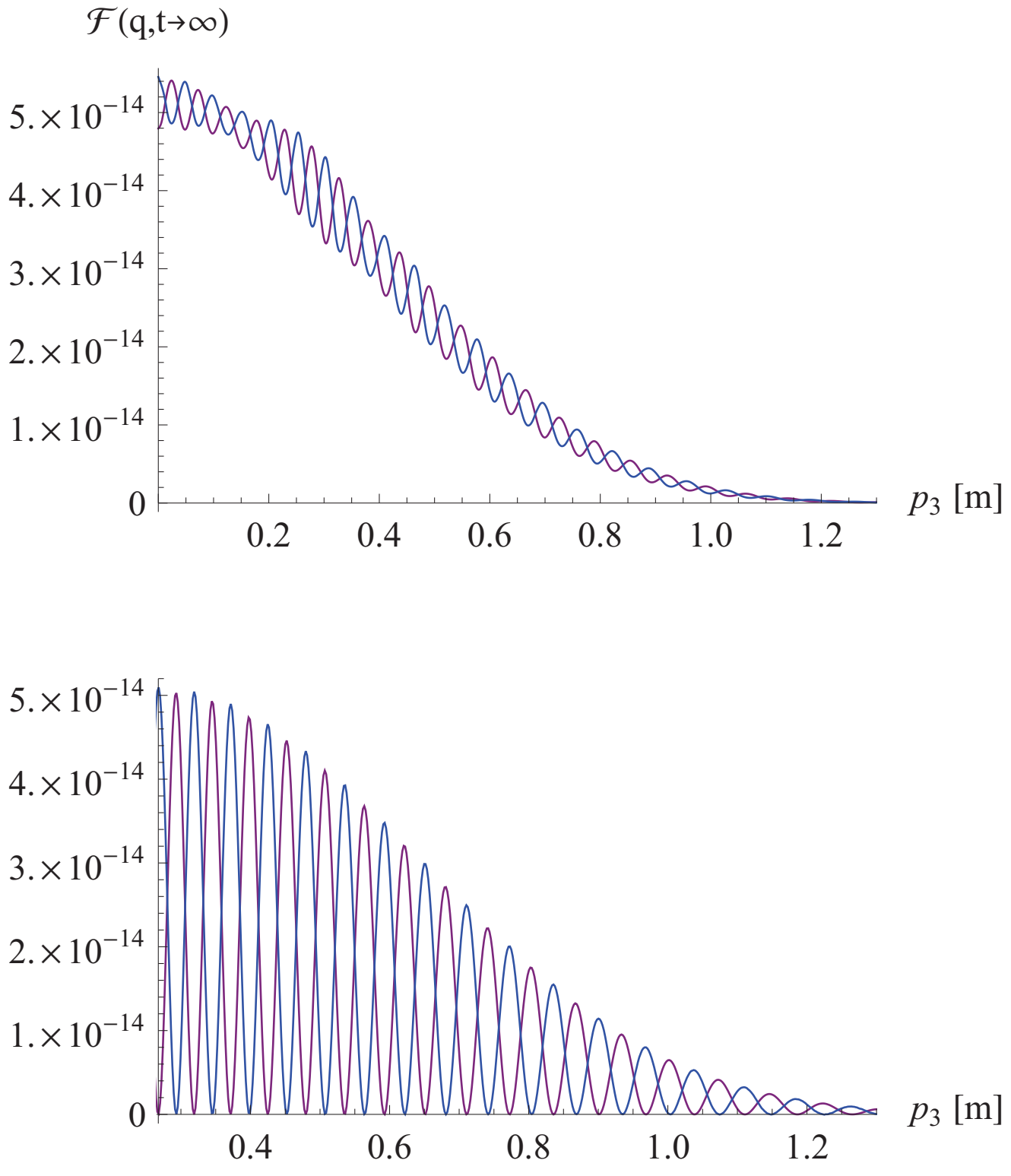

Figure 4.13: Comparison of $\mathcal{F}_{\text{sp}}(\mathbf{q}, t \to \infty)$ (blue) with $\mathcal{F}_{\text{sc}}(\mathbf{q}, t \to \infty)$ (purple) for $\varphi = 0$ (*top*) and $\varphi = \frac{\pi}{2}$ (*bottom*). The momentum distributions are only shown for kinetic momenta $p_3 > p_3^{pk}$. All other parameters are given in the text.

### 4.4.2 Quantum statistics effect

So far, all numerical simulations have been based upon the quantum Vlasov equation in the framework of QED:[56]

$$\dot{\mathcal{F}}_{\text{sp}}(\mathbf{q}, t) = \tfrac{1}{2} Q_{\text{sp}}(\mathbf{q}, t) \int_{t_{\text{vac}}}^{t} Q_{\text{sp}}(\mathbf{q}, t')[1 - 2\mathcal{F}_{\text{sp}}(\mathbf{q}, t')] \cos\left[2\Theta(\mathbf{q}, t', t)\right] . \tag{4.178}$$

[56]Note, that the quantum Vlasov equation has now been written in terms of the one-particle distribution function for one spin direction $\mathcal{F}_{\text{sp}}(\mathbf{q}, t)$ in order to recognize the close similarity between QED and sQED later on. Note that Eq. (4.33) is recovered according to:

$$\mathcal{F}(\mathbf{q}, t) \equiv 2\mathcal{F}_{\text{sp}}(\mathbf{q}, t) .$$

The analogous quantum Vlasov equation in the framework of sQED reads [29]:

$$\dot{\mathcal{F}}_{\rm sc}(\mathbf{q},t) = \frac{1}{2} Q_{\rm sc}(\mathbf{q},t) \int\limits_{t_{\rm vac}}^{t} Q_{\rm sc}(\mathbf{q},t')[1+2\mathcal{F}_{\rm sc}(\mathbf{q},t')] \cos\left[2\Theta(\mathbf{q},t',t)\right] \ , \tag{4.179}$$

with:

$$Q_{\rm sp}(\mathbf{q},t) = \frac{eE(t)\epsilon_\perp}{\omega^2(\mathbf{q},t)} \qquad \text{and} \qquad Q_{\rm sc}(\mathbf{q},t) = \frac{eE(t)\pi_3(q_3,t)}{\omega^2(\mathbf{q},t)} \ . \tag{4.180}$$

There are two major differences between the quantum Vlasov equations in the framework of QED and sQED, respectively: First of all, a Bose enhancement factor appears for scalars $[1 + 2\mathcal{F}_{\rm sc}]$ whereas a Pauli-blocking factor is found for spinors $[1-2\mathcal{F}_{\rm sp}]$. Moreover, the different prefactors $Q_{\rm sp}(\mathbf{q},t)$ and $Q_{\rm sc}(\mathbf{q},t)$ account for the difference in the phase space occupation between fermions and bosons.

In Fig. 4.13, the asymptotic one-particle distribution function $\mathcal{F}_{\rm sp}(\mathbf{q},t\to\infty)$ for spinor particles is compared with the asymptotic one-particle distribution function $\mathcal{F}_{\rm sc}(\mathbf{q},t\to\infty)$ for scalar particles. The averaged momentum distribution is in fact the same for both QED and sQED, however, quantum statistics plays a crucial rule for the detailed shape: The momentum distribution for spinor particles shows a local maximum at those momentum values at which the momentum distribution for scalar particles shows a local minimum, and vice versa. This out-of-phase behavior between QED and sQED is in fact found for any value of $\varphi$. As for the explanation of the oscillatory structure of the momentum distribution, this out-of-phase behavior between QED and sQED has very recently been explained in detail in an extended WKB approximation [38, 40].

Chapter 5

# Schwinger effect for $\mathbf{E}(\mathbf{x}, t)$

In this chapter I present the results of an ab initio simulation of the Schwinger effect in the presence of a simple space- and time-dependent electric field. This investigation represents in fact substantial progress as there have not been any rigorous studies of the Schwinger effect in space- and time-dependent electric fields so far. Note, however, that a solution of the equations of motion for the equal-time Wigner components Eq. (3.36) – (3.43) in the framework of $3+1$ dimensional QED ($\text{QED}_{3+1}$) is currently not feasible for computational reasons. Accordingly, I have to restrict myself on $1+1$ dimensional QED ($\text{QED}_{1+1}$) within this thesis.[1]

In Section 5.1, I briefly introduce the equal-time Wigner formalism in $\text{QED}_{1+1}$ upon which the subsequent simulations are based. In Section 5.2, various strategies for solving the equations of motion for the equal-time Wigner components are presented, including both an exact and an approximate scheme as well as a local density approximation. Eventually, in Section 5.3, I discuss the numerical results concerning the momentum distribution, the charge distribution as well as the total number of created particles. Moreover, a comparison between the exact solution and the approximate ones is drawn.

## 5.1 Equal-time Wigner formalism in $\text{QED}_{1+1}$

It has been stated that the equal-time Wigner formalism in mean field approximation is complete in the sense that knowledge of the equal-time Wigner components suffices to calculate any physical observable. Accordingly, the investigation of the Schwinger effect can be considered as completed from a theoretical point of view.

[1] In order to roughly estimate the required memory for a direct numerical simulation in $\text{QED}_{3+1}$, consider a discretization of the PDE system Eq. (3.36) – (3.43) on a finite grid, with the number of grid points in each direction being of the order of $n_i \sim \mathcal{O}(10^2)$. Accordingly, in order to fully characterize the state of the system at a given instant of time, the required memory is of the order of:

$$m \sim \mathcal{O}(10^{13}\text{B}) \sim \mathcal{O}(10\text{TB}) \; .$$

Actually, this estimate is far too conservative: The experience in simulating vacuum pair creation in $1+1$ dimensional systems shows that the number of grid points has to be chosen to be $n_i = \mathcal{O}(10^3) - \mathcal{O}(10^4)$ in order to yield both high accuracy and good convergence. Accordingly, the required memory is far beyond the scope of current technology.

Nonetheless, the numerical simulation of the PDE system Eq. (3.36) – Eq. (3.43) poses still a serious problem from a practical point of view: This PDE system is formulated in terms of 16 equal-time Wigner components $\mathbb{w}(\mathbf{x}, \mathbf{p}, t)$ which themselves depend on six phase space variables $\{\mathbf{x}, \mathbf{p}\}$ as well as the time variable $t$. Accordingly, any numerical simulation is doomed to fail for computational reasons for the time being.

As a matter of fact, one could restrict oneself to a highly symmetric electromagnetic field $F^{\mu\nu}(\mathsf{x})$ in order to reduce the number of phase space variables by a proper choice of the coordinate system. Choosing a cylindrical symmetric configuration, for instance, the polar angle could most probably be eliminated for symmetry reasons.[2] Nonetheless, four phase space variables corresponding to the directions parallel $\{x_{\|}, p_{\|}\}$ and perpendicular $\{x_{\perp}, p_{\perp}\}$ to the symmetry axis, respectively, would still remain.

In order to avoid all these difficulties I restrict myself to $\mathrm{QED}_{1+1}$ from the very beginning.[3] Accordingly, one is left with only two phase space variables $\{x, p\}$. Moreover, the number of equal-time Wigner components $\mathbb{w}(x, p, t)$ can be reduced from 16 to 4 as well.[4] Note, however, that the notion of a magnetic field does not exist in $\mathrm{QED}_{1+1}$ in contrast to $\mathrm{QED}_{3+1}$ so that fermions have to be considered as spinless particles. Nonetheless, this peculiarity of $\mathrm{QED}_{1+1}$ should not be of severe consequence: As the focus of this thesis lies on vacuum pair creation in electric fields, magnetic effects would have been neglected anyway. Accordingly, even if the results from $\mathrm{QED}_{1+1}$ cannot be transferred one-to-one to $\mathrm{QED}_{3+1}$, it can still be hoped that they contain at least the main features of the Schwinger effect in the presence of space- and time-dependent electric fields in $\mathrm{QED}_{3+1}$.

Moreover, it has to be emphasized that the experience which is gained by the numerical simulation of the Schwinger effect in $\mathrm{QED}_{1+1}$ can most probably be carried over to the investigation of the Schwinger effect in $\mathrm{QED}_{3+1}$ as well.

[2] The generalization of the equal-time Wigner formalism to curvilinear coordinates has not been worked out in this thesis. This extension of the current work is surely worthwhile to consider in future investigations.

[3] For massless fermions this is called the Schwinger model, which can be solved analytically [121, 122]. For details on the massive Schwinger model see [123, 124].

[4] Accordingly, the required memory for a direct numerical simulation on a grid with $n_i \sim \mathcal{O}(10^4)$ grid points in each direction is feasible with current technology:

$$m \sim \mathcal{O}(10^9\mathrm{B}) \sim \mathcal{O}(1\mathrm{GB}) \ .$$

### 5.1.1 QED in $1+1$ dimensions

The Lagrangian of $\mathrm{QED}_{1+1}$ looks alike the Lagrangian of $\mathrm{QED}_{3+1}$ given in Eq. (3.1):

$$\mathcal{L}(\Psi,\bar{\Psi},\mathsf{A}) = \tfrac{1}{2}\Big(\bar{\Psi}\gamma^{\mu}[i\partial_{\mu} - eA_{\mu}]\Psi - [i\partial_{\mu} + eA_{\mu}]\bar{\Psi}\gamma^{\mu}\Psi\Big) - m\bar{\Psi}\Psi - \tfrac{1}{4}F^{\mu\nu}F_{\mu\nu}\,. \quad (5.1)$$

Note, however, that space-time is only two-dimensional so that $\mu = \{0,1\}$. Accordingly, the Dirac algebra is composed of two Dirac gamma matrices only:[5]

$$\{\gamma^{\mu},\gamma^{\nu}\} = 2g^{\mu\nu} \qquad \text{with} \qquad [\gamma^0]^{\dagger} = \gamma^0\ ,\ [\gamma^1]^{\dagger} = -\gamma^1\ . \quad (5.2)$$

As a consequence, the Dirac field $\Psi(\mathsf{x})$ and the adjoint field $\bar{\Psi}(\mathsf{x})$ posses only two components. Moreover, it is again possible to define a chirality matrix:[6]

$$\gamma^5 \equiv \gamma^0\gamma^1 \qquad \text{with} \qquad [\gamma^5]^{\dagger} = \gamma^5\ . \quad (5.3)$$

Additionally, the field strength tensor $F^{\mu\nu}(\mathsf{x})$ possesses only one non-trivial component which can be identified with the electric field:[7]

$$F^{\mu\nu}(\mathsf{x}) = \partial^{\mu}A^{\nu}(\mathsf{x}) - \partial^{\nu}A^{\mu}(\mathsf{x}) \qquad \text{with} \qquad F^{10}(\mathsf{x}) \equiv E(\mathsf{x})\ . \quad (5.4)$$

### 5.1.2 Covariant Wigner formalism

The derivation of the equation of motion for the equal-time Wigner function in $\mathrm{QED}_{1+1}$ is along the same lines as the corresponding derivation in the framework of $\mathrm{QED}_{3+1}$, which has been discussed in detail in Chapter 3. Starting off with the covariant Wigner operator:

$$\hat{\mathcal{W}}^{(2)}(\mathsf{x},\mathsf{p}) \equiv \frac{1}{2}\int \mathrm{d}^2y\, e^{i\mathsf{p}\cdot\mathsf{y}}\, e^{ie\int_{-1/2}^{1/2}\mathrm{d}\xi \mathsf{A}(\mathsf{x}+\xi\mathsf{y})\cdot\mathsf{y}}\, \left[\bar{\Psi}(\mathsf{x}-\tfrac{\mathsf{y}}{2}),\Psi(\mathsf{x}+\tfrac{\mathsf{y}}{2})\right], \quad (5.5)$$

the covariant Wigner function is again defined as the vacuum expectation value:

$$\mathcal{W}^{(2)}(\mathsf{x},\mathsf{p}) = \langle\Omega|\hat{\mathcal{W}}^{(2)}(\mathsf{x},\mathsf{p})|\Omega\rangle\ . \quad (5.6)$$

[5] The Dirac algebra in two dimensions can be represented in terms of the first two Pauli matrices:

$$\gamma^0 \equiv \sigma_1 \quad \text{and} \quad \gamma^1 \equiv -i\sigma_2\ .$$

[6] With the previous choice of Dirac gamma matrices, the chirality matrix is given by the third Pauli matrix:

$$\gamma^5 = \sigma_3\ .$$

[7] In this chapter, sans serif variables denote two-vectors, for example $\mathsf{p} \equiv p^{\mu} = (p^0, p)$

The equation of motion for the covariant Wigner function is then again found by taking advantage of the Dirac equation and its adjoint. As the BBGKY hierarchy is again truncated at the one-body level in mean field approximation, one finally obtains the self-consistent PDE system:

$$\frac{1}{2}D_\mu\left[\gamma^\mu\,,\mathcal{W}^{(2)}\right] - i\Pi_\mu\left\{\gamma^\mu\,,\mathcal{W}^{(2)}\right\} \;=\; -2im\mathcal{W}^{(2)}\;, \tag{5.7}$$

$$\frac{1}{2}D_\mu\left\{\gamma^\mu\,,\mathcal{W}^{(2)}\right\} - i\Pi_\mu\left[\gamma^\mu\,,\,\mathcal{W}^{(2)}\right] \;=\; 0\;, \tag{5.8}$$

and

$$\partial^{\mathsf{x}}_\mu F^{\mu\nu}(\mathsf{x}) = \langle\Omega|\tfrac{e}{2}\left[\bar\Psi(\mathsf{x}),\gamma^\nu\Psi(\mathsf{x})\right]|\Omega\rangle = e\int\frac{\mathrm{d}^2p}{(2\pi)^2}\,\mathrm{tr}[\gamma^\nu\mathcal{W}^{(2)}(\mathsf{x},\mathsf{p})]\;, \tag{5.9}$$

with:

$$D_\mu(\mathsf{x},\mathsf{p}) \;\equiv\; \partial^{\mathsf{x}}_\mu - e\int_{-\frac{1}{2}}^{\frac{1}{2}}\mathrm{d}\xi\,F_{\mu\nu}(\mathsf{x}-i\xi\partial_{\mathsf{p}})\partial^\nu_{\mathsf{p}}\;, \tag{5.10}$$

$$\Pi_\mu(\mathsf{x},\mathsf{p}) \;\equiv p_\mu - ie\int_{-\frac{1}{2}}^{\frac{1}{2}}\mathrm{d}\xi\xi F_{\mu\nu}(\mathsf{x}-i\xi\partial_{\mathsf{p}})\partial^\nu_{\mathsf{p}}\,. \tag{5.11}$$

Decomposing the covariant Wigner function into its Dirac bilinears:

$$\mathcal{W}^{(2)}(\mathsf{x},\mathsf{p}) = \tfrac{1}{2}\left[\mathbb{S} + i\gamma^5\mathbb{P} + \gamma^\mu\mathbb{V}_\mu\right]\;, \tag{5.12}$$

one may derive equations of motion for the covariant Wigner components $\mathbb{W}(\mathsf{x},\mathsf{p})$. Adopting both the commutator and the anticommutator between $\gamma^\mu$ and the Dirac bilinears:[8]

| | $\{\gamma^\mu,\cdot\}$ | $[\gamma^\mu,\cdot]$ |
|---|---|---|
| I | $2\gamma^\mu$ | $0$ |
| $\gamma_5$ | $0$ | $-2\epsilon^{\mu\alpha}\gamma_\alpha$ |
| $\gamma^\nu$ | $2g^{\mu\nu}$ | $2\epsilon^{\mu\nu}\gamma_5$ |

(5.13)

the resulting equations of motion split again into a set of inhomogeneous equations:

$$\Pi^\mu\mathbb{V}_\mu = \quad m\mathbb{S} \quad , \tag{5.14}$$

$$\Pi^\mu\mathbb{S} - \tfrac{1}{2}\epsilon^{\mu\nu}D_\nu\mathbb{P} = \quad m\mathbb{V}^\mu \quad , \tag{5.15}$$

$$D^\mu\mathbb{V}^\nu - D^\nu\mathbb{V}^\mu = -2m\epsilon^{\mu\nu}\mathbb{P}\,, \tag{5.16}$$

[8]Again, the usual convention $\epsilon^{01} = 1$ for the totally antisymmetric tensor is used.

as well as into a set of homogenous equations:

$$D^\mu \mathbb{V}_\mu = 0\,, \tag{5.17}$$

$$D^\mu \mathbb{S} + 2\epsilon^{\mu\nu}\Pi_\nu \mathbb{P} = 0\,, \tag{5.18}$$

$$\Pi^\mu \mathbb{V}^\nu - \Pi^\nu \mathbb{V}^\mu = 0\,. \tag{5.19}$$

### 5.1.3 Equal-time Wigner formalism

The equal-time Wigner function is then again defined as the energy average of the covariant Wigner function:

$$\mathcal{W}^{(1)}(x,p,t) \equiv \int \frac{\mathrm{d}p_0}{(2\pi)} \mathcal{W}^{(2)}(\mathsf{x},\mathsf{p})\,. \tag{5.20}$$

Decomposing the equal-time Wigner function into its Dirac bilinears:

$$\mathcal{W}^{(1)}(x,p,t) = \tfrac{1}{2}\left[\mathbb{s} + i\gamma_5 \mathbb{p} + \gamma^\mu \mathbb{v}_\mu\right]\,, \tag{5.21}$$

the equations of motion for the equal-time Wigner components $\mathbb{w}(x,p,t)$ are again found by taking the energy average of Eq. (5.14) – (5.19). The *time-evolution equations* are then given by:[9]

$$D_t\,\mathbb{s} - 2p\,\mathbb{p} = 0\,, \tag{5.22}$$

$$D_t\,\mathbb{v}_0 + \tfrac{\partial}{\partial x}\mathbb{v} = 0\,, \tag{5.23}$$

$$D_t\,\mathbb{v} + \tfrac{\partial}{\partial x}\mathbb{v}_0 = -2m\,\mathbb{p}\,, \tag{5.24}$$

$$D_t\,\mathbb{p} + 2p\,\mathbb{s} = 2m\,\mathbb{v}\,, \tag{5.25}$$

with

$$D_t(x,p,t) \equiv \tfrac{\partial}{\partial t} + e\int_{-\frac{1}{2}}^{\frac{1}{2}} \mathrm{d}\xi\, E\big(x + i\xi\tfrac{\partial}{\partial p}, t\big)\tfrac{\partial}{\partial p}\,. \tag{5.26}$$

The *constraint equations*, on the other hand, read:

$$\mathbb{s}^{[1]} + \Pi_t\,\mathbb{s} - \tfrac{1}{2}\tfrac{\partial}{\partial x}\mathbb{p} = m\,\mathbb{v}_0\,, \tag{5.27}$$

$$\mathbb{v}_0^{[1]} + \Pi_t\,\mathbb{v}_0 - p\,\mathbb{v} = m\,\mathbb{s}\,, \tag{5.28}$$

$$\mathbb{v}^{[1]} + \Pi_t\,\mathbb{v} - p\,\mathbb{v}_0 = 0\,, \tag{5.29}$$

$$\mathbb{p}^{[1]} + \Pi_t\,\mathbb{p} + \tfrac{1}{2}\tfrac{\partial}{\partial x}\mathbb{s} = 0\,, \tag{5.30}$$

[9] The pseudoscalar equal-time Wigner component $\mathbb{p}(x,p,t)$ in $\mathrm{QED}_{1+1}$ takes on the role of the tensorial equal-time Wigner component $\mathbb{t}_1(\mathbf{x},\mathbf{p},t)$ in $\mathrm{QED}_{3+1}$ as $\sigma^{10} = i\gamma^5$ in $1+1$ dimensions.

with

$$\Pi_t(x,p,t) = ie\int_{-\frac{1}{2}}^{\frac{1}{2}} \mathrm{d}\xi\xi\, E\big(x + i\xi\tfrac{\partial}{\partial p}, t\big)\tfrac{\partial}{\partial p}\ . \tag{5.31}$$

It can be shown again that the transport equations for all higher energy moments $\mathbb{w}^{[n]}(x,p,t)$ are identically fulfilled in mean field approximation once the equal-time Wigner components obey Eq. (5.22) – (5.25). Along with appropriate vacuum initial conditions:

$$\mathbb{s}_{\text{vac}}(p) = -\frac{m}{\omega(p)} \qquad \text{and} \qquad \mathbb{v}_{\text{vac}}(p) = -\frac{p}{\omega(p)}\ , \tag{5.32}$$

with $\omega(p) = \sqrt{m^2+p^2}$, the equal-time Wigner formalism is complete in the sense that $\mathbb{w}(x,p,t)$ encodes the information of all higher energy moments.

It has to be emphasized that the equal-time Wigner formalism in $\text{QED}_{1+1}$ represents a drastic simplification compared to the equal-time Wigner formalism in $\text{QED}_{3+1}$ as the number of equal-time Wigner components is reduced from 16 to 4. Nonetheless, it can still be hoped that the Schwinger effect in $\text{QED}_{1+1}$ contains at least the main features of the Schwinger effect in $\text{QED}_{3+1}$.

### 5.1.4 Observables and marginal distributions

The observables in this theory are again found from an analysis of the conservation laws. Noether's theorem proves again the conservation of the total charge $\mathcal{Q}$, the total energy $\mathcal{E}$ and the total momentum $\mathcal{P}$ as well as the Lorentz boost operator $\mathcal{K}$:[10]

$$\mathcal{Q} = e\int \mathrm{d}\Gamma\, \mathbb{v}_0(x,p,t)\ , \tag{5.33}$$

$$\mathcal{E} = \int \mathrm{d}\Gamma [m\mathbb{s}(x,p,t) + p\,\mathbb{v}(x,p,t)] + \tfrac{1}{2}\int \mathrm{d}x\, |E(x)|^2\ , \tag{5.34}$$

$$\mathcal{P} = \int \mathrm{d}\Gamma\, p\,\mathbb{v}_0(x,p,t)\ , \tag{5.35}$$

$$\mathcal{K} = t\mathcal{P} - \int \mathrm{d}\Gamma\, x\,[m\mathbb{s}(x,p,t) + p\,\mathbb{v}(x,p,t)] - \tfrac{1}{2}\int \mathrm{d}x\, x|E(x)|^2\ , \tag{5.36}$$

with $\mathrm{d}\Gamma = \mathrm{d}x\mathrm{d}p/(2\pi)$ being the phase space volume element. These conserved quantities suggest to define various marginal distributions, most notably the *real space charge density* $q(x,t)$ and the *momentum space charge density* $q(p,t)$:

$$q(x,t) \equiv \int \frac{\mathrm{d}p}{(2\pi)}\mathbb{v}_0(x,p,t) \qquad \text{and} \qquad q(p,t) \equiv \int \frac{\mathrm{d}x}{(2\pi)}\mathbb{v}_0(x,p,t)\ , \tag{5.37}$$

[10]The notion of angular momentum does not exist in $1+1$ dimensions.

as well as the *real space energy density* $\epsilon(x,t)$ and the *momentum space energy density* $\epsilon(p,t)$ of Dirac particles:

$$\epsilon(x,t) \equiv \int \frac{\mathrm{d}p}{(2\pi)}\big[m\mathbb{s}(x,p,t) + p\,\mathbb{v}(x,p,t)\big] \;, \tag{5.38}$$

$$\epsilon(p,t) \equiv \int \frac{\mathrm{d}x}{(2\pi)}\big[m\mathbb{s}(x,p,t) + p\,\mathbb{v}(x,p,t)\big] \;. \tag{5.39}$$

The total energy of the Dirac particles $\mathcal{E}_D(t)$ is in fact composed of a vacuum contribution as well as a contribution due to created particles. It is hence indicated to introduce the total energy of created Dirac particles $\mathcal{E}_D^v(t)$:

$$\mathcal{E}_D^v(t) \equiv \mathcal{E}_D(t) - \mathcal{E}_{D,\mathrm{vac}} \;, \tag{5.40}$$

with:

$$\mathcal{E}_D^v(t) = \int \mathrm{d}\Gamma\big[m[\mathbb{s}(x,p,t) - \mathbb{s}_{\mathrm{vac}}(p)] + p\,[\mathbb{v}(x,p,t) - \mathbb{v}_{\mathrm{vac}}(p)]\big] \;. \tag{5.41}$$

On the other hand, $\mathcal{E}_D^v(t)$ should also be calculable by integrating a *particle number quasi-distribution* $n(x,p,t)$ times the one-particle energy $\omega(p)$:

$$\mathcal{E}_D^v(t) \stackrel{!}{=} \int \mathrm{d}\Gamma \omega(p) n(x,p,t) \;. \tag{5.42}$$

Correspondingly one defines the *real space particle number density* $n(x,t)$ and the *momentum space particle number density* $n(p,t)$:[11]

$$n(x,t) \equiv \int \frac{\mathrm{d}p}{(2\pi)} \frac{m[\mathbb{s}(x,p,t) - \mathbb{s}_{\mathrm{vac}}(p)] + p\,[\mathbb{v}(x,p,t) - \mathbb{v}_{\mathrm{vac}}(p)]}{\omega(p)} \;, \tag{5.43}$$

$$n(p,t) \equiv \int \frac{\mathrm{d}x}{(2\pi)} \frac{m[\mathbb{s}(x,p,t) - \mathbb{s}_{\mathrm{vac}}(p)] + p\,[\mathbb{v}(x,p,t) - \mathbb{v}_{\mathrm{vac}}(p)]}{\omega(p)} \;. \tag{5.44}$$

The total number of created particles is then calculated in terms of the particle number densities:

$$\mathcal{N}(t) = \int \mathrm{d}p\, n(p,t) = \int \mathrm{d}x\, n(x,t) \;. \tag{5.45}$$

The charge densities, the particle number densities as well as the total number of created particles are investigated in detail in Section 5.3.

[11] The momentum space particle number density $n(p,t)$ is related to the one-particle distribution function $\mathcal{F}(p,t)$ in the limit of a spatially homogeneous, time-dependent electric field $E(t)$ according to Eq. (4.67):

$$\mathcal{F}(p,t) = \frac{m[\mathbb{s}(p,t) - \mathbb{s}_{\mathrm{vac}}(p)] + p\,[\mathbb{v}(p,t) - \mathbb{v}_{\mathrm{vac}}(p)]}{2\omega(p)} = \lim_{L\to\infty} \frac{\pi}{L} n(p,t) \;.$$

## 5.2 Solution strategies

The subsequent investigation of the Schwinger effect in space- and time-dependent electric fields $E(x,t)$ is based on the equations of motion for the equal-time Wigner components Eq. (5.22) – (5.25), which are solved in order to calculate the charge densities, the particle number densities and the total number of created particles.

As it is in general not possible to find an analytic solution for the equal-time Wigner components $\mathbb{w}(x,p,t)$, these calculations have to be based on numerical simulations. It has to be emphasized that the PDE system Eq. (5.22) – (5.25) is in fact linear and first order in the time derivative $\frac{\partial}{\partial t}$ as well as in the spatial derivative $\frac{\partial}{\partial x}$, however, due to the appearance of the pseudo-differential operator:

$$\mathit{\Delta}(x,p,t) \equiv e \int_{-\frac{1}{2}}^{\frac{1}{2}} \mathrm{d}\xi \, E\big(x + i\xi \tfrac{\partial}{\partial p}, t\big) \tfrac{\partial}{\partial p} \,, \tag{5.46}$$

arbitrarily high momentum derivatives have to be taken into account.[12] Accordingly, it is not possible to numerically solve this hyperbolic PDE system without further approximations or manipulations.

Before actually discussing various solution strategies, it is convenient to transform the homogeneous PDE system Eq. (5.22) – (5.25) into an *inhomogeneous* one by defining *modified equal-time Wigner components*:

$$\mathbb{w}^v(x,p,t) \equiv \mathbb{w}(x,p,t) - \mathbb{w}_{\mathrm{vac}}(p) \ , \tag{5.47}$$

with:

$$\mathbb{w}^v(x,p,t_{\mathrm{vac}}) = 0 \ . \tag{5.48}$$

The equations of motion for the modified equal-time Wigner components $\mathbb{w}^v(x,p,t)$ are accordingly given by:

$$\big[\tfrac{\partial}{\partial t} + \mathit{\Delta}(x,p,t)\big] \, \mathbb{s}^v - 2p \, \mathbb{p}^v = \mathit{\Delta}(x,p,t) \tfrac{m}{\omega(p)} \ , \tag{5.49}$$

$$\big[\tfrac{\partial}{\partial t} + \mathit{\Delta}(x,p,t)\big] \, \mathbb{v}_0^v + \tfrac{\partial}{\partial x} \mathbb{v}^v = 0 \ , \tag{5.50}$$

$$\big[\tfrac{\partial}{\partial t} + \mathit{\Delta}(x,p,t)\big] \, \mathbb{v}^v + \tfrac{\partial}{\partial x} \mathbb{v}_0^v + 2m \, \mathbb{p}^v = \mathit{\Delta}(x,p,t) \tfrac{p}{\omega(p)} \ , \tag{5.51}$$

$$\big[\tfrac{\partial}{\partial t} + \mathit{\Delta}(x,p,t)\big] \, \mathbb{p}^v + 2p \, \mathbb{s}^v - 2m \, \mathbb{v}^v = 0 \ . \tag{5.52}$$

[12]It is assumed that the electric field $E(x,t)$ can be expanded in a Taylor series with respect to the spatial coordinate.

### 5.2.1 Derivative expansion in $p$-space

The expansion of the pseudo-differential operator Eq. (5.46) in a Taylor series with respect to the spatial coordinate reads:

$$\mathit{\Delta}(x,p,t) = \sum_{n=0}^{\infty} \frac{(-1)^n}{4^n(2n+1)2n!} E^{(2n)}(x,t) \frac{\partial^{2n+1}}{\partial p^{2n+1}} \,, \tag{5.53}$$

where $E^{(k)}(x,t)$ denotes the k-th derivative with respect to the spatial coordinate. Assuming that:

$$\Big| E(x,t) \frac{\partial\, \mathbb{w}^v(x,p,t)}{\partial p} \Big| \gg \Big| \frac{E^{(2)}(x,t)}{24} \frac{\partial^3 \mathbb{w}^v(x,p,t)}{\partial p^3} \Big| \,, \tag{5.54}$$

it is well justified to neglect the higher momentum derivatives so that $\mathit{\Delta}(x,p,t)$ might be approximated according to:

$$\mathit{\Delta}(x,p,t) \simeq eE(x,t) \frac{\partial}{\partial p} \,. \tag{5.55}$$

Note that this expression becomes exact in the limit of a spatially homogeneous electric field. Additionally, due to the fact that the k-th derivative with respect to the spatial coordinate scales as:

$$E^{(k)}(x,t) \sim \left[ \frac{\lambda_C}{\lambda} \right]^k \,, \tag{5.56}$$

this approximation is assumed to be valid for weakly varying electric fields with the spatial variation scale being much larger than the Compton wavelength, $\lambda \gg \lambda_C$. On the other hand, it is clear that this approximation is expected to fail once spatial variation scales of the order of the Compton wavelength $\lambda \sim \mathcal{O}(\lambda_C)$ are under consideration.

Accordingly, the equations of motion for the modified equal-time Wigner components Eq. (5.49) – (5.52) turn into a linear, first order hyperbolic PDE system in $p$-space within this approximation:

$$[\tfrac{\partial}{\partial t} + eE(x,t)\tfrac{\partial}{\partial p}]\, \mathbb{s}^v \quad - 2p\, \mathbb{p}^v \quad = -eE(x,t)\tfrac{mp}{\omega^3(p)} \,, \tag{5.57}$$

$$[\tfrac{\partial}{\partial t} + eE(x,t)\tfrac{\partial}{\partial p}]\, \mathbb{v}_0^v + \tfrac{\partial}{\partial x}\mathbb{v}^v \quad = \quad 0 \,, \tag{5.58}$$

$$[\tfrac{\partial}{\partial t} + eE(x,t)\tfrac{\partial}{\partial p}]\, \mathbb{v}^v + \tfrac{\partial}{\partial x}\mathbb{v}_0^v \quad +2m\, \mathbb{p}^v = \quad eE(x,t)\tfrac{m^2}{\omega^3(p)} \,, \tag{5.59}$$

$$[\tfrac{\partial}{\partial t} + eE(x,t)\tfrac{\partial}{\partial p}]\, \mathbb{p}^v \quad + 2p\, \mathbb{s}^v \; -2m\, \mathbb{v}^v = \quad 0 \,, \tag{5.60}$$

with $\{x,p,t\} \in \mathbb{R}$ as well as $\mathbb{w}^v(x,p,t_{\mathrm{vac}}) = 0$.

### 5.2.2 Full solution in $y$-space

The range of validity of the derivative expansion is restricted as it is expected to fail for spatial variation scales of the order of the Compton wavelength $\lambda \sim \mathcal{O}(\lambda_C)$. Accordingly, a full solution of Eq. (5.49) – (5.52) is still needed in order to determine the actual range of validity of the derivative expansion as well as to check the quality of this approximate solution. Note that this full solution can surely be not obtained in $p$-space as arbitrarily high momentum derivatives occur when expanding the pseudo-differential operator $\mathit{\Delta}(x,p,t)$ for a general electric field $E(x,t)$.

Due to the fact that the momentum variable $p$ appears linearly in the equations of motion for the modified equal-time Wigner components Eq. (5.49) – (5.52), it seems to be advantageous to transform these equations to conjugate $y$-space and solve them there. Along with the definition of the Fourier transformation:

$$\mathrm{FT}[\mathbb{w}^v(x,p,t)] = \tilde{\mathbb{w}}^v(x,y,t) \equiv \int \frac{\mathrm{d}p}{(2\pi)} e^{ipy} \mathbb{w}^v(x,p,t) \ , \tag{5.61}$$

momentum derivatives $\frac{\partial}{\partial p}$ transform into linear factors of $y$ and vice versa:

$$\mathrm{FT}\left[p\,\mathbb{w}^v(x,p,t)\right] = -i\tfrac{\partial}{\partial y}\tilde{\mathbb{w}}^v(x,y,t) \ , \tag{5.62}$$
$$\mathrm{FT}\left[\tfrac{\partial}{\partial p}\mathbb{w}^v(x,p,t)\right] = -iy\,\tilde{\mathbb{w}}^v(x,y,t) \ . \tag{5.63}$$

Note that:

$$\tilde{\mathbb{w}}^v(x,y,t) = [\tilde{\mathbb{w}}^v(x,-y,t)]^* \ , \tag{5.64}$$

due to the fact that $\mathbb{w}^v(x,p,t) \in \mathbb{R}$. Accordingly, one can restrict oneself to $y \in \mathbb{R}_+$. Additionally, the pseudo-differential operator Eq. (5.46) turns into a function of $y$ upon Fourier transforming it:[13]

$$\mathrm{FT}\left[\mathit{\Delta}(x,p,t)\right] = -iey \int_{-\frac{1}{2}}^{\frac{1}{2}} \mathrm{d}\xi\, E(x+\xi y,t) \equiv -ie\tilde{\mathcal{E}}(x,y,t) \ . \tag{5.65}$$

Finally, it is also necessary to Fourier transform the inhomogeneous part in order to formulate the PDE system Eq. (5.49) – (5.52) in $y$-space. This seems impossible on a first view as the Fourier transformation of $\mathbb{v}_{\mathrm{vac}}(p)$ does not exist. Note, however, that the relevant quantity is in fact:

$$\mathrm{FT}\left[\mathit{\Delta}(x,p,t)\mathbb{w}_{\mathrm{vac}}(p)\right] = \frac{e\tilde{\mathcal{E}}(x,y,t)}{y}\,\mathrm{FT}\left[\tfrac{\partial}{\partial p}\mathbb{w}_{\mathrm{vac}}(p)\right] \ , \tag{5.66}$$

[13] The function $\tilde{\mathcal{E}}(x,y,t)$ has been introduced for later use. Note that the parameter integral over $\xi$ cannot be carried out analytically in general.

so that the remaining Fourier transformation can indeed be calculated in terms of modified Bessel functions of the second kind:[14]

$$\text{FT}\left[\tfrac{\partial}{\partial p}\tfrac{m}{\omega(p)}\right] = -i\frac{my}{\pi}K_0\big(m|y|\big)\ , \tag{5.67}$$

$$\text{FT}\left[\tfrac{\partial}{\partial p}\tfrac{p}{\omega(p)}\right] = \frac{m|y|}{\pi}K_1\big(m|y|\big)\ . \tag{5.68}$$

Therefore, the equations of motion for the modified equal-time Wigner components Eq. (5.49) – (5.52) turn into a linear, first order hyperbolic PDE system in $y$-space:

$$\begin{aligned}
&[\tfrac{\partial}{\partial t} - ie\tilde{\mathcal{E}}(x,y,t)]\,\tilde{\mathbb{s}}^v + 2i\tfrac{\partial}{\partial y}\,\tilde{\mathbb{p}}^v = -\,i\tfrac{em}{\pi}\tilde{\mathcal{E}}(x,y,t)K_0(my)\ , && (5.69)\\
&[\tfrac{\partial}{\partial t} - ie\tilde{\mathcal{E}}(x,y,t)]\,\tilde{\mathbb{v}}_0^v + \tfrac{\partial}{\partial x}\tilde{\mathbb{v}}^v = 0\ , && (5.70)\\
&[\tfrac{\partial}{\partial t} - ie\tilde{\mathcal{E}}(x,y,t)]\,\tilde{\mathbb{v}}^v + \tfrac{\partial}{\partial x}\tilde{\mathbb{v}}_0^v + 2m\,\tilde{\mathbb{p}}^v = \tfrac{em}{\pi}\tilde{\mathcal{E}}(x,y,t)K_1(my)\ , && (5.71)\\
&[\tfrac{\partial}{\partial t} - ie\tilde{\mathcal{E}}(x,y,t)]\,\tilde{\mathbb{p}}^v - 2i\tfrac{\partial}{\partial y}\,\tilde{\mathbb{s}}^v - 2m\,\tilde{\mathbb{v}}^v = 0\ , && (5.72)
\end{aligned}$$

with $y \in \mathbb{R}_+$ and $\{x,t\} \in \mathbb{R}$ as well as $\tilde{\mathbb{w}}^v(x,y,t_{\text{vac}}) = 0$. It has to be emphasized that the solution of this PDE system is indeed a full one as compared to the approximate solution of the PDE system Eq. (5.57) – (5.60).[15]

Note that the PDE systems which are obtained either from the Fourier transformation to $y$-space or from a leading order derivative expansion in $p$-space are both linear and first order. Nonetheless, the numerical cost for actually solving them is quite different: As both of them are hyperbolic PDE systems they describe in fact propagation phenomena in $\{x,p\}$-space and in $\{x,y\}$-space, respectively. The speed of propagation can be deduced from the prefactors of the partial derivatives:

$$\begin{aligned}
y-\text{space}:&\quad \tfrac{\partial}{\partial x} \quad \text{and} \quad 2\tfrac{\partial}{\partial y}\ ,\\
p-\text{space}:&\quad \tfrac{\partial}{\partial x} \quad \text{and} \quad eE(x,t)\tfrac{\partial}{\partial p}\ .
\end{aligned}$$

Considering the PDE system in $y$-space, there is in fact continued propagation in both $x$- and $y$-directions so that it is necessary to adopt an expanding grid in both directions for actually solving the PDE system numerically. For the PDE system in $p$-space, however, there is continued propagation only in $x$-direction whereas propagation in $p$-direction takes place only when the electric field $E(x,t)$ is non-vanishing. Accordingly, an expanding grid is only needed in $x$-direction whereas one can stick to a fixed grid in $p$-direction.

[14] For details on Bessel functions see [113], chapter 9.

[15] An inverse Fourier transformation has to be performed in order to recover the modified equal-time Wigner components $\mathbb{w}^v(x,p,t)$ in $p$-space.

### 5.2.3 Local density approximation

The local density approximation is in fact no solution of the PDE system Eq. (5.49) – (5.52) but rather gives an approximation of the momentum space particle number density $n(p,t)$ as well as the total number of created particles $\mathcal{N}(t)$. This approximation is based on the one-particle distribution function $\mathcal{F}(p,t)$ which is obtained from solving the quantum Vlasov equation Eq. (4.36) – (4.38) in the presence of a spatially homogeneous, time-dependent electric field $E(t)$.

The idea is the following: Given that the spatial variation scale is much larger than the Compton wavelength, $\lambda \gg \lambda_C$, it is well justified to describe the Schwinger effect at any point $x_i$ independently. Considering a separable space- and time-dependent electric field:

$$E(x,t) = E_0 g(x) h(t) \ , \tag{5.73}$$

one can repeatedly solve the quantum Vlasov equation for a spatially homogeneous electric field $E(t)$ with field strength $E_0 g(x_i)$ at any point $x_i$. This yields the one-particle distribution functions $\mathcal{F}(p,t;E_0 g(x_i))$, respectively, where the dependence on the field strength has been explicitly indicated. Correspondingly, one defines the particle number quasi-distribution in local density approximation:

$$n_{\text{loc}}(x,p,t) \equiv 2\mathcal{F}(p,t;E_0 g(x)) \ . \tag{5.74}$$

Integrating over real space then gives the momentum space particle number density in local density approximation:

$$n_{\text{loc}}(p,t) = \int \frac{\mathrm{d}x}{(2\pi)} n_{\text{loc}}(x,p,t) \ , \tag{5.75}$$

whereas integration over the whole phase space yields the total number of created particles in local density approximation:

$$\mathcal{N}_{\text{loc}}(t) = \int \mathrm{d}\Gamma \, n_{\text{loc}}(x,p,t) \ . \tag{5.76}$$

It is clear that this approximation has a restricted range of validity as one has to assume that $\lambda \gg \lambda_C$ so that the pair creation process at any point $x_i$ can be considered as taking place in a spatially homogeneous, time-dependent electric field with field strength $E_0 g(x_i)$. Additionally, it has to be emphasized that no predictions on the charge densities can be made as there is local charge neutrality in the case of a spatially homogeneous, time-dependent electric field $E(t)$.

## 5.3 Numerical results

In this section I finally present the results of an ab initio simulation of the Schwinger effect in a simple space- and time-dependent electric field $E(x,t)$ in $\mathrm{QED}_{1+1}$. Solving the equations of motion for the modified equal-time Wigner components allows to analyze the time evolution of the system and to investigate various observable quantities such as the charge densities, the particle number densities as well as the total number of created particles. This study should consequently shed some light on the effect of spatial and temporal variations on the pair creation process.

As this is the first study of this kind, I restrict myself to a rather simple space- and time-dependent electric field:

$$E(x,t) = E_0 g(x) h(t) = E_0 \exp\left(-\frac{x^2}{2\lambda^2}\right) \operatorname{sech}^2\left(\frac{t}{\tau}\right) \ , \tag{5.77}$$

where $\tau$ determines the temporal extent of the pulse whereas $\lambda$ describes the characteristic length scale of the problem. The PDE system Eq. (5.57) – (5.60) originating from a leading order derivative expansion in $p$-space as well as the PDE system Eq. (5.69) – (5.72) derived from the Fourier transformation to $y$-space are then solved by means of a finite difference scheme with second order accuracy [125, 126].[16]

In order to actually solve the PDE system in $y$-space later on, one additionally needs to calculate $\tilde{\mathcal{E}}(x,y,t)$ as defined in Eq. (5.65). The parameter integral of the spatial dependence over $\xi$ is given by:

$$\int_{-\frac{1}{2}}^{\frac{1}{2}} \mathrm{d}\xi \, \exp\left(\frac{[x+\xi y]^2}{2\lambda^2}\right) = \sqrt{\frac{\pi}{2}} \lambda \frac{\operatorname{Erf}\left(\frac{y+2x}{\sqrt{8}\lambda}\right) + \operatorname{Erf}\left(\frac{y-2x}{\sqrt{8}\lambda}\right)}{y} \ , \tag{5.78}$$

so that:[17]

$$\tilde{\mathcal{E}}(x,y,t) = \sqrt{\frac{\pi}{2}} E_0 \lambda \left[\operatorname{Erf}\left(\frac{y+2x}{\sqrt{8}\lambda}\right) + \operatorname{Erf}\left(\frac{y-2x}{\sqrt{8}\lambda}\right)\right] \operatorname{sech}^2\left(\frac{t}{\tau}\right) \ . \tag{5.79}$$

The parameters are chosen in such a way that the pulse length parameter takes the value $\tau = \frac{10}{m}$ and the field strength parameter is given by $\epsilon = 0.5$. Accordingly, the Keldysh parameter $\gamma = 0.2$ is still in the non-perturbative regime. The spatial variation scale $\lambda$ is considered for both $\lambda \gg \lambda_C$ as well as for $\lambda \sim \mathcal{O}(\lambda_C)$.

[16] Details can be found in Appendix B.

[17] In order to solve the PDE system in $y$-space efficiently, it is necessary to choose the spatial dependence in such a way that the parameter integral over $\xi$ can be carried out analytically.

### 5.3.1 Time evolution

This part is dedicated to the investigation of the time evolution of the marginal distributions for $\lambda = 10\lambda_C$.[18] These observables are calculated by numerically solving the PDE system Eq. (5.69) – (5.72) in $y$-space and subsequently performing an inverse Fourier transformation:[19]

$$\mathbb{w}^v(x,p,t) = \int \mathrm{d}y\, e^{-ipy} \tilde{\mathbb{w}}^v(x,y,t) \ . \tag{5.80}$$

In Fig. 5.1, the time evolution of the momentum space marginal distributions is shown. These observables show several features which are partly known from the investigation of the Schwinger effect in spatially homogeneous, time-dependent electric fields $E(t)$:

First of all, the pair creation process shows a non-trivial momentum dependence. Most notably, field excitations are created with momenta around $p = 0$ and are then accelerated by the electric field to higher momenta. It has again to be emphasized that the interpretation of the particle number density $n(p,t)$ as momentum distribution of real particles is only possible at asymptotic times $t \to \infty$. At intermediate times, on the other hand, it can only be considered as mixture between real and virtual excitations.

Moreover, the particle number density $n(p,t)$ takes a fixed value once the electric field vanishes at asymptotic times $t \to \infty$ since there are no other forces acting than the electric.[20] Note that this behavior would change as soon as the self-induced electric field due to the pair creation process is taken into account.

Additionally, one would naively expect that particles with charge $e$ are accelerated to positive momenta whereas particles with charge $-e$ are accelerated to negative momenta. It is a peculiarity of the equal-time Wigner formalism, however, that both types of particles are accelerated to positive momenta. Consequently, a particle with charge $-e$ and momentum $p$ has to be interpreted as an antiparticle with *physical momentum* $-p$.[21] This is further discussed when considering the time evolution of the real space marginal distributions in a moment.

Finally, the charge density $q(p,t)$ vanishes identically at all times. This can in fact be traced back to the choice of a symmetric spatial dependence $g(x) = g(-x)$.

---

[18] It would also be possible to investigate the time evolution of the equal-time Wigner components. It is, however, much more enlightening to consider observable quantities instead of auxiliary ones.

[19] As the modified equal-time Wigner components are calculated on a finite grid in both $x$- and $y$-directions, one has to approximate the inverse Fourier transformation by a finite sum.

[20] Note that $t = 3\tau$ is already very close to $t \to \infty$ for the temporal dependence $h(t) = \mathrm{sech}^2\left(\frac{t}{\tau}\right)$.

[21] This peculiarity is well known from relativistic quantum mechanics, where a negative energy solution with momentum $p$ is considered as an antiparticle with physical momentum $-p$ as well.

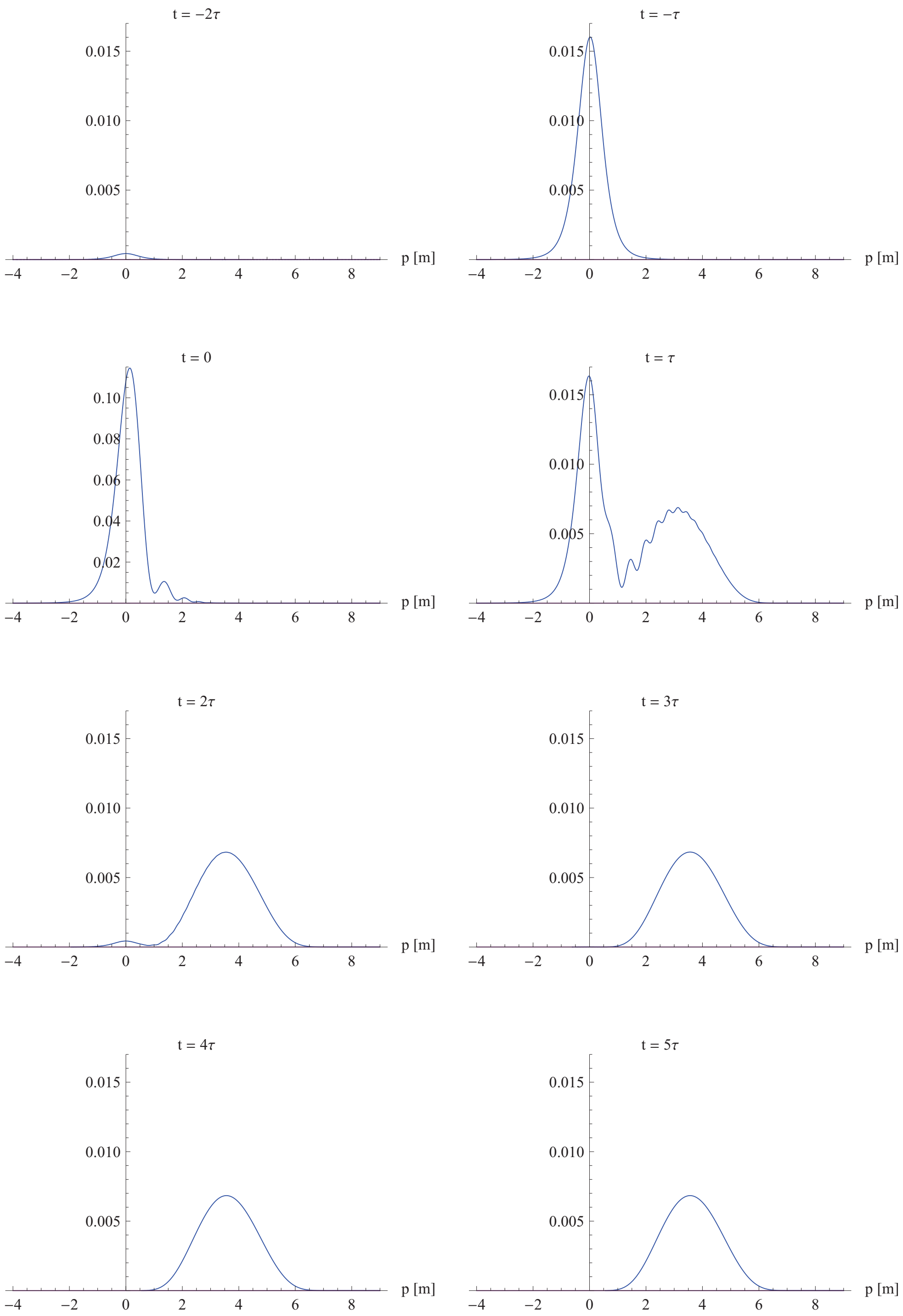

Figure 5.1: Time evolution of the momentum space particle number density $n(p,t)$ (blue) and charge density $q(p,t)$ (purple). Note that another scale is used at $t = 0$. All other parameters are given in the text.

In Fig. 5.2, the time evolution of the real space marginal distributions is shown. Along with the discussion of the momentum space marginal distributions, a coherent picture of the Schwinger effect in space- and time-dependent electric fields $E(x, t)$ can be drawn:

First of all, the investigation of the real space marginal distribution at early times $t < 0$ shows that the pair creation process takes only place in the space region around $x = 0$ where the electric field acts. It has to be emphasized that the spatial extent is actually determined by the spatial variation scale $\lambda$.[22]

Moreover, the particle number density $n(x, t)$ is peaked around $x = 0$ whereas the charge density $q(x, t)$ vanishes there at early times $t < 0$. This can be interpreted as local charge neutrality in the center of the pulse whereas one observes charge separation at the edges of the pulse. The effect of charge separation is in fact driven by the electric field which accelerates excitations with charge $e$ into the positive $x$-direction whereas excitations with charge $-e$ are accelerated into the opposite direction.

The effect of charge separation can be seen even better at later times $t > 0$. Owing to the acceleration in the electric field, one bunch of excitations with charge $e$ propagates into the positive $x$-direction whereas another bunch of excitations with charge $-e$ propagates into the opposite direction. These bunches can in fact be identified with particles and antiparticles, respectively, once the electric field vanishes either at asymptotic times $t \to \infty$ or beyond the region where the electric field acts.

It has been mentioned when discussing the momentum space marginal distributions that both excitations with charge $e$ and excitations with charge $-e$ are accelerated to positive momenta against naive expectations. Nonetheless, they actually move apart from each other in opposite directions even though both types of excitations do have positive momenta. This again motivates why a particle with charge $-e$ and momentum $p$ has rather to be interpreted as an antiparticle with *physical momentum* $-p$.

Finally, the particle number density $n(x, t)$ is symmetric with respect to reflections at $x = 0$ whereas the charge density $q(x, t)$ is antisymmetric. This symmetry properties of the real space marginal distributions can again be traced back to the choice of a symmetric spatial dependence $g(x) = g(-x)$.

[22] For the spatial dependence $g(x) = \exp(-\frac{x^2}{2\lambda^2})$, the electric field has dropped to approximately 1% of its maximum value at $|x| \sim 3\lambda$.

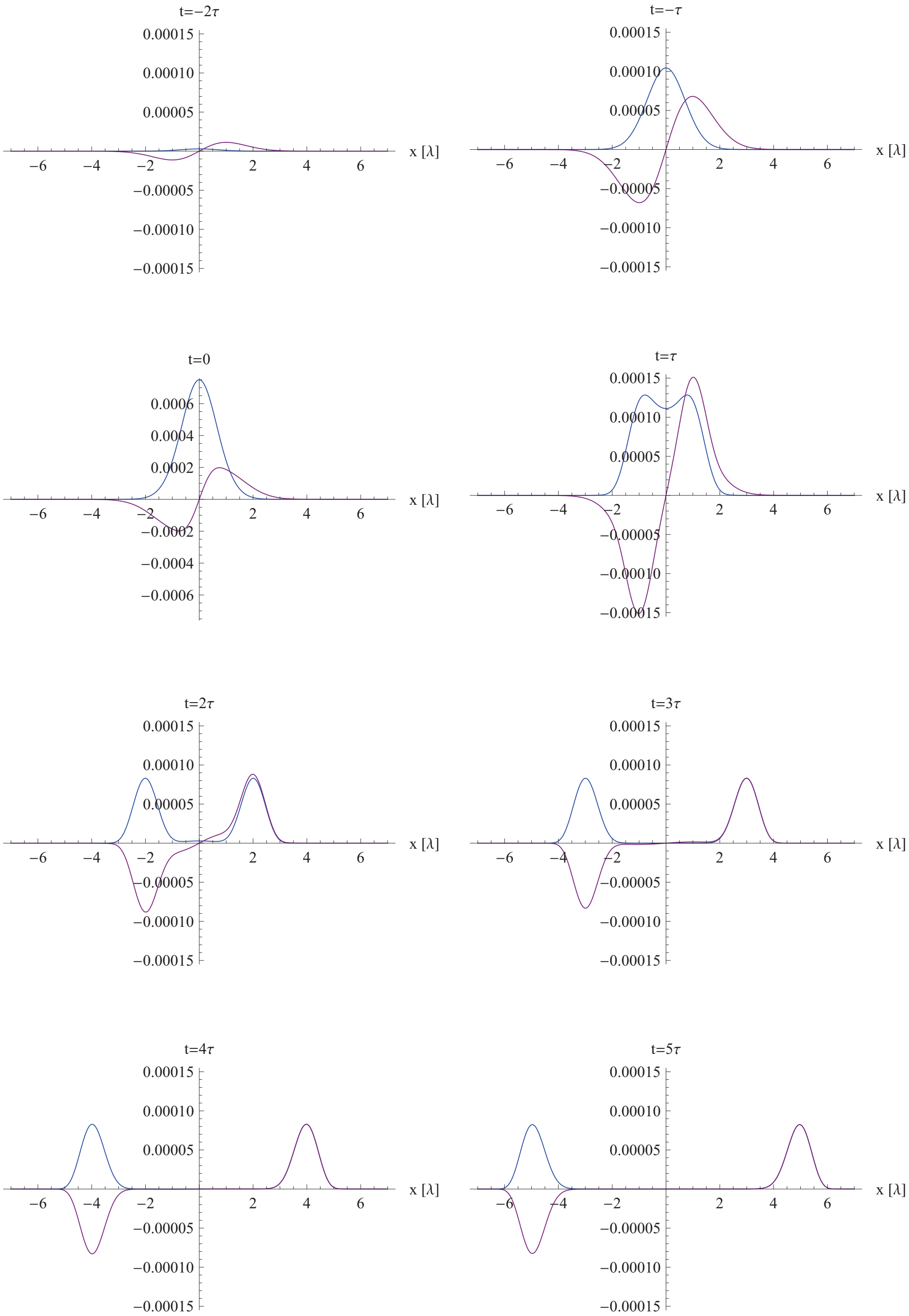

Figure 5.2: Time evolution of the real space particle number density $n(x,t)$ (blue) and charge density $q(x,t)$ (purple). Note that another scale is used at $t = 0$. All other parameters are given in the text.

### 5.3.2 Particle number density

The time evolution of the marginal distributions has been considered in order to gain insight into the dynamics of the pair creation process in the presence of a space- and time-dependent electric field $E(x,t)$. The investigation of the real space marginal distributions made in fact plain that bunches of particles and antiparticles move apart from each other in opposite directions.

Note, however, that the dynamics of this motion is in fact governed by the momentum of the particles which is encoded in the momentum space marginal distributions. Due to the fact that the momentum space charge density $q(p,t)$ vanishes identically at all times, it is indicated to focus on the *momentum space particle number density* $n(p,t)$ in the following.

#### 5.3.2.1 Dependence on $\lambda$

This part is dedicated to the investigation of the dependence of the particle number density $n(p,t)$ on the spatial variation scale $\lambda$. To this end, the time evolution of $n(p,t)$ for different values of $\lambda$ is investigated. Again, these results are obtained by numerically solving the PDE system Eq. (5.69) – (5.72) in $y$-space and performing an inverse Fourier transformation afterwards. For the sake of better comparability, it is in fact more convenient to investigate the *reduced particle number density* $\bar{n}(p,t)$ instead of $n(p,t)$:[23]

$$\bar{n}(p,t) \equiv \frac{n(p,t)}{\lambda} \ . \tag{5.81}$$

In Fig. 5.3, the time evolution of $\bar{n}(p,t)$ is displayed for different values of $\lambda$, whereas the asymptotic value $\bar{n}(p,\infty)$ is shown in more detail in Fig. 5.4. Disregarding the trivial scaling effect it turns out that $\bar{n}(p,t)$ shows various remarkable features as a function of $\lambda$:

First of all, the behavior of $\bar{n}(p,t)$ for different values of $\lambda$ is rather similar at early times $t<0$, at least for $\lambda \gtrsim 5\lambda_C$: Field excitations are created with momenta around $p=0$ and are then accelerated by the electric field to higher momenta. Depending on the actual value of $\lambda$, however, the shape of the asymptotic reduced particle number density $\bar{n}(p,\infty)$ changes in various respects.

---

[23]Decreasing the spatial variation scale $\lambda$ results in a diminishment of the total energy of the electromagnetic field $\mathcal{E}_{em}(t)$ as well:

$$\frac{\mathcal{E}_{\rm em}(t;\lambda_1)}{\mathcal{E}_{\rm em}(t;\lambda_2)} = \frac{\lambda_1}{\lambda_2} \quad \text{with} \quad \mathcal{E}_{\rm em}(t) = \frac{1}{2}\int {\rm d}x \, |E(x,t)|^2 \ .$$

Accordingly, there is less energy available and fewer particle-antiparticle pairs are created. This trivial effect of the decline of $n(p,t)$ with decreasing $\lambda$ is in fact accounted for by considering $\bar{n}(p,t)$ instead of $n(p,t)$.

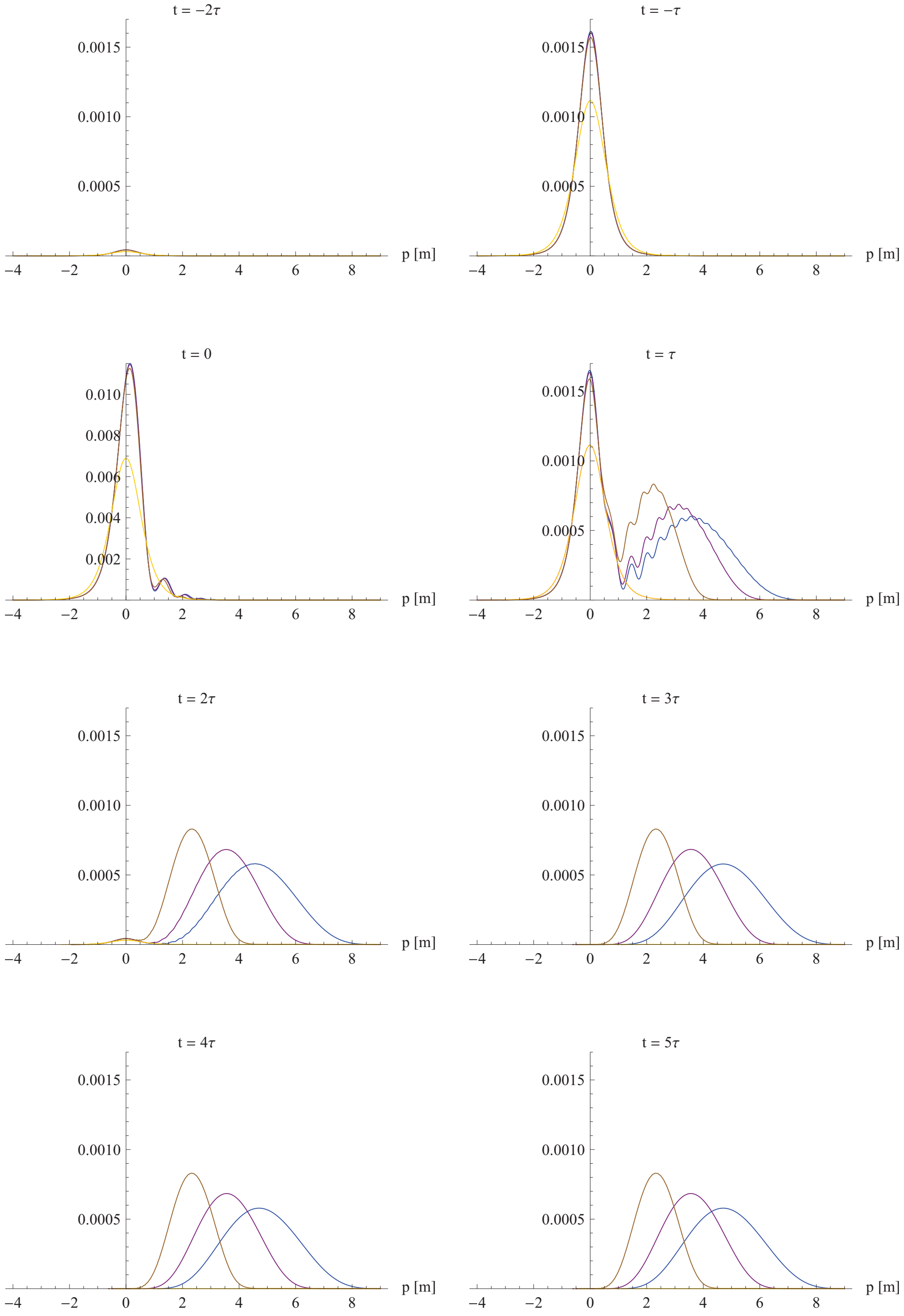

Figure 5.3: Time evolution of the reduced particle number density $\bar{n}(p,t)$ for various values of $\lambda$: $100\lambda_C$ (blue), $10\lambda_C$ (purple), $5\lambda_C$ (brown), $\lambda_C$ (yellow). Note that another scale is used at $t = 0$. All other parameters are given in the text.

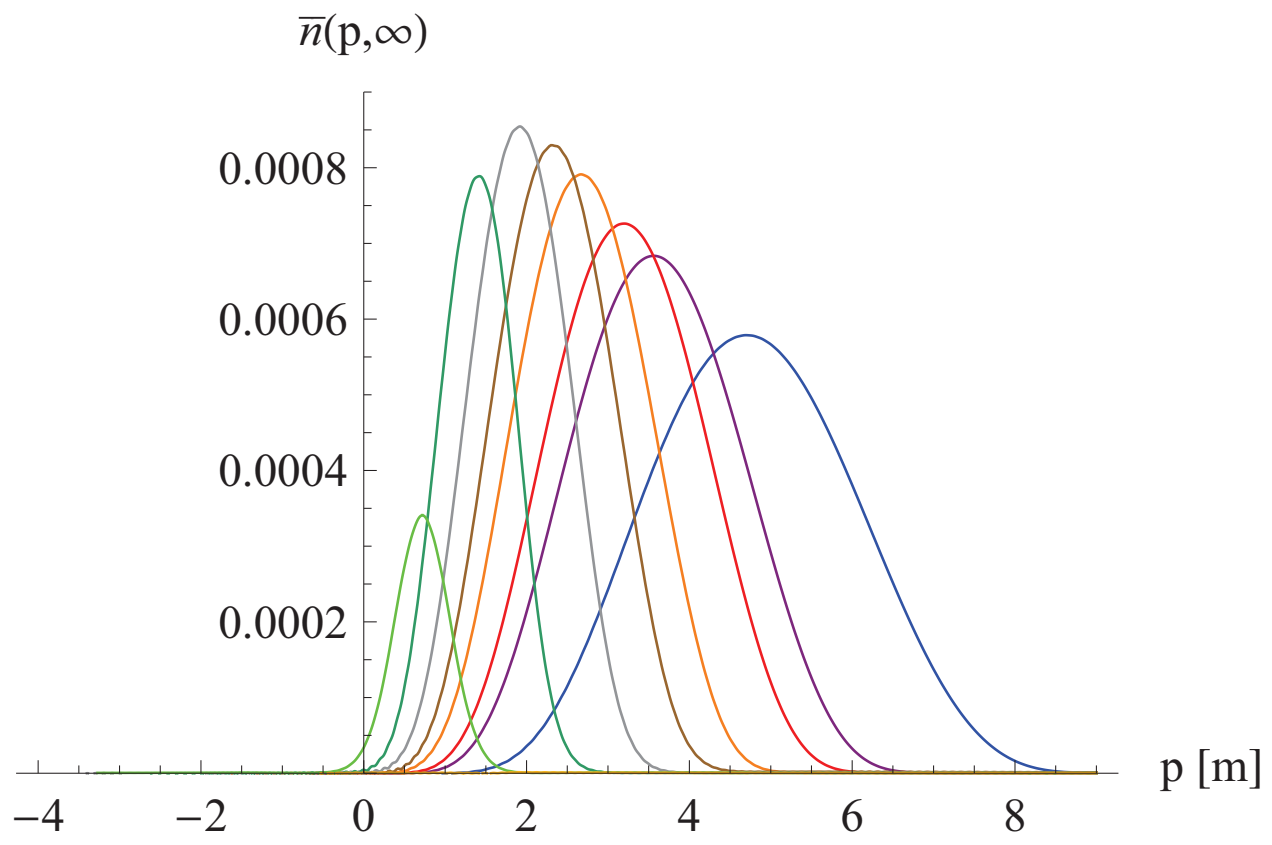

Figure 5.4: Asymptotic reduced particle number density $\bar{n}(p,\infty)$ for various values of $\lambda$: $100\lambda_C$ (blue), $10\lambda_C$ (purple), $8\lambda_C$ (red), $6\lambda_C$ (orange), $5\lambda_C$ (brown), $4\lambda_C$ (gray), $3\lambda_C$ (dark green), $2\lambda_C$ (light green), $\lambda_C$ (yellow). All other parameters are given in the text.

Most notably, a decreasing value of $\lambda$ involves a shift of the peak momentum $p_\lambda^{pk}$ to smaller values:

$$p_{\lambda_1}^{pk} < p_{\lambda_2}^{pk} \qquad \text{for} \qquad \lambda_1 < \lambda_2 \, . \tag{5.82}$$

This behavior is understood in the following way: The field excitations are accelerated in such a way that the peak momentum takes a certain value $p_\infty^{pk}$ in the case of a spatially homogeneous, time-dependent electric field $E(t)$. For a space- and time-dependent electric field $E(x,t) = E(t)g(x)$, however, the value of the acceleration by the electric field depends on the actual position $x$. Most notably, the field excitations are less accelerated compared to the spatially homogeneous case as $|g(x)| \leq 1$. This finally results in a shift of the peak momentum $p_\lambda^{pk}$ to smaller values for decreasing values of $\lambda$.

Moreover, the shape of $\bar{n}(p,\infty)$ becomes higher and narrower for decreasing values of $\lambda$, at least for $\lambda \gtrsim 4\lambda_C$. This is kind of a self-focussing effect which is caused by the spatial inhomogeneity. Excitations which are already created with high momenta are accelerated for a shorter period compared to excitations which are created with small momenta, which are then accelerated for a longer period. Consequently, the created particles are bunched into a smaller phase space volume.

This behavior changes again for $\lambda \lesssim 4\lambda_C$ when $\bar{n}(p,\infty)$ becomes in fact even narrower, however, the height of the peak momentum $p_\lambda^{pk}$ decreases again. This is a direct consequence of the decreasing values of $\lambda$ as more and more excitations gain too less energy in order to finally turn into real particles. This behavior is most clearly seen for $\lambda = \lambda_C$ when the energy content of the electric field is too small so that none of the field excitations turns into real particles at the end.

#### 5.3.2.2 Full solution vs. approximate solutions

As it has become clear that $\bar{n}(p,t)$ shows a strong dependence on the spatial variation scale $\lambda$ especially for small values of $\lambda$, it is about time to compare the full solution with the approximate ones. This comparison is actually done for two reasons:

On the one hand, the comparison of $\bar{n}_{[y]}(p,t)$, which is obtained by solving the PDE system Eq. (5.69) – (5.72) in $y$-space, with $\bar{n}_{\text{loc}}(p,t)$ in local density approximation is suggested from a physical point of view: A deviation of $\bar{n}_{[y]}(p,t)$ from $\bar{n}_{\text{loc}}(p,t)$ indicates a change in the pair creation behavior in the sense that the pair creation process at any point $x_i$ cannot be considered as taking place in a spatially homogeneous, time-dependent electric field $E(t)g(x_i)$ anymore.

On the other hand, the comparison of $\bar{n}_{[y]}(p,t)$ with $\bar{n}_{[p]}(p,t)$, which is obtained by solving the PDE system Eq. (5.57) – (5.60) originating from a leading order derivative expansion, is rather important from a computational point of view as the numerical solution in $y$-space is much more expensive than the numerical solution in $p$-space.

In Fig. 5.5, the full solution $\bar{n}_{[y]}(p,\infty)$ is compared with the leading order derivative expansion result $\bar{n}_{[p]}(p,\infty)$ as well as with the local density approximation result $\bar{n}_{\text{loc}}(p,\infty)$ for different values of $\lambda$:

Most notably, the difference between the various results is rather minor for $\lambda \gtrsim 100\lambda_C$. Accordingly, the pair creation process at any point $x_i$ can indeed be considered as taking place in a spatially homogeneous, time-dependent electric field $E(t)g(x_i)$ for large values of $\lambda$.

For decreasing values of $\lambda$, however, the different results differ from each other: Most notably, the leading order derivative expansion result $\bar{n}_{[p]}(p,\infty)$ shows truncation artefacts for momenta around $p=0$ compared to the full solution $\bar{n}_{[y]}(p,\infty)$. Additionally, a shift of the peak momentum $p_\lambda^{pk} < p_{\text{loc}}^{pk}$ occurs as discussed in detail previously.

As expected, the leading order derivative expansion result $\bar{n}_{[p]}(p,\infty)$ becomes worse for decreasing values of $\lambda$. However, it has not been anticipated that the leading order derivative expansion would fail for momenta around $p=0$ as a previous study had estimated that it would rather fail for large momenta [33].[24] The leading order derivative expansion eventually breaks down once the truncation artefacts become even larger than the reduced particle number density itself.

[24] In fact, a different type of electric field had been considered in this investigation:

$$E(x,t) = E_0 \cos(\tfrac{x}{\lambda}) \operatorname{sech}^2(\tfrac{t}{\tau}) \ .$$

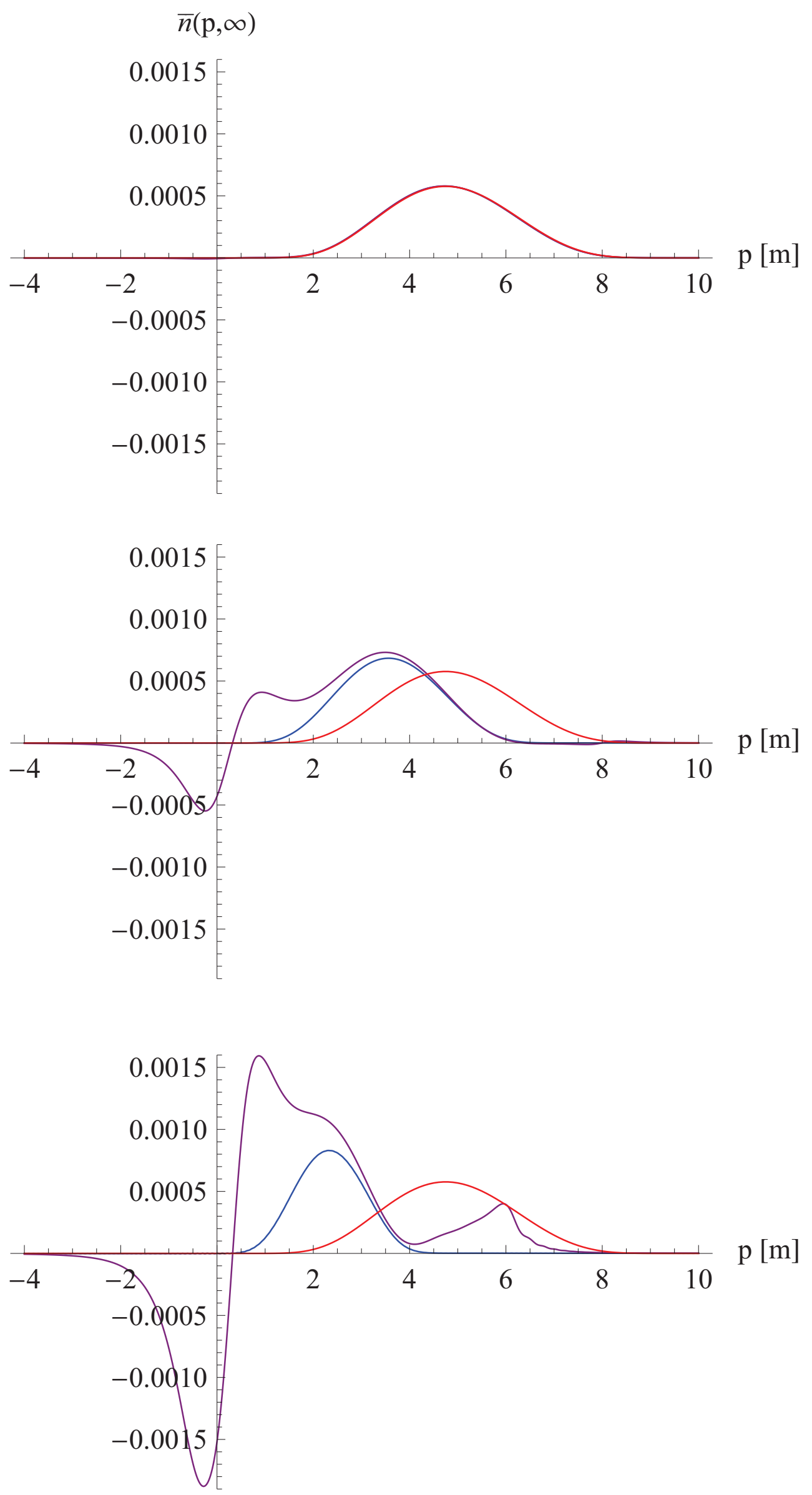

Figure 5.5: Comparison of $\bar{n}_{[\mathrm{y}]}(p,\infty)$ (blue) with $\bar{n}_{[\mathrm{p}]}(p,\infty)$ (purple) as well as $\bar{n}_{\mathrm{loc}}(p,\infty)$ (red). *Top:* $\lambda = 100\lambda_C$. *Middle:* $\lambda = 10\lambda_C$. *Bottom:* $\lambda = 5\lambda_C$. All other parameters are given in the text.

#### 5.3.2.3 Asymmetry of $n_{[y]}(p,\infty)$

Even though $\bar{n}_{[y]}(p,\infty)$ seems to be symmetric around the peak momentum $p_\lambda^{pk}$, it turns out that there is in fact a small asymmetry which is displayed in Fig. 5.6 for two different values of $\lambda$:

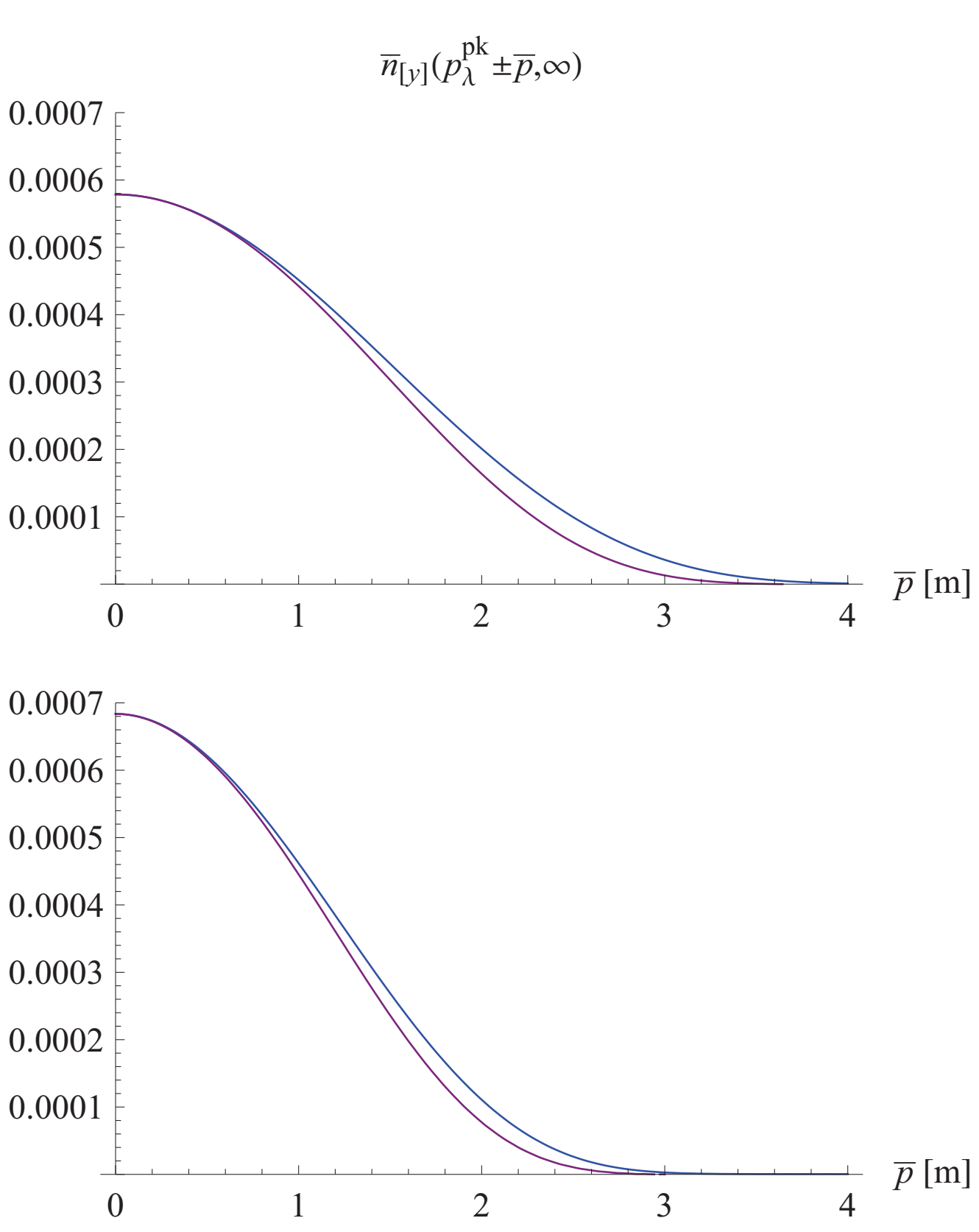

Figure 5.6: Asymmetry of the reduced particle number density around the peak momentum $p_\lambda^{pk}$ with $\bar{n}_{[y]}(p_\lambda^{pk} + \bar{p}, \infty)$ (blue) and $\bar{n}_{[y]}(p_\lambda^{pk} - \bar{p}, \infty)$ (purple). *Top:* $\lambda = 100\lambda_C$. *Bottom:* $\lambda = 10\lambda_C$. All other parameters are given in the text.

First of all, this asymmetry seems to be a peculiarity of $\bar{n}_{[y]}(p, \infty)$ as it appears for all spatial variation scales $\lambda$ in such a way that:

$$\bar{n}_{[y]}(p_\lambda^{pk} + \bar{p}, \infty) \geq \bar{n}_{[y]}(p_\lambda^{pk} - \bar{p}, \infty) \qquad \text{with} \qquad \bar{p} > 0 \ . \tag{5.83}$$

Additionally, one nicely observes again that the main peak becomes higher and narrower for decreasing values of $\lambda$. This has already been discussed in detail previously.

In order to better understand the origin of this asymmetry, it is convenient to consider the local density approximation for a moment. Due to the fact that $\bar{n}_{[y]}(p, \infty)$ and $\bar{n}_{\rm loc}(p, \infty)$ nearly coincide for $\lambda \gtrsim 100\lambda_C$, it should be possible to give an explanation already by considering the analytic expression for the asymptotic

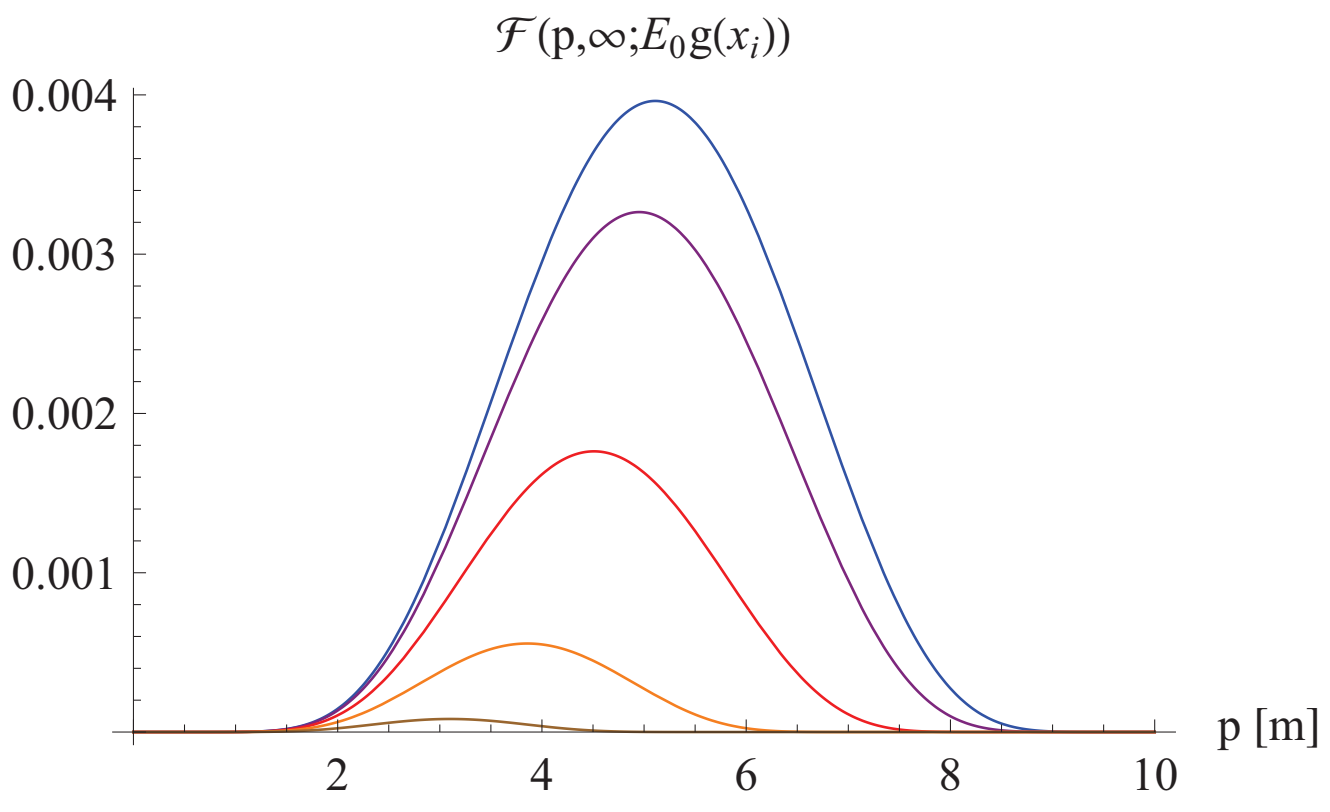

Figure 5.7: Asymptotic one-particle distribution function $\mathcal{F}(p,\infty;E_0 g(x_i))$ for $x_i = 0$ (blue), $x_i = \frac{\lambda}{4}$ (purple), $x_i = \frac{\lambda}{2}$ (red), $x_i = \frac{3\lambda}{4}$ (orange) and $x_i = \lambda$ (brown).

one-particle distribution function Eq. (4.145):[25]

$$\mathcal{F}(p,\infty;E_0) = \frac{\sinh\left(\frac{\pi\tau}{2}\left[2eE_0\tau+\omega_0(p)-\omega_1(p)\right]\right)\sinh\left(\frac{\pi\tau}{2}\left[2eE_0\tau-\omega_0(p)+\omega_1(p)\right]\right)}{\sinh\left(\pi\tau\omega_0(p)\right)\sinh\left(\pi\tau\omega_1(p)\right)}\ , \tag{5.84}$$

with

$$\omega_0(p) = \sqrt{m^2+[p-2eE_0\tau]^2} \qquad \text{and} \qquad \omega_1(p) = \sqrt{m^2+p^2}\ . \tag{5.85}$$

Accordingly, the local density approximation result $\bar{n}_{\mathrm{loc}}(p,\infty)$ might be approximated by a Riemann sum:

$$\bar{n}_{\mathrm{loc}}(p,\infty) \sim \frac{2}{\lambda}\frac{\Delta x}{(2\pi)}\left[\mathcal{F}(p,\infty;E_0) + 2\sum_{i=1}^{N}\mathcal{F}(p,\infty;E_0 g(i\Delta x))\right]. \tag{5.86}$$

Note that each individual contribution $\mathcal{F}(p,\infty;E_0 g(x_i))$ is in fact symmetric around its corresponding peak momentum:

$$p^{pk}_{\mathcal{F}}(x_i) = eE_0 g(x_i)\tau\ . \tag{5.87}$$

Moreover, the peak momentum $p^{pk}_{\mathcal{F}}(x_i)$ of each individual contribution is shifted to a smaller value for increasing values of $x_i$ as displayed in Fig. 5.7:

$$p^{pk}_{\mathcal{F}}(x_1) < p^{pk}_{\mathcal{F}}(x_2) \qquad \text{for} \qquad x_1 > x_2\ . \tag{5.88}$$

[25] Note that the asymptotic one-partice distribution function is now expressed in terms of the phase-space kinetic momentum $p$. Additionally, the dependence on $E_0$ has been indicated explicitly. Moreover, the factor 2 has been dropped since fermions are spinless particles in $1+1$ dimensions.

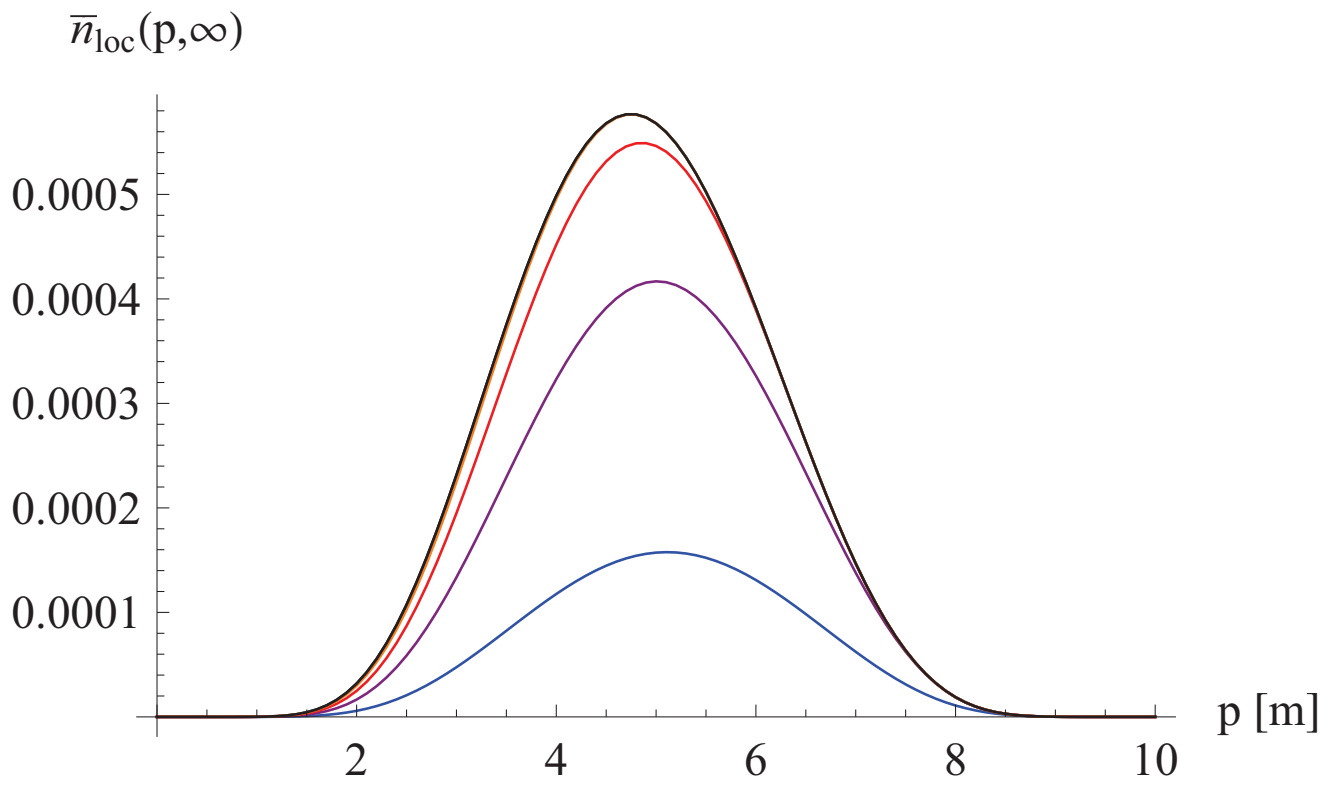

Figure 5.8: Summing up the individual contributions $\mathcal{F}(p,\infty;E_0 g(x_i))$ of the Riemann sum with $\Delta x = \frac{\lambda}{4}$ and $N = 0$ (blue), $N = 1$ (purple), $N = 2$ (red), $N = 3$ (orange) finally gives $\bar{n}_{\text{loc}}(p,\infty)$ (black).

Performing the Riemann sum as displayed in Fig. 5.8, the peak momentum $p_{\text{loc}}^{pk}$ is shifted to a smaller value compared to the peak momentum $p_{\mathcal{F}}^{pk}(0)$ of the one-particle distribution function $\mathcal{F}(p,\infty;E_0)$:

$$p_{\text{loc}}^{pk} < p_{\mathcal{F}}^{pk}(0) \; . \tag{5.89}$$

Accordingly, $\bar{n}_{\text{loc}}(p,\infty)$ becomes asymmetric around $p_{\text{loc}}^{pk}$ as the individual contributions $\mathcal{F}(p,\infty;E_0 g(x_i))$ are in fact asymmetric around $p_{\text{loc}}^{pk}$ as well.

### 5.3.3 Total number of created particles

I finally turn to the investigation of the number of created particles $\mathcal{N}(t)$. Actually, as the particle interpretation of the field excitations is only possible at asymptotic times $t \to \infty$, I restrict myself to the asymptotic number of created particles $\mathcal{N}(\infty)$.

In Fig. 5.9, the full solution $\mathcal{N}_{[\text{y}]}(\infty)$ is compared with the leading order derivative expansion result $\mathcal{N}_{[\text{p}]}(\infty)$ as well as with the local density approximation result $\mathcal{N}_{\text{loc}}(\infty)$ for different values of $\lambda$. Note that it is again more convenient to consider the *reduced number of created particles*:

$$\overline{\mathcal{N}}(\infty) \equiv \frac{\mathcal{N}(\infty)}{\lambda} \; , \tag{5.90}$$

so that the trivial scaling effect with respect to $\lambda$ is disregarded. These results perfectly fit into the picture of the Schwinger effect in space- and time-dependent electric fields $E(x,t)$ as concluded from the analysis of the marginal distributions:

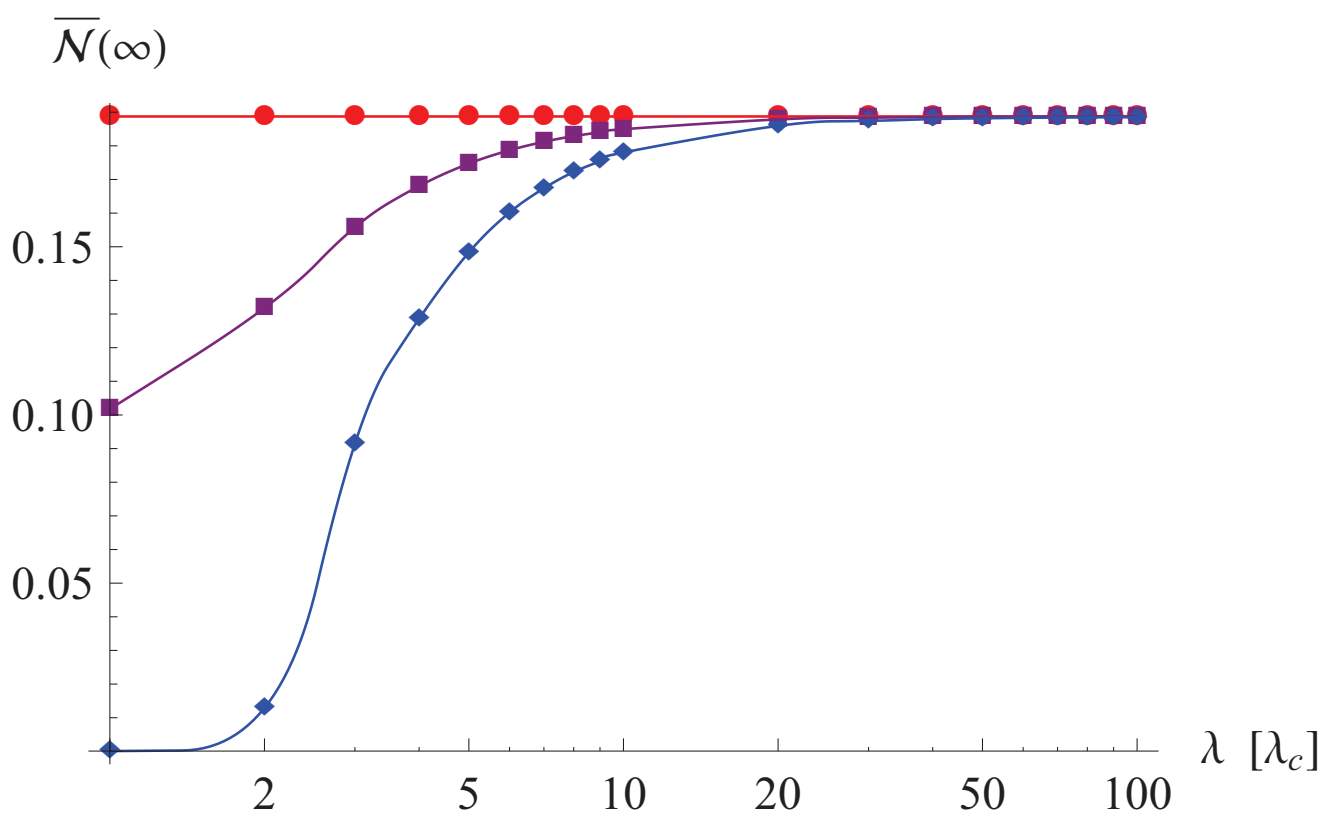

Figure 5.9: Comparison of $\overline{\mathcal{N}}_{[y]}(\infty)$ (blue) with $\overline{\mathcal{N}}_{[p]}(\infty)$ (purple) and $\overline{\mathcal{N}}_{\rm loc}(\infty)$ (red) for different values of $\lambda$ in a lin-log plot. All other parameters are given in the text.

First of all, $\mathcal{N}_{\rm loc}(\infty)$ shows an exact scaling behavior as function of $\lambda$. This exact scaling exhibits itself as a straight line in the lin-log plot. The outcome of either method is then nearly the same for large values of $\lambda$. This good agreement between the full solution with the approximate ones for large values of $\lambda$ has already been pointed out previously when discussing the results for the reduced particle number density $\bar{n}(p,\infty)$ within the various schemes.

For smaller values of $\lambda$, however, the various solutions differ significantly. Most notably, both the full solution $\overline{\mathcal{N}}_{[y]}(\infty)$ and the leading order derivative expansion result $\overline{\mathcal{N}}_{[p]}(\infty)$ do not show the exact scaling behavior of $\overline{\mathcal{N}}_{\rm loc}(\infty)$. Moreover, both of them predict a decrease in the total number of created particles compared to the local density approximation result. This behavior can be traced back to the fact that for a decreasing value of $\lambda$ more and more excitations gain too less energy in order to finally turn into real particles.

Nonetheless, there is still a difference between the full solution $\overline{\mathcal{N}}_{[y]}(\infty)$ and the leading order derivative expansion result $\overline{\mathcal{N}}_{[p]}(\infty)$ for small values of $\lambda$: Both of them show a significant decrease compared to $\overline{\mathcal{N}}_{\rm loc}(\infty)$, however, $\overline{\mathcal{N}}_{[y]}(\infty)$ actually drops to zero for $\lambda \sim \lambda_C$ whereas $\overline{\mathcal{N}}_{[p]}(\infty)$ does not. Actually, this is no surprise as it has already been shown that the leading order derivative expansion fails for small values of $\lambda$.

The sharp drop of $\overline{\mathcal{N}}_{[y]}(\infty)$ for small values of $\lambda$ is in fact in good agreement with previous studies of the Schwinger effect in space-dependent electric fields $E(x)$ [22, 23, 109]. It has already been mentioned that the pair creation process terminates once the work done by the electric field over its spatial extent is smaller than twice

the electron mass:

$$e \int E(x)\mathrm{d}x < 2m \ . \tag{5.91}$$

Specifically, for the space-dependent electric field:

$$E(x) = E_0 \exp\left(-\frac{x^2}{2\lambda^2}\right) \ , \tag{5.92}$$

this condition reads:

$$\sqrt{2\pi} e E_0 \lambda < 2m \ . \tag{5.93}$$

Accordingly, considering this space-dependent electric field $E(x)$ for a field strength parameter $\epsilon = 0.5$, the pair creation process terminates for spatial variation scales:

$$\lambda < \frac{1}{\epsilon}\sqrt{\frac{2}{\pi}}\lambda_C \sim 1.595\lambda_C \ . \tag{5.94}$$

This estimate is in fact in good agreement with the numerical result for $\overline{\mathcal{N}}_{[\mathrm{y}]}(\infty)$ in the presence of a space- and time-dependent electric field $E(x,t)$ as well.

Chapter 6

# Conclusions and outlook

The main issue of this thesis was to investigate various aspects of the Schwinger effect in inhomogeneous electric fields. As a matter of fact, vacuum pair creation in the presence of simple field configurations such as static, sinusoidal or pulsed electric fields has been investigated since the early days of quantum mechanics, however, it was not until now that it becomes possible to study the Schwinger effect in realistic electric fields showing both temporal and spatial variations. This recent progress is due to a better theoretical understanding of the mechanism behind the pair creation process as well as owing to the rapid development of computer technology which makes extensive numerical simulations feasible today. These advances in the theoretical description of the Schwinger effect are in fact urgently needed as a new generation of high-intensity laser systems such as the European XFEL or the Extreme Light Infrastructure are already in the starting blocks to shed some light on the strong-field regime of QED soon.

The first part of this thesis was dedicated to the theoretical description of the Schwinger effect by means of quantum kinetic methods and briefly reviewed the Wigner formalism as well as the quantum Vlasov equation. In Section 4.2 it was eventually shown that the quantum Vlasov equation in its nowadays widely used form is in fact equivalent to the equal-time Wigner formalism in the presence of a spatially homogeneous, time-dependent electric field. This is quite remarkable as the quantum Vlasov equation is derived from canonical quantization whereas the equal-time Wigner formalism is formulated in phase space. As a consequence, it was even possible to calculate analytic expressions for the one-particle distribution function as well as for the equal-time Wigner components in the presence of both a static and a pulsed electric field in Section 4.3.

It was a shortcoming of previous studies that the Schwinger effect has only been considered for simple time-dependent electric fields. This gap has been partly closed by the investigation of the pair creation process in the presence of a pulsed electric field with sub-cycle structure in Section 4.4. The corresponding results indicate that the momentum distribution of created particles is extremely sensitive to the various laser parameters, resulting in a number of new observables such as the width of the momentum distribution or its oscillatory structure. As these characteristics might become crucial for the observation of the Schwinger effect in upcoming high-intensity laser experiments, future investigations should focus on the issue of pulse shaping in order to maximize the particle yield and provide clear signatures.

A second drawback of previous investigations was that there have not been any rigorous studies of the Schwinger effect in the presence of space- and time-dependent electric fields. In this respect, Section 5 provides substantial progress: Based on the equal-time Wigner formalism in $\text{QED}_{1+1}$, an ab initio simulation of the Schwinger effect in such a type of electric field has been conducted for the first time, allowing for the calculation of the time evolution of various observable quantities such as the particle number density, the charge density or the number of created particles. These results provide in fact deeper insight into the pair creation process, however, they also raise further questions:

The comparison between the full solution with a local density approximation showed that the momentum distribution strongly depends on the spatial variation scale $\lambda$. Moreover, it turned out that the momentum distribution of particles is identical to that of antiparticles owing to the choice of a symmetric spatial variation $g(x) = g(-x)$. In contrast to that, particles and antiparticles should behave differently in the presence of an electric field which does not show this symmetry. It is, however, not clear a priori which consequences the choice of such an electric field would have. Accordingly, it would surely be worthwile to consider this question in future investigations as well.

The comparison between the full solution with an approximate solution, which has been based upon a leading-order derivative expansion of the occurring pseudo-differential operator, was another issue of investigation. This study in fact showed that the approximate solution deviates significantly from the full solution already at comparatively large spatial variation scales $\lambda$. Quite surprisingly, this deviation occurs at small kinetic momentum values and leads to a break down of the leading-order derivative expansion for spatial variation scales $\lambda$ which are still somewhat beyond the Compton wavelength $\lambda_C$.

Even though this first investigation of the Schwinger effect in a space- and time-dependent electric field in $\text{QED}_{1+1}$ represents substantial progress, one has to admit that this study has to be viewed only as the first step towards a full description of the pair creation process in upcoming high-intensity laser experiments. As a matter of fact, there are still various issues which should be tackled in future investigations:

First of all, the space- and time-dependent electric field which has been under consideration cannot be considered as realistic in the sense of representing a field configuration to be realized at high-intensity laser facilities. As a next step one could in fact try to consider a more realistic time dependence such as a pulsed electric field with sub-cycle structure again.

Secondly, the numerical simulation which has been performed was based on the equal-time Wigner formalism in $\mathrm{QED}_{1+1}$. Accordingly, the notion of a magnetic field did not even exist and Dirac fermions had to be considered as spinless particles. In order to avoid these shortcomings one should rather consider the equal-time Wigner formalism in $\mathrm{QED}_{3+1}$ in future investigations. However, as it does not seem feasible to solve the equations of motion for the equal-time Wigner components for an arbitrary field configuration for computational reasons in the near future, one should restrict oneself to highly symmetric configurations for now.

Finally, as soon as it becomes doable to perform numerical simulations of the Schwinger effect in the presence of space- and time-dependent electromagnetic fields in 3+1 dimensions, one should also reconsider the backreaction issue. As a start one could in fact be satisfied with a mean field description, however, the ultimate goal should be to consistently describe the pair creation process beyond the mean field level by taking into account photon corrections to the background electromagnetic field.

Appendix A

# Specific calculations

This appendix is devoted to giving the details of some calculations.

## A.1 Covariant Wigner formalism: Equation of motion

The starting point for the derivation of the equation of motion for the covariant Wigner operator $\hat{\mathcal{W}}(\mathsf{x},\mathsf{p})$ is the underlying gauge-invariant density matrix $\hat{\mathcal{C}}(\mathsf{A};\mathsf{x},\mathsf{y})$:

$$\hat{\mathcal{C}}(\mathsf{A};\mathsf{x},\mathsf{y}) = \mathcal{U}(\mathsf{A};\mathsf{x},\mathsf{y})\left[\bar{\Psi}(\mathsf{x}_1 = \mathsf{x} - \tfrac{\mathsf{y}}{2}), \Psi(\mathsf{x}_2 = \mathsf{x} + \tfrac{\mathsf{y}}{2})\right] \ . \tag{A.1}$$

Note that the derivatives with respect to $\mathsf{x}_1$ and $\mathsf{x}_2$, respectively, can be expressed as:

$$\partial_\mu^{\mathsf{x}_1} = \tfrac{1}{2}\partial_\mu^{\mathsf{x}} - \partial_\mu^{\mathsf{y}} \quad \text{and} \quad \partial_\mu^{\mathsf{x}_2} = \tfrac{1}{2}\partial_\mu^{\mathsf{x}} + \partial_\mu^{\mathsf{y}} \ . \tag{A.2}$$

Taking the derivative of $\hat{\mathcal{C}}(\mathsf{A};\mathsf{x},\mathsf{y})$ with respect to $\mathsf{x}_2$ will eventually result in the equation of motion for the Wigner operator Eq. (3.10). Note that taking the derivative with respect to $\mathsf{x}_1$ instead would result in the adjoint equation of motion Eq. (3.11):

$$\gamma^\mu \partial_\mu^{\mathsf{x}_2}\hat{\mathcal{C}}(\mathsf{A};\mathsf{x},\mathsf{y}) = \overbrace{\partial_\mu^{\mathsf{x}_2}\mathcal{U}(\mathsf{A};\mathsf{x},\mathsf{y})}^{[\mathrm{I}]}\gamma^\mu\left[\bar{\Psi}(\mathsf{x}_1), \Psi(\mathsf{x}_2)\right] + \mathcal{U}(\mathsf{A};\mathsf{x},\mathsf{y})\overbrace{\gamma^\mu\left[\bar{\Psi}(\mathsf{x}_1), \partial_\mu^{\mathsf{x}_2}\Psi(\mathsf{x}_2)\right]}^{[\mathrm{II}]} \ . \tag{A.3}$$

In order to calculate part [ I ], one uses Eq. (A.2) and performs the derivatives with respect to $\mathsf{x}$ and $\mathsf{y}$, respectively:

$$\partial_\mu^{\mathsf{x}_2}\mathcal{U}(\mathsf{A};\mathsf{x},\mathsf{y}) = ie\int_{-\frac{1}{2}}^{\frac{1}{2}} \mathrm{d}\xi\,\left\{(\tfrac{1}{2}+\xi)y^\nu\partial_\mu A_\nu(\mathsf{x}+\xi\mathsf{y}) + A_\mu(\mathsf{x}+\xi\mathsf{y})\right\}\mathcal{U}(\mathsf{A};\mathsf{x},\mathsf{y}) \ . \tag{A.4}$$

Using the definition of the field strength tensor $F_{\mu\nu}(\mathsf{x}) = \partial_\mu A_\nu(\mathsf{x}) - \partial_\nu A_\mu(\mathsf{x})$ along with the identity $\frac{\mathrm{d}}{\mathrm{d}\xi}A_\mu(\mathsf{x}+\xi\mathsf{y}) = y^\nu\partial_\nu A_\mu(\mathsf{x}+\xi\mathsf{y})$, the integral can be rewritten as:

$$ie\int_{-\frac{1}{2}}^{\frac{1}{2}} \mathrm{d}\xi\,\left\{(\tfrac{1}{2}+\xi)F_{\mu\nu}(\mathsf{x}+\xi\mathsf{y})y^\nu + \tfrac{\mathrm{d}}{\mathrm{d}\xi}\left[(\tfrac{1}{2}+\xi)A_\mu(\mathsf{x}+\xi\mathsf{y})\right]\right\}. \tag{A.5}$$

Accordingly, part [ I ] is given by:

$$\partial_\mu^{\mathsf{x}_2}\mathcal{U}(\mathsf{A};\mathsf{x},\mathsf{y}) = ie\left[A_\mu(\mathsf{x}_2) + \int_{-\frac{1}{2}}^{\frac{1}{2}} \mathrm{d}\xi(\tfrac{1}{2}+\xi)F_{\mu\nu}(\mathsf{x}+\xi\mathsf{y})y^\nu\right]\mathcal{U}(\mathsf{A};\mathsf{x},\mathsf{y})\,. \quad \text{(A.6)}$$

Part [ II ], on the other hand, can be calculated immediately by taking into account the Dirac equation Eq. (3.2):

$$\gamma^\mu\left[\bar{\Psi}(\mathsf{x}_1), \partial_\mu^{\mathsf{x}_2}\Psi(\mathsf{x}_2)\right] = -im\left[\bar{\Psi}(\mathsf{x}_1), \Psi(\mathsf{x}_2)\right] - ie\gamma^\mu A_\mu(\mathsf{x}_2)\left[\bar{\Psi}(\mathsf{x}_1), \Psi(\mathsf{x}_2)\right]\,. \quad \text{(A.7)}$$

Adding up the both parts [ I ] and [ II ], one obtains the equation of motion for the gauge-invariant density matrix:[1]

$$\left[\tfrac{1}{2}\tilde{D}_\mu(\mathsf{x},\mathsf{y}) + \tilde{\Pi}_\mu(\mathsf{x},\mathsf{y})\right]\gamma^\mu\hat{\mathcal{C}}(\mathsf{A};\mathsf{x},\mathsf{y}) = -im\hat{\mathcal{C}}(\mathsf{A};\mathsf{x},\mathsf{y})\,, \quad \text{(A.8)}$$

with the operators $\tilde{D}_\mu(\mathsf{x},\mathsf{y})$ and $\tilde{\Pi}(\mathsf{x},\mathsf{y})$ being given by:

$$\tilde{D}_\mu(\mathsf{x},\mathsf{y}) \equiv \partial_\mu^{\mathsf{x}} - ie\int_{-\frac{1}{2}}^{\frac{1}{2}}\mathrm{d}\xi\, F_{\mu\nu}(\mathsf{x}+\xi\mathsf{y})y^\nu\,, \quad \text{(A.9)}$$

$$\tilde{\Pi}_\mu(\mathsf{x},\mathsf{y}) \equiv \partial_\mu^{\mathsf{y}} - ie\int_{-\frac{1}{2}}^{\frac{1}{2}}\mathrm{d}\xi\xi F_{\mu\nu}(\mathsf{x}+\xi\mathsf{y})y^\nu\,. \quad \text{(A.10)}$$

In order to switch to momentum space, one performs a Fourier transformation with respect to $\mathsf{y}$. Accordingly, derivatives with respect to $\mathsf{y}$ become linear factors in $\mathsf{p}$ and vice versa:

$$\partial_\mu^{\mathsf{y}} \to -ip_\mu \qquad \text{and} \qquad y_\mu \to -i\partial_\mu^{\mathsf{p}}\,. \quad \text{(A.11)}$$

This immediately results in the equation of motion for the covariant Wigner operator Eq. (3.10):[2]

$$\left[\tfrac{1}{2}D_\mu(\mathsf{x},\mathsf{p}) - i\Pi_\mu(\mathsf{x},\mathsf{p})\right]\gamma^\mu\hat{\mathcal{W}}^{(4)}(\mathsf{x},\mathsf{p}) = -im\hat{\mathcal{W}}^{(4)}(\mathsf{x},\mathsf{p})\,. \quad \text{(A.12)}$$

[1] Taking the derivative of $\hat{\mathcal{C}}(\mathsf{x},\mathsf{y})$ with respect to $\mathsf{x}_1$ instead of $\mathsf{x}_2$ yields:

$$\left[\tfrac{1}{2}\tilde{D}_\mu(\mathsf{x},\mathsf{y}) - \tilde{\Pi}_\mu(\mathsf{x},\mathsf{y})\right]\hat{\mathcal{C}}(\mathsf{A};\mathsf{x},\mathsf{y})\gamma^\mu = im\hat{\mathcal{C}}(\mathsf{A};\mathsf{x},\mathsf{y})\,.$$

[2] Taking the Fourier transformation of the adjoint equation yields the equation of motion Eq. (3.11):

$$\left[\tfrac{1}{2}D_\mu(\mathsf{x},\mathsf{p}) + i\Pi_\mu(\mathsf{x},\mathsf{p})\right]\hat{\mathcal{W}}^{(4)}(\mathsf{x},\mathsf{p})\gamma^\mu = im\hat{\mathcal{W}}^{(4)}(\mathsf{x},\mathsf{p})\,.$$

## A.2 Equal-time Wigner formalism: Equation of motion

An energy average is performed in order to switch from the covariant Wigner formalism to the equal-time Wigner formalism. Considering the pseudo-differential operators Eq. (3.12) – (3.13):[3]

$$D_0(\mathsf{x},\mathsf{p}) \equiv \partial_{x_0} + e\int_{-\frac{1}{2}}^{\frac{1}{2}} \mathrm{d}\xi\, \mathbf{E}(\mathsf{x} - i\xi\partial_{\mathsf{p}}) \cdot \nabla_{\mathbf{p}}\,, \tag{A.13}$$

$$\mathbf{D}(\mathsf{x},\mathsf{p}) \equiv \nabla_{\mathbf{x}} + e\int_{-\frac{1}{2}}^{\frac{1}{2}} \mathrm{d}\xi\,[\mathbf{E}(\mathsf{x} - i\xi\partial_{\mathsf{p}})\partial_{p_0} + \mathbf{B}(\mathsf{x} - i\xi\partial_{\mathsf{p}}) \times \nabla_{\mathbf{p}}]\,, \tag{A.14}$$

$$\Pi_0(\mathsf{x},\mathsf{p}) \equiv p_0 + ie\int_{-\frac{1}{2}}^{\frac{1}{2}} \mathrm{d}\xi\xi\mathbf{E}(\mathsf{x} - i\xi\partial_{\mathsf{p}}) \cdot \nabla_{\mathbf{p}}\,, \tag{A.15}$$

$$\mathbf{\Pi}(\mathsf{x},\mathsf{p}) \equiv \mathbf{p} - ie\int_{-\frac{1}{2}}^{\frac{1}{2}} \mathrm{d}\xi\xi\,[\mathbf{E}(\mathsf{x} - i\xi\partial_{\mathsf{p}})\partial_{p_0} + \mathbf{B}(\mathsf{x} - i\xi\partial_{\mathsf{p}}) \times \nabla_{\mathbf{p}}]\,, \tag{A.16}$$

one assumes that the electromagnetic field can be expanded in a Taylor series with respect to the temporal coordinate:

$$F_{\mu\nu}(\mathbf{x} + i\xi\nabla_{\mathbf{p}}, x_0 - i\xi\partial_{p_0}) = \sum_{n=0}^{\infty} \tfrac{1}{n!}F_{\mu\nu}^{(n)}(\mathbf{x} + i\xi\nabla_{\mathbf{p}}, x_0)(-i\xi\partial_{p_0})^n\,. \tag{A.17}$$

Here, $F_{\mu\nu}^{(n)}$ denotes the n-th derivative with respect to the temporal coordinate. Upon taking the energy average of any of these operators acting on the covariant Wigner components $\mathbb{W}(\mathsf{x},\mathsf{p})$, one finds:

$$\int \frac{\mathrm{d}p_0}{(2\pi)} D_0(\mathsf{x},\mathsf{p})\mathbb{W}(\mathsf{x},\mathsf{p}) = D_t(\mathbf{x},\mathbf{p},t)\mathbb{w}(\mathbf{x},\mathbf{p},t) \quad , \tag{A.18}$$

$$\int \frac{\mathrm{d}p_0}{(2\pi)} \mathbf{D}(\mathsf{x},\mathsf{p})\mathbb{W}(\mathsf{x},\mathsf{p}) = \mathbf{D}(\mathbf{x},\mathbf{p},t)\mathbb{w}(\mathbf{x},\mathbf{p},t) \quad , \tag{A.19}$$

$$\int \frac{\mathrm{d}p_0}{(2\pi)} \Pi_0(\mathsf{x},\mathsf{p})\mathbb{W}(\mathsf{x},\mathsf{p}) = \Pi_t(\mathbf{x},\mathbf{p},t)\mathbb{w}(\mathbf{x},\mathbf{p},t) + \mathbb{w}^{[1]}(\mathbf{x},\mathbf{p},t)\,, \tag{A.20}$$

$$\int \frac{\mathrm{d}p_0}{(2\pi)} \mathbf{\Pi}(\mathsf{x},\mathsf{p})\mathbb{W}(\mathsf{x},\mathsf{p}) = \mathbf{\Pi}(\mathbf{x},\mathbf{p},t)\mathbb{w}(\mathbf{x},\mathbf{p},t) \quad , \tag{A.21}$$

given that the covariant Wigner components and all its derivatives with respect to $p_0$ vanish for $|p_0| \to \infty$. The transport equations Eq. (3.36) – (3.43) and the constraint equations Eq. (3.48) – (3.55) are accordingly found by taking the energy average of the equations of motion for the covariant Wigner components Eq. (3.22) – (3.31).

[3]Note that the operator $\mathbf{D}(\mathsf{x},\mathsf{p})$ corresponds to the *covariant* components $D_i(\mathsf{x},\mathsf{p})\mathbf{e}_i$ whereas $\mathbf{\Pi}(\mathsf{x},\mathsf{p})$ is related to the *contravariant* components $\Pi^i(\mathsf{x},\mathsf{p})\mathbf{e}_i$.

## A.3 Hierarchy truncation

In order to derive the equations of motion for the first energy moments $\mathbb{w}^{[1]}(\mathbf{x},\mathbf{p},t)$, one first multiplies the equations of motion for the covariant Wigner components Eq. (3.22) – (3.31) by a factor $p_0$ and subsequently takes the energy average:

$$
\begin{aligned}
&D_t\,\mathbb{s}^{[1]} + D_t^{[1]}\,\mathbb{s} && -2\mathbf{\Pi}\cdot\mathbb{t}_1^{[1]} -2\mathbf{\Pi}^{[1]}\cdot\mathbb{t}_1 = 0\ , && \text{(A.22)}\\
&D_t\,\mathbb{p}^{[1]} + D_t^{[1]}\,\mathbb{p} && +2\mathbf{\Pi}\cdot\mathbb{t}_2^{[1]} +2\mathbf{\Pi}^{[1]}\cdot\mathbb{t}_2 = -2m\,\mathbb{a}_0^{[1]}\ , && \text{(A.23)}\\
&D_t\,\mathbb{v}_0^{[1]} + D_t^{[1]}\,\mathbb{v}_0 + \mathbf{D}\cdot\mathbb{v}^{[1]} + \mathbf{D}^{[1]}\cdot\mathbb{v} && = 0\ , && \text{(A.24)}\\
&D_t\,\mathbb{a}_0^{[1]} + D_t^{[1]}\,\mathbb{a}_0 + \mathbf{D}\cdot\mathbb{a}^{[1]} + \mathbf{D}^{[1]}\cdot\mathbb{a} && = 2m\,\mathbb{p}^{[1]}\ , && \text{(A.25)}\\
&D_t\,\mathbb{v}^{[1]} + D_t^{[1]}\,\mathbb{v} + \mathbf{D}\,\mathbb{v}_0^{[1]} + \mathbf{D}^{[1]}\,\mathbb{v}_0 && +2\mathbf{\Pi}\times\mathbb{a}^{[1]} +2\mathbf{\Pi}^{[1]}\times\mathbb{a} = -2m\,\mathbb{t}_1^{[1]}\ , && \text{(A.26)}\\
&D_t\,\mathbb{a}^{[1]} + D_t^{[1]}\,\mathbb{a} + \mathbf{D}\,\mathbb{a}_0^{[1]} + \mathbf{D}^{[1]}\,\mathbb{a}_0 && +2\mathbf{\Pi}\times\mathbb{v}^{[1]} +2\mathbf{\Pi}^{[1]}\times\mathbb{v} = 0\ , && \text{(A.27)}\\
&D_t\,\mathbb{t}_1^{[1]} + D_t^{[1]}\,\mathbb{t}_1 + \mathbf{D}\times\mathbb{t}_2^{[1]} + \mathbf{D}^{[1]}\times\mathbb{t}_2 && +2\mathbf{\Pi}\,\mathbb{s}^{[1]} +2\mathbf{\Pi}^{[1]}\,\mathbb{s} = 2m\,\mathbb{v}^{[1]}\ , && \text{(A.28)}\\
&D_t\,\mathbb{t}_2^{[1]} + D_t^{[1]}\,\mathbb{t}_2 - \mathbf{D}\times\mathbb{t}_1^{[1]} - \mathbf{D}^{[1]}\times\mathbb{t}_1 && -2\mathbf{\Pi}\,\mathbb{p}^{[1]} -2\mathbf{\Pi}^{[1]}\,\mathbb{p} = 0\ . && \text{(A.29)}
\end{aligned}
$$

Here, the new operators $D_t^{[1]}(\mathbf{x},\mathbf{p},t)$, $\mathbf{D}^{[1]}(\mathbf{x},\mathbf{p},t)$ and $\mathbf{\Pi}^{[1]}(\mathbf{x},\mathbf{p},t)$ are given by:

$$D_t^{[1]}(\mathbf{x},\mathbf{p},t) \equiv ie\int_{-\frac{1}{2}}^{\frac{1}{2}} d\xi\,\xi\,\mathbf{E}^{(1)}(\mathbf{x}+i\xi\nabla_{\mathbf{p}},t)\cdot\nabla_{\mathbf{p}}\,, \tag{A.30}$$

$$\mathbf{D}^{[1]}(\mathbf{x},\mathbf{p},t) \equiv ie\int_{-\frac{1}{2}}^{\frac{1}{2}} d\xi\big[\,\xi\,\mathbf{B}^{(1)}(\mathbf{x}+i\xi\nabla_{\mathbf{p}},t)\times\nabla_{\mathbf{p}} + i\xi\mathbf{E}(\mathbf{x}+i\xi\nabla_{\mathbf{p}},t)\big]\,, \tag{A.31}$$

$$\mathbf{\Pi}^{[1]}(\mathbf{x},\mathbf{p},t) \equiv e\int_{-\frac{1}{2}}^{\frac{1}{2}} d\xi\big[\xi^2\mathbf{B}^{(1)}(\mathbf{x}+i\xi\nabla_{\mathbf{p}},t)\times\nabla_{\mathbf{p}} + i\xi\mathbf{E}(\mathbf{x}+i\xi\nabla_{\mathbf{p}},t)\big]\,. \tag{A.32}$$

One can then express $\mathbb{w}^{[1]}(\mathbf{x},\mathbf{p},t)$ in terms of $\mathbb{w}(\mathbf{x},\mathbf{p},t)$ by using the constraint equations Eq. (3.48) – (3.55). Various commutators are needed in order to simplify the resulting expressions when pulling through the various operators. Additionally, the homogeneous Maxwell's equations:

$$\nabla_{\mathbf{x}}\cdot\mathbf{B}(\mathbf{x},t) = 0 \quad \text{and} \quad \mathbf{B}^{(1)}(\mathbf{x},t) + \nabla_{\mathbf{x}}\times\mathbf{E}(\mathbf{x},t) = 0\ , \tag{A.33}$$

as well as the representation of the electromagnetic field at shifted arguments are used:[4]

$$F^{\mu\nu}(\mathbf{x}+i\xi\nabla_{\mathbf{p}},t) = e^{i\xi\triangle}F^{\mu\nu}(\mathbf{x},t) = \sum_{n=0}^{\infty}\frac{(i\xi\triangle)^n}{n!}F^{\mu\nu}(\mathbf{x},t)\ . \tag{A.34}$$

[4]The triangle operator is a shorthand notation for $\triangle = \nabla_{\mathbf{x}}\cdot\nabla_{\mathbf{p}}$ .

The relevant commutators including $D_t(\mathbf{x},\mathbf{p},t)$ are given by:

$$\big[D_t\,,\Pi_t\big] \;=\; D_t^{[1]}\;, \tag{A.35}$$

$$\big[D_t\,,\mathbf{D}_i\big] \;=\; -\,e\int_{-\frac{1}{2}}^{\frac{1}{2}} \mathrm{d}\xi\,\triangle\,\mathbf{E}_i(\mathbf{x}+i\xi\nabla_{\mathbf{p},t})\;, \tag{A.36}$$

$$\big[D_t\,,\mathbf{\Pi}_i\big] = \nabla_{\mathbf{x},i}\,\Pi_t - \mathbf{D}_i^{[1]}\;. \tag{A.37}$$

The commutators including $\Pi_t(\mathbf{x},\mathbf{p},t)$ read:

$$\big[\Pi_t\,,\mathbf{D}_i\big] \;=\; -\,\nabla_{\mathbf{x},i}\,\Pi_t\;, \tag{A.38}$$

$$\big[\Pi_t\,,\mathbf{\Pi}_i\big] \;=\; -\,\mathbf{\Pi}_i^{[1]} + e\int_{-\frac{1}{2}}^{\frac{1}{2}} \mathrm{d}\xi\big[2i\xi-\xi^2\triangle\big]\mathbf{E}_i(\mathbf{x}+i\xi\nabla_{\mathbf{p},t})\;. \tag{A.39}$$

Finally, the commutators between $\mathbf{D}(\mathbf{x},\mathbf{p},t)$ and $\mathbf{\Pi}(\mathbf{x},\mathbf{p},t)$ are given by:[5]

$$\big[\mathbf{D}_i\,,\mathbf{D}_j\big] \;=\; e\int_{-\frac{1}{2}}^{\frac{1}{2}} \mathrm{d}\xi\,\epsilon_{ijk}\triangle\mathbf{B}_k(\mathbf{x}+i\xi\nabla_{\mathbf{p}},t)\;, \tag{A.40}$$

$$\big[\mathbf{\Pi}_i\,,\mathbf{D}_j\big] \;=\; -e\int_{-\frac{1}{2}}^{\frac{1}{2}} \mathrm{d}\xi\,\epsilon_{ijk}\,[1+i\xi\triangle]\,\mathbf{B}_k(\mathbf{x}+i\xi\nabla_{\mathbf{p}},t)\;, \tag{A.41}$$

$$\big[\mathbf{\Pi}_i\,,\mathbf{\Pi}_j\big] \;=\; e\int_{-\frac{1}{2}}^{\frac{1}{2}} \mathrm{d}\xi\,\epsilon_{ijk}\,\big[2i\xi-\xi^2\triangle\big]\,\mathbf{B}_k(\mathbf{x}+i\xi\nabla_{\mathbf{p}},t)\;. \tag{A.42}$$

As a matter of fact, one finds:

$$e\int_{-\frac{1}{2}}^{\frac{1}{2}} \mathrm{d}\xi\big[2i\xi-\xi^2\triangle+\tfrac{1}{4}\triangle\big]F^{\mu\nu}(\mathbf{x}+i\xi\nabla_{\mathbf{p}},t) = 0\;, \tag{A.43}$$

which can be easily checked by plugging in the Taylor expansion Eq. (A.34) and calculating order by order the coefficients of $\triangle^n$. Accordingly, the following identities hold:

$$\mathbf{\Pi}_i^{[1]} + \big[\Pi_t\;,\mathbf{\Pi}_i\big] - \tfrac{1}{4}\big[D_t\,,\mathbf{D}_i\big] \;=\; 0\;, \tag{A.44}$$

$$\big[\mathbf{\Pi}_i\,,\mathbf{\Pi}_j\big] + \tfrac{1}{4}\big[\mathbf{D}_i\,,\mathbf{D}_j\big] \;=\; 0\;. \tag{A.45}$$

[5]In fact, the actual value of the commutator $[\mathbf{\Pi}_i,\mathbf{D}_j]$ is only needed insofar as one needs to know:

$$[\mathbf{\Pi}_i,\mathbf{D}_i]=0 \quad \text{and} \quad [\mathbf{\Pi}_i,\mathbf{D}_j]=[\mathbf{D}_i,\mathbf{\Pi}_j]\;.$$

In order to simplify notation afterwards, I recall the equations of motion for the equal-time Wigner components Eq. (3.36) – (3.43) at this point:[6]

$$
\begin{aligned}
&D_t\,\mathbb{s} && &&- 2\mathbf{\Pi}\cdot\mathbb{t}_1 && &&= 0\,, && \text{(A.46)}\\
&D_t\,\mathbb{p} && &&+ 2\mathbf{\Pi}\cdot\mathbb{t}_2 &&+ 2m\mathbb{a}_0 &&= 0\,, && \text{(A.47)}\\
&D_t\,\mathbb{v}_0 &&+ \mathbf{D}\cdot\mathbb{v} && && &&= 0\,, && \text{(A.48)}\\
&D_t\,\mathbb{a}_0 &&+ \mathbf{D}\cdot\mathbb{a} && &&- 2m\mathbb{p} &&= 0\,, && \text{(A.49)}\\
&D_t\,\mathbb{v} &&+ \mathbf{D}\,\mathbb{v}_0 &&+ 2\mathbf{\Pi}\times\mathbb{a} &&+ 2m\mathbb{t}_1 &&= 0\,, && \text{(A.50)}\\
&D_t\,\mathbb{a} &&+ \mathbf{D}\,\mathbb{a}_0 &&+ 2\mathbf{\Pi}\times\mathbb{v} && &&= 0\,, && \text{(A.51)}\\
&D_t\,\mathbb{t}_1 &&+ \mathbf{D}\times\mathbb{t}_2 &&+ 2\mathbf{\Pi}\,\mathbb{s} &&- 2m\mathbb{v} &&= 0\,, && \text{(A.52)}\\
&D_t\,\mathbb{t}_2 &&- \mathbf{D}\times\mathbb{t}_1 &&- 2\mathbf{\Pi}\,\mathbb{p} && &&= 0\,. && \text{(A.53)}
\end{aligned}
$$

Taking into account all the identities Eq. (A.35) – (A.45) as well as the equations of motions for the equal-time Wigner components Eq. (A.46) – (A.53), the equations of motion for the first energy moments can be written as:

$$
\begin{aligned}
&-\Pi_t\,[\text{lhs Eq. }(A.46)] + \tfrac{1}{2}\mathbf{D}\cdot[\text{lhs Eq. }(A.52)] && &&= 0, && \text{(A.54)}\\
&-\Pi_t\,[\text{lhs Eq. }(A.47)] - \tfrac{1}{2}\mathbf{D}\cdot[\text{lhs Eq. }(A.53)] && &&= 0, && \text{(A.55)}\\
&-\Pi_t\,[\text{lhs Eq. }(A.48)] && +\mathbf{\Pi}\cdot[\text{lhs Eq. }(A.50)] &&= 0, && \text{(A.56)}\\
&-\Pi_t\,[\text{lhs Eq. }(A.49)] && +\mathbf{\Pi}\cdot[\text{lhs Eq. }(A.51)] &&= 0, && \text{(A.57)}\\
&-\Pi_t\,[\text{lhs Eq. }(A.50)] - \tfrac{1}{2}\mathbf{D}\times[\text{lhs Eq. }(A.51)] && +\mathbf{\Pi}\,[\text{lhs Eq. }(A.48)] &&= 0, && \text{(A.58)}\\
&-\Pi_t\,[\text{lhs Eq. }(A.51)] - \tfrac{1}{2}\mathbf{D}\times[\text{lhs Eq. }(A.50)] && +\mathbf{\Pi}\,[\text{lhs Eq. }(A.49)] &&= 0, && \text{(A.59)}\\
&-\Pi_t\,[\text{lhs Eq. }(A.52)] - \tfrac{1}{2}\mathbf{D}\,[\text{lhs Eq. }(A.46)] && +\mathbf{\Pi}\times[\text{lhs Eq. }(A.53)] &&= 0, && \text{(A.60)}\\
&-\Pi_t\,[\text{lhs Eq. }(A.53)] + \tfrac{1}{2}\mathbf{D}\,[\text{lhs Eq. }(A.47)] && -\mathbf{\Pi}\times[\text{lhs Eq. }(A.52)] &&= 0. && \text{(A.61)}
\end{aligned}
$$

Accordingly, it turns out that all these equations are trivially fulfilled once the equal-time Wigner components obey the equations of motion Eq. (A.46) – (A.53). Consequently, one does not have to solve the infinite hierarchy but it suffices to use the constraint equations in order to calculate all higher energy moments.[7]

[6] Note that the mass terms have been brought to the left hand side (denoted as 'lhs' in the following) compared to Eq. (3.36) – (3.43).

[7] This calculation for first energy moments $\mathbb{w}^{[1]}(\mathbf{x},\mathbf{p},t)$ served only as demonstration. As a matter of fact, it has been proven generally that the equations of motion for all the higher energy moments $\mathbb{w}^{[n]}(\mathbf{x},\mathbf{p},t)$ are trivially fulfilled as well [93].

## A.4 Vacuum value of the equal-time Wigner function

In order to calculate the vacuum value of the equal-time Wigner function one canonically quantizes the free Dirac field as discussed in Section 4.1. Accordingly, one has:

$$\Psi(\mathbf{x},t) = \int \frac{\mathrm{d}^3 q}{(2\pi)^3}\, e^{i\mathbf{q}\cdot\mathbf{x}} \sum_{r=1}^{2} \Big[u_r(\mathbf{q},t) a_r(\mathbf{q}) + v_r(-\mathbf{q},t) b_r^\dagger(-\mathbf{q})\Big] \ , \tag{A.62}$$

with:[8]

$$u_r(\mathbf{q},t) = \big[\ \gamma^0\omega(\mathbf{q}) - \boldsymbol{\gamma}\cdot\mathbf{q} + m\big] g_{\text{vac}}^{(+)}(\mathbf{q},t) R_r \ , \tag{A.63}$$
$$v_r(-\mathbf{q},t) = \big[-\gamma^0\omega(\mathbf{q}) - \boldsymbol{\gamma}\cdot\mathbf{q} + m\big] g_{\text{vac}}^{(-)}(\mathbf{q},t) R_r \ . \tag{A.64}$$

The properly normalized positive and negative energy plane wave solutions $g_{\text{vac}}^{(\pm)}(\mathbf{q},t)$ are given by:

$$g_{\text{vac}}^{(\pm)}(\mathbf{q},t) = \frac{e^{\mp i\omega(\mathbf{q})t}}{\sqrt{2\omega(\mathbf{q})[\omega(\mathbf{q}) \mp q_3]}} \ . \tag{A.65}$$

The spinors $u_r(\mathbf{q},t)$ and $v_r(\mathbf{q},t)$ obey the following completeness relations:[9]

$$\sum_{r=1}^{2} u_r(\mathbf{q},t)\bar{u}_r(\mathbf{q},\mathbf{t}) = \frac{1}{2\omega(\mathbf{q})}\big[\gamma^0\omega(\mathbf{q}) - \boldsymbol{\gamma}\cdot\mathbf{q} + m\big] \ , \tag{A.66}$$
$$\sum_{r=1}^{2} v_r(\mathbf{q},t)\,\bar{v}_r(\mathbf{q},\mathbf{t}) = \frac{1}{2\omega(\mathbf{q})}\big[\gamma^0\omega(\mathbf{q}) - \boldsymbol{\gamma}\cdot\mathbf{q} - m\big] \ . \tag{A.67}$$

The equal-time Wigner function for the free Dirac field is defined as:

$$\mathcal{W}_{\text{vac}}^{(3)}(\mathbf{x},\mathbf{p},t) = \frac{1}{2}\int \mathrm{d}^3 y e^{-i\mathbf{p}\cdot\mathbf{y}} \langle\Omega|\left[\bar{\Psi}(\mathbf{x}-\tfrac{\mathbf{y}}{2},t), \Psi(\mathbf{x}+\tfrac{\mathbf{y}}{2},t)\right]|\Omega\rangle \ . \tag{A.68}$$

Plugging in the explicit expressions Eq. (A.62) as well as using the non-vanishing anticommutators:

$$\big\{a_r(\mathbf{q}), a_s^\dagger(\mathbf{q}')\big\} = \big\{b_r(\mathbf{q})\,,\, b_s^\dagger(\mathbf{q}')\big\} = (2\pi)^3\delta_{rs}\delta(\mathbf{q}-\mathbf{q}') \ , \tag{A.69}$$

[8]For simplicity, the same spinor representation as the one chosen in Section 4.1 is used. Note, however, that any other complete set of spinors would do the job equally well.

[9]Note that the common factor $\frac{1}{2\omega(\mathbf{q})}$ is non-standard but a direct consequence of the chosen spinor normalization:

$$u_r^\dagger(\mathbf{q},t) u_s(\mathbf{q},t) = v_r^\dagger(\mathbf{q},t) v_s(\mathbf{q},t) = \delta_{rs} \ .$$

one finds:

$$\mathcal{W}^{(3)}_{\text{vac}}(\mathbf{x},\mathbf{p},t) = \frac{1}{2}\sum_{r=1}^{2}\big[v_r(-\mathbf{p},t)\bar{v}_r(-\mathbf{p},t) - u_r(\mathbf{p},t)\bar{u}_r(\mathbf{p},t)\big] \ . \tag{A.70}$$

Taking advantage of the completeness relations Eq. (A.66) – (A.67), this finally yields:

$$\mathcal{W}^{(3)}_{\text{vac}}(\mathbf{x},\mathbf{p},t) = \frac{1}{2\omega(\mathbf{q})}\big[\boldsymbol{\gamma}\cdot\mathbf{p} - m\big] \ . \tag{A.71}$$

Accordingly, the vacuum value of the equal-time Wigner function is spatially homogeneous and time-independent as it should be. Moreover, it turns out that $\mathbb{s}_{\text{vac}}(\mathbf{p})$ and $\mathbb{v}_{\text{vac}}(\mathbf{p})$ are the only non-vanishing equal-time Wigner components:

$$\mathbb{s}_{\text{vac}}(\mathbf{p}) = -\frac{2m}{\omega(\mathbf{p})} \qquad \text{and} \qquad \mathbb{v}_{\text{vac}}(\mathbf{p}) = -\frac{2\mathbf{p}}{\omega(\mathbf{p})} \ . \tag{A.72}$$

## A.5 Basis expansion

In order to expand $\mathbb{z}(\mathbf{q},t)$ in the basis $\{\mathbb{e}_1, ..., \mathbb{e}_{10}\}$ as indicated in Eq. (4.57), it is convenient to choose the first basis vector in such a way that $\mathcal{E}_1(\mathbf{q}, t_{\text{vac}}) = 1$ is the only non-vanishing coefficient in the vacuum:

$$\mathbb{e}_1(\mathbf{q},t) = \frac{1}{\omega(\mathbf{q},t)}\begin{pmatrix} m \\ \boldsymbol{\pi}(\mathbf{q},t) \\ \mathbf{0} \\ \mathbf{0} \end{pmatrix} \ . \tag{A.73}$$

Acting with $\mathbb{M}(\mathbf{q},t)$ as well as with the total time derivative on this first basis vector $\mathbb{e}_1(\mathbf{q},t)$, one finds:

$$\mathbb{M}(\mathbf{q},t)\mathbb{e}_1(\mathbf{q},t) = 0 \qquad \text{and} \qquad \dot{\mathbb{e}}_1(\mathbf{q},t) = -Q(\mathbf{q},t)\mathbb{e}_2(\mathbf{q},t) \ , \tag{A.74}$$

with the second basis vector $\mathbb{e}_2(\mathbf{q},t)$ being defined as:

$$\mathbb{e}_2(\mathbf{q},t) = \frac{1}{\omega(\mathbf{q},t)\epsilon_\perp}\begin{pmatrix} m\pi_3(q_3,t) \\ \boldsymbol{\pi}(\mathbf{q},t)\pi_3(q_3,t) - \omega^2(\mathbf{q},t)\mathbf{e}_3 \\ \mathbf{0} \\ \mathbf{0} \end{pmatrix} \ . \tag{A.75}$$

Repeating this procedure with the second basis vector $\mathbb{e}_2(\mathbf{q},t)$, one finds:

$$\mathbb{M}(\mathbf{q},t)\mathbb{e}_2(\mathbf{q},t) = 2\omega(\mathbf{q},t)\mathbb{e}_3(\mathbf{q},t) \qquad \text{and} \qquad \dot{\mathbb{e}}_2(\mathbf{q},t) = Q(\mathbf{q},t)\mathbb{e}_1(\mathbf{q},t) \ , \tag{A.76}$$

with the third basis vector $\mathbb{e}_3(\mathbf{q},t)$ being given by:

$$\mathbb{e}_3(\mathbf{q},t) = \frac{1}{\epsilon_\perp}\begin{pmatrix} 0 \\ \mathbf{0} \\ \boldsymbol{\pi}(\mathbf{q},t)\times\mathbf{e}_3 \\ -m\mathbf{e}_3 \end{pmatrix} . \tag{A.77}$$

Continuing with the third basis vector $\mathbb{e}_3(\mathbf{q},t)$, it turns out that the system closes:[10]

$$\mathbb{M}(\mathbf{q},t)\mathbb{e}_3(\mathbf{q},t) = -2\omega(\mathbf{q},t)\mathbb{e}_2(\mathbf{q},t) \qquad \text{and} \qquad \dot{\mathbb{e}}_2(\mathbf{q},t) = 0 . \tag{A.78}$$

As $\mathcal{E}_1(\mathbf{q},t_{\text{vac}}) = 1$ is the only non-vanishing coefficient in the vacuum, one does in fact not need to calculate the remaining basis vectors. Accordingly, the system is fully described by means of the first three expansion coefficients $\mathcal{E}_1(\mathbf{q},t)$, $\mathcal{E}_2(\mathbf{q},t)$ and $\mathcal{E}_3(\mathbf{q},t)$.

## A.6 Series reversion

One needs to invert the variable transformation in order to calculate the WKB instanton action $2\mathcal{S}_{\text{inst.}}(\mathbf{q})$:

$$T = \sqrt{\frac{\pi}{2}}e^{-\frac{\sigma^2}{2}}\tau\,\text{Erf}\,\Big(\frac{t}{\sqrt{2}\tau} + \frac{i\sigma}{\sqrt{2}}\Big) + c.c. . \tag{A.79}$$

Due to the fact that the series expansion of the error function $\text{Erf}(x+iy)$ is given by [113]:

$$\begin{aligned}\text{Erf}(x+iy) &= \frac{2}{\sqrt{\pi}}\sum_{n=0}^{\infty}\frac{(-1)^n}{n!(2n+1)}(x+iy)^{2n+1} = \\ &= \frac{2}{\sqrt{\pi}}\sum_{n=0}^{\infty}\sum_{m=0}^{2n+1}\frac{(-1)^n}{n!(2n+1)}\binom{2n+1}{m}x^{2n+1-m}(iy)^m ,\end{aligned} \tag{A.80}$$

the series expansion of $T$ can be written as:

$$\begin{aligned}T(t) &= e^{-\frac{\sigma^2}{2}}\sum_{n=0}^{\infty}\sum_{m=0}^{n}\frac{(-1)^{n+m}}{2^n n!(2n+1)}\frac{\sigma^{2m}}{\tau^{2n-2m}}\binom{2n+1}{2m}t^{2n+1-2m} = \\ &= t - \frac{1+\sigma^2}{6\tau^2}t^3 + \frac{3+6\sigma^2+\sigma^4}{120\tau^4}t^5 + \mathcal{O}(t^7) .\end{aligned} \tag{A.81}$$

[10] As a matter of fact, the basis vectors $\mathbb{e}_i(\mathbf{q},t)$ with $i = \{1,2,3\}$ form an orthonormalized set.

Given the series expansion of any function $y(x)$ with $y(0) = 0$ as well as $y'(0) \neq 0$:

$$y(x) = \sum_{n=1}^{\infty} c_n x^n \quad \text{with} \quad c_1 \neq 0 \ , \tag{A.82}$$

the inverse series expansion can be calculated term by term:

$$x(y) = \sum_{n=1}^{\infty} d_n y^n \ , \tag{A.83}$$

with the first few coefficients being determined via [113]:

$$\begin{aligned}
c_1 d_1 &= 0 \ , \\
c_1^3 d_2 &= -c_2 \ , \\
c_1^5 d_3 &= 2c_2^2 - c_1 c_3 \ , \\
c_1^7 d_4 &= 5c_1 c_2 c_3 - c_1^2 c_4 - 5c_2^3 \ , \\
c_1^9 d_5 &= 6c_1^2 c_2 c_4 + 3c_1^2 c_3^2 + 14c_2^4 - c_1^3 c_5 - 21 c_1 c_2^2 c_3 \ .
\end{aligned}$$

Accordingly, the inverse series $t(T)$ can be calculated term by term by taking into account the series expansion Eq. (A.81),

$$t(T) = T + \frac{1 + \sigma^2}{6\tau^2} T^3 + \frac{7 + 14\sigma^2 + 9\sigma^4}{120\tau^4} T^5 + \mathcal{O}(T^7) \ . \tag{A.84}$$

Appendix B

# Modified Lax-Wendroff scheme

In order to solve the PDE systems Eq. (5.57) – (5.60) and Eq. (5.69) – (5.72), respectively, a modified Lex-Wendroff scheme for the solution of hyperbolic PDE systems with source terms has been used [125, 126]. A linear, first order PDE system in two dimensions reads in general:[1]

$$\frac{\partial}{\partial t}\mathbf{w}(\alpha,\beta,t) = A_{[\alpha]}\frac{\partial}{\partial \alpha}\mathbf{w}(\alpha,\beta,t) + B_{[\beta]}\frac{\partial}{\partial \beta}\mathbf{w}(\alpha,\beta,t) + \mathbf{s}(\alpha,\beta,t;\mathbf{w}) \ , \tag{B.1}$$

with $\mathbf{w}(\alpha,\beta,t)$ being a $n$-component vector of field variables and $\mathbf{s}(\alpha,\beta,t;\mathbf{w})$ denoting the $n$-component source vector:

$$\mathbf{s}(\alpha,\beta,t;\mathbf{w}) = C\mathbf{w}(\alpha,\beta,t) + \bar{\mathbf{s}}(\alpha,\beta,t) \ . \tag{B.2}$$

The PDE system Eq. (B.1) is in fact hyperbolic given that all linear combinations of the $n \times n$ matrices $A_{[\alpha]}$ and $B_{[\beta]}$ are diagonalizable with real eigenvalues.[2] The domain is then divided into rectangular cells in the $\alpha$-$\beta$ plane in order to actually solve this PDE system by means of a *finite difference scheme*. Moreover, it is convenient to introduce the following abbreviation for the vector of field variables at a grid point $(\alpha_i, \beta_j)$ at a given time $t_n$:

$$\mathbf{w}^n_{i,j} \equiv \mathbf{w}(\alpha_i,\beta_j,t_n) \ . \tag{B.3}$$

[1] The $\alpha$- and $\beta$-direction correspond to the $x$- and $p$-direction for Eq. (5.57) – (5.60) as well as to the $x$- and the $y$-direction for Eq. (5.69) – (5.72), respectively.

[2] It can be checked easily that the PDE system Eq. (5.57) – (5.60) with:

$$A_{[x]} = \begin{pmatrix} 0 & 0 & 0 & 0 \\ 0 & 0 & -1 & 0 \\ 0 & -1 & 0 & 0 \\ 0 & 0 & 0 & 0 \end{pmatrix} \quad \text{and} \quad B_{[p]} = \begin{pmatrix} -eE(x,t) & 0 & 0 & 0 \\ 0 & -eE(x,t) & 0 & 0 \\ 0 & 0 & -eE(x,t) & 0 \\ 0 & 0 & 0 & -eE(x,t) \end{pmatrix}$$

as well as the PDE system Eq. (5.69) – (5.72) with:

$$A_{[x]} = \begin{pmatrix} 0 & 0 & 0 & 0 \\ 0 & 0 & -1 & 0 \\ 0 & -1 & 0 & 0 \\ 0 & 0 & 0 & 0 \end{pmatrix} \quad \text{and} \quad B_{[y]} = \begin{pmatrix} 0 & 0 & 0 & -2i \\ 0 & 0 & 0 & 0 \\ 0 & 0 & 0 & 0 \\ 2i & 0 & 0 & 0 \end{pmatrix}$$

fulfill this hyperbolicity condition.

The numerical solution at a subsequent time $t_{n+1} = t_n + \Delta t$ can for instance be found by adopting an *explicit two-step scheme* for discretizing the PDE system.[3] Introducing the abbreviations:

$$\mathbf{a}_{i,j}^{n} = A_{i,j;[\alpha]}^{n}\mathbf{w}_{i,j}^{n} , \tag{B.4}$$

$$\mathbf{b}_{i,j}^{n} = B_{i,j;[\beta]}^{n}\mathbf{w}_{i,j}^{n} , \tag{B.5}$$

$$\mathbf{s}_{i,j}^{n} = C_{i,j}^{n}\mathbf{w}_{i,j}^{n} + \bar{\mathbf{s}}_{i,j}^{n} , \tag{B.6}$$

the first step utilizes the data $\mathbf{w}^n$ in order to calculate:

$$\begin{aligned}\mathbf{w}_{i+1/2,j+1/2}^{n+1/2} = & \ \tfrac{1}{4}\left[\mathbf{w}_{i+1,j+1}^{n} + \mathbf{w}_{i,j+1}^{n} + \mathbf{w}_{i+1,j}^{n} + \mathbf{w}_{i,j}^{n}\right] \\ & + \tfrac{\lambda_{[\alpha]}}{4}\left[\mathbf{a}_{i+1,j+1}^{n} - \mathbf{a}_{i,j+1}^{n} + \mathbf{a}_{i+1,j}^{n} - \mathbf{a}_{i,j}^{n}\right] \\ & + \tfrac{\lambda_{[\beta]}}{4}\left[\mathbf{b}_{i+1,j+1}^{n} + \mathbf{b}_{i,j+1}^{n} - \mathbf{b}_{i+1,j}^{n} - \mathbf{b}_{i,j}^{n}\right] \\ & + \tfrac{\Delta t}{8}\left[\mathbf{s}_{i+1,j+1}^{n} + \mathbf{s}_{i,j+1}^{n} + \mathbf{s}_{i+1,j}^{n} + \mathbf{s}_{i,j}^{n}\right] ,\end{aligned} \tag{B.7}$$

with:

$$\lambda_{[\alpha]} = \tfrac{\Delta t}{\Delta\alpha} \qquad \text{and} \qquad \lambda_{[\beta]} = \tfrac{\Delta t}{\Delta\beta} . \tag{B.8}$$

The second step, on the other hand, utilizes the intermediate data $\mathbf{w}^{n+1/2}$ in order to calculate:[4]

$$\begin{aligned}\mathbf{w}_{i,j}^{n+1} = \mathbf{w}_{i,j}^{n} & + \tfrac{\lambda_{[\alpha]}}{2}\left[\mathbf{a}_{i+1/2,j+1/2}^{n+1/2} - \mathbf{a}_{i-1/2,j+1/2}^{n+1/2} + \mathbf{a}_{i+1/2,j-1/2}^{n+1/2} - \mathbf{a}_{i-1/2,j-1/2}^{n+1/2}\right] \\ & + \tfrac{\lambda_{[\beta]}}{2}\left[\mathbf{b}_{i+1/2,j+1/2}^{n+1/2} + \mathbf{b}_{i-1/2,j+1/2}^{n+1/2} - \mathbf{b}_{i+1/2,j-1/2}^{n+1/2} - \mathbf{b}_{i-1/2,j-1/2}^{n+1/2}\right] \\ & + \tfrac{\Delta t}{4}\left[\mathbf{s}_{i+1/2,j+1/2}^{n+1/2} + \mathbf{s}_{i-1/2,j+1/2}^{n+1/2} + \mathbf{s}_{i+1/2,j-1/2}^{n+1/2} + \mathbf{s}_{i-1/2,j-1/2}^{n+1/2}\right] .\end{aligned} \tag{B.9}$$

This two-step scheme is in fact consistent and of second-order accuracy. The stability condition reads:

$$c\lambda \leq 1 \qquad \text{with} \qquad \lambda = \max\left(\lambda_{[\alpha]}, \lambda_{[\beta]}\right) , \tag{B.10}$$

with $c$ denoting the absolute maximum of the propagation speeds which in turn is calculated by determining the absolute maximum of the eigenvalues of:

$$A_{[\alpha]}\cos(\phi) + B_{[\beta]}\sin(\phi) \qquad \text{with} \qquad |\phi| \leq \pi . \tag{B.11}$$

[3] The initial value problem is set up by choosing appropriate initial conditions at a certain initial time $t_0$.

[4] Note that it is necessary to choose appropriate numerical boundary conditions at the edge of the domain in order to solve the initial value problem on a finite grid $(\alpha_i, \beta_j)$.

# Acknowledgments

It is finally about time to express my gratitude to those who advised and supported me throughout the last years:

First and foremost I would like to thank my advisor Reinhard Alkofer for giving me the opportunity to write this thesis. I am very grateful for his advice and support during the last years. Furthermore, I am grateful to Holger Gies for his warm hospitality during my stay in Jena as well as the fruitful and enlightening discussions.

Moreover, I am indebted to my remaining collaborators: Gerald Dunne actually brought up the idea of investigating the Schwinger effect in a pulsed electric field with sub-cycle structure. Markus Orthaber was good company during his time as diploma student. Eventually, it was great fun to study the Wigner formalism on the light front in company with Anton Ilderton, Mattias Marklund and Jens Zamanian.

Additionally, I would like to show my gratitude to Bernd Thaller and Herbert Egger who helped me out when mathematics tried to play tricks on me. Furthermore I would like to thank Jens Braun, Antonio DiPiazza, Babette Döbrich, Julian Grond, Markus Huber, Mario Mitter, Carsten Müller and Selym Villalba-Chávez for each and every discussion. Finally, I thank all members of the doctoral program 'Hadrons in vacuum, nuclei and stars' for being good company during the last years.

I am indebted to the funding commission of the Austrian Academy of Sciences for granting me a DOC fellowship for the period of two years. Moreover, I thank the speaker and the principal investigators of the doctoral program 'Hadrons in vacuum, nuclei and stars' for covering my travel expenses during that time.

Finally, it is an honor for me to say 'thank you' to my parents for their ongoing support and encouragement: Without you I would not be where I am today. Many thanks also to my best friends Christof Brunner, Thomas Eichmann, Patrick Freidl, Manfred Gruber, Rupert Kügerl and Markus Scheer: It is always a pleasure to spend time with you.

Last but not least I owe my deepest gratitude to my partner Cornelia Müller for her love and understanding: There is nothing like coming home and seeing the smile on your face. I love you.